\documentclass{statsoc}
\usepackage[a4paper]{geometry}
\usepackage{amsmath, amssymb,mathtools}
\usepackage{amsfonts, multirow, epsfig, floatrow}
\usepackage{graphicx, pdflscape, verbatim, enumerate, colortbl, setspace}
\usepackage{setspace, color,bm}
\usepackage[normalem]{ulem}
\usepackage{cite}
\usepackage{multirow}
\usepackage{booktabs,array}
\usepackage{hyperref}
\usepackage{float}
\usepackage{caption}
\usepackage{subcaption}
\usepackage{tikz}
\usepackage{algorithm, algorithmicx, algpseudocode}
\let\emptyset\varnothing

\newtheorem{Proposition}{Proposition}
\newtheorem{Condition}{Condition}
\newtheorem{Lemma}{Lemma}
\newtheorem{Definition}{Definition}
\newtheorem{Theorem}{Theorem}
\newtheorem{Example}{Example} 
\newtheorem{Remark}{Remark}

\let\emptyset\varnothing

\newcommand*{\tabindent}{ \hspace{3mm}}
\DeclareMathOperator*{\argmax}{arg\,max}

\newcommand{\yes}{$\checkmark$}
\newcommand{\no}{ $\text{\sffamily X}$}
\newcommand{\TSHT}{\texttt{TSHT}}
\newcommand{\CIIV}{\texttt{CIIV}}
\newcommand{\VTSHT}{\widehat{\mathcal{V}}^{\TSHT}}
\newcommand{\VCIIV}{\widehat{\mathcal{V}}^{\CIIV}}

\newcommand{\Cov}{{\bf \rm Cov}}

\newcommand{\Gammab}{\Gammap}
\newcommand{\gammab}{\gammap}
\newcommand{\pib}{\pip}

\newcommand{\pz}{p_{\rm z}}
\newcommand{\px}{p_{\rm x}}

\newcommand{\piesta}{\widehat{{\pi}}^{[j]}}

\newcommand{\R}{\mathbb{R}}
\newcommand{\E}{{\mathbf{E}}}

\newcommand{\cip}{\overset{p}{\to}}
\newcommand{\cid}{\overset{d}{\to}}

\newcommand{\betatemp}{\widehat{\beta}^{[j]}}
\newcommand{\gammap}{\gamma^*}

\newcommand{\pip}{\pi^*}
\newcommand{\betap}{\beta^*}
\newcommand{\Gammap}{\Gamma^*}

\newcommand{\V}{{\rm \bf V}}
\newcommand{\C}{{\rm \bf C}}
\newcommand{\m}{[m]}
\newcommand{\bz}{{\bf z}}
\newcommand{\Shat}{\widehat{\mathcal{S}}}
\newcommand{\Vhat}{\widehat{\mathcal{V}}}
\newcommand{\PP}{\mathbb{P}}
\newcommand{\RR}{\mathbf{R}}
\newcommand{\CIsearchdef}{{\rm CI}^{\rm sear}}
\newcommand{\CIsearch}{\widehat{\rm CI}^{\rm sear}}
\newcommand{\CIsample}{{\rm CI}^{\rm samp}}
\newcommand{\CIsamplenull}{{\rm CI}_0^{\rm samp}}
\newcommand{\err}{ {\rm err}}

\newcommand{\Data}{ \mathcal{O}}
\newcommand{\Diff}{\mathbf{T}}
\newcommand{\sep}{{\rm sep}(n)}
\title[Uniform Inference with Invalid Instruments]{Causal Inference with Invalid Instruments: Post-selection Problems and A Solution Using Searching and Sampling}

\author[Guo]{Zijian Guo}
\address{Rutgers University, Piscataway, USA}

\begin{document}

\maketitle 

\begin{abstract}
Instrumental variable methods are among the most commonly used causal inference approaches to deal with unmeasured confounders in observational studies. The presence of invalid instruments is the primary concern for practical applications, and a fast-growing area of research is inference for the causal effect with possibly invalid instruments. This paper illustrates that the existing confidence intervals may undercover when the valid and invalid instruments are hard to separate in a data-dependent way. To address this, we construct uniformly valid confidence intervals that are robust to the mistakes in separating valid and invalid instruments. We propose to search for a range of treatment effect values that lead to sufficiently many valid instruments. We further devise a novel sampling method, which, together with searching, leads to a more precise confidence interval. Our proposed searching and sampling confidence intervals are uniformly valid and achieve the parametric length under the finite-sample majority and plurality rules. We apply our proposal to examine the effect of education on earnings. The proposed method is implemented in the R package \texttt{RobustIV} available from CRAN.
\end{abstract}

\section{Introduction}
Unmeasured confounders are a major concern for causal inference with observational studies. The instrumental variable (IV) method is one of the most commonly used causal inference approaches to deal with unmeasured confounders. The IVs are required to satisfy three identification conditions: conditioning on the baseline covariates,
\vspace{-1mm}
\begin{enumerate}
\item[(A1)] the IVs are associated with the treatment;
\vspace{-1.5mm}
\item[(A2)] the IVs are independent with the unmeasured confounders;
\vspace{-1.5mm}
\item[(A3)] the IVs have no direct effect on the outcome.
\vspace{-1mm}
\end{enumerate}
The main challenge of IV-based methods is identifying instruments satisfying (A1), (A2), and (A3) simultaneously.
Assumptions (A2) and (A3) are crucial for identifying the causal effect as they assume that the IVs can only affect the outcome through the treatment. 
However, assumptions (A2) and (A3) may be violated in applications and cannot even be tested in a data-dependent way. We define an IV as ``invalid" if it violates assumptions (A2) or (A3). If an invalid IV is mistakenly taken as valid, it generally leads to a biased estimator of the causal effect. A fast-growing literature is to conduct causal inference with possibly invalid IVs \citep[e.g.]{bowden2015mendelian,Bowden16,Kole15,Kang16,guo2018confidence,windmeijer2019use,tchetgen2017genius,kang2020two,fan2020endogenous}. Many of these works are motivated by Mendelian Randomization studies using genetic variants as IVs \citep{Dylan17}. The adopted genetic variants can be invalid since they may affect both treatment and outcome due to the pleiotropy effect \citep{Davey03}.

{ The current paper focuses on the linear outcome model with heteroscedastic errors and multiple possibly invalid IVs. Under this multiple IV framework, the effect identification requires extra conditions, such as the majority rule 
\citep{Kang16} and plurality rule \citep{guo2018confidence}, assuming that a sufficient proportion of  IVs are valid, but the validity of any IV is unknown a priori. The existing works \citep[e.g.]{Kang16,guo2018confidence,windmeijer2019use,windmeijer2019confidence} leveraged the majority and plurality rules to select valid IVs in a data-dependent way. The selected valid IVs were used for the following-up causal inference, with the invalid IVs being included as the baseline covariates. However, there are chances that we make mistakes in separating valid and invalid IVs.  Certain invalid IVs can be hard to detect in applications with the given amount of data. We refer to such invalid IVs as ``locally invalid IVs" and provide a formal definition in the following Definition \ref{def: locally invalid}.  In Section \ref{sec: post-selection}, we demonstrate that, when there exist locally invalid IVs, the existing inference methods $\TSHT$ \citep{guo2018confidence} and $\CIIV$ \citep{windmeijer2019confidence} may produce unreliable confidence intervals.

 The current paper proposes uniformly valid confidence intervals (CIs) robust to IV selection error. To better accommodate finite-sample inferential properties, we introduce the finite-sample majority and plurality rule in the following Conditions \ref{cond: majority-finite} and \ref{cond: plurality-finite}, respectively. We start with the finite-sample majority rule and explain the \textit{searching} idea under this setting. For every value of the treatment effect, we implement a hard thresholding step to decide which candidate IVs are valid. We propose to search for a range of treatment effect values such that the majority of candidate IVs can be taken as valid. We further propose a novel \textit{sampling} method to improve the precision of the searching CI. For the plurality rule setting, we first construct an initial estimator $\Vhat$ of the set of valid IVs and then apply the searching and sampling method over $\Vhat$. Our proposed searching CI works even if $\Vhat$ does not correctly recover the set of valid IVs. 

Our proposed searching and sampling CIs are shown to achieve the desired coverage under the finite-sample majority or plurality rule. The CIs are uniformly valid in the sense that the coverage is guaranteed even in the presence of locally invalid IVs. We also establish that the searching and sampling CIs achieve the $1/\sqrt{n}$ length. The proposed CIs are computationally efficient as the searching method searches over one-dimension space, and we only resample the reduced-form estimators instead of the entire data. 

We discuss other related works on invalid IVs in the following. \citet{kang2020two} proposed the \texttt{union} CI, which takes a union of intervals being constructed by a given number of valid IVs and passing the Sargan test \citep{sargan1958estimation}. The \texttt{union} CI requires an upper bound for the number of invalid IVs, while our proposed CI does not rely on such information. Our proposed searching and sampling CIs are typically much shorter and computationally more efficient than the \texttt{union} CI.}  Different identifiability conditions have been proposed to identify the causal effect when the  IV assumptions (A2) and (A3) fail to hold. \citet{bowden2015mendelian} and \citet{Kole15} assumed that the IVs' direct effect on the outcome and the IVs' association with the treatment are nearly orthogonal. 
In addition, there has been progress in identifying the treatment effect when all IVs are invalid; for instance, \citet{lewbel2012using,tchetgen2017genius,liu2020mendelian} leveraged heteroscedastic covariance of regression errors, while \citet{guo2022two} relied on identifying nonlinear treatment models with machine learning methods. { \citet{goh2022causal} emphasized the importance of accounting for the uncertainty of choosing valid IVs by the penalized methods \citep{Kang16} and proposed a Bayesian approach to construct a credible interval with possibly invalid IVs. However, no theoretical justification exists for this credible interval being a valid CI.} In Mendelian Randomization studies, much progress has been made in inference with summary statistics, which is not the main focus of the current paper; see \citet{bowden2015mendelian,Bowden16,zhao2020statistical} for examples.
In the GMM setting,
\citet{liao2013adaptive,cheng2015select,caner2018adaptive} leveraged a set of pre-specified valid moment conditions for the model identification and further tested the validity of another set of moment conditions. The current paper is entirely different in the sense that there is no prior knowledge of the validity of any given IV. 

The construction of uniformly valid CIs after model selection is a major focus in statistics under the name of post-selection inference. Many methods \citep{berk2013valid,lee2016exact,leeb2005model,zhang2014confidence,javanmard2014confidence,van2014asymptotically,chernozhukov2015post,cai2017confidence, xie2022repro} have been proposed, and the focus is on (but not limited to) inference for regression coefficients after some variables or sub-models are selected. This paper considers a different problem, post-selection inference for causal effect with possibly invalid instruments. To our best knowledge, this problem has not been carefully investigated in the post-selection inference literature. Furthermore, our proposed sampling method differs from other existing post-selection inference methods.

\noindent {\bf Notations.} For a set $S$ and a vector $x\in \R^{p}$, $S^{c}$ denotes the complement of $S$, $\left|S\right|$ denotes the cardinality of $S$, and $x_{S}$ is the sub-vector of $x$ with indices in $S$. For sets $B\subset A$, we define $A\backslash B=A\cap B^{c}$. The $\ell_q$ norm of a vector $x$ is defined as $\|x\|_{q}=\left(\sum_{l=1}^{p}|x_l|^q\right)^{\frac{1}{q}}$ for $q \geq 0$ with $\|x\|_0=\left|\{1\leq l\leq p: x_l \neq 0\}\right|$ and $\|x\|_{\infty}=\max_{1\leq l \leq p}|x_l|$. We use ${\bf 0}_{q}$ and ${\bf 1}_{q}$ to denote the $q$-dimension vector with all entries equal to $0$ and $1$, respectively. For a matrix $X$, $X_{i \cdot},$ $X_{\cdot j}$ and $X_{ij}$ are used to denote its $i$-th row, $j$-th column and $(i,j)$ entry, respectively. 
For a sequence of random variables $X_n$ indexed by $n$, we use $X_{n} \cid X$ to denote that $X_n$ converges to $X$ in distribution. We use $c$ and $C$ to denote generic positive constants that may vary from place to place.
For two positive sequences $a_n$ and $b_n$,  $a_n \lesssim b_n$ means that $\exists C > 0$ such that $a_n \leq C b_n$ for all $n$;
$a_n \asymp b_n $ if $a_n \lesssim b_n$ and $b_n \lesssim a_n$, and $a_n \ll b_n$ if $\limsup_{n\rightarrow\infty} {a_n}/{b_n}=0$. For a matrix $A$, we use $\|A\|_2$ and $\|A\|_{\infty}$ to denote its spectral and element-wise maximum norm, respectively. For a symmetric matrix $A$, we use $\lambda_{\max}(A)$ and $\lambda_{\min}(A)$ to denote its maximum and minimum eigenvalues, respectively.

\section{Models and Reduced-form Estimators}
\label{sec: model}
We consider the i.i.d. data $\{Y_i, D_i, X_{i\cdot}, Z_{i\cdot}\}_{1\leq i\leq n}$, where $Y_i\in \R$, $D_i\in \R$ and $X_{i\cdot}\in \R^{\px}$ and $Z_{i\cdot}\in \R^{\pz}$ denote the outcome, the treatment, the baseline covariates, and candidate IVs, respectively. { We consider the following outcome model with possibly invalid IVs \citep{small2007sensitivity,Kang16}},
\begin{equation}
Y_i=D_i\betap+ Z_{i\cdot}^{\intercal}\pi^*+X_{i\cdot}^{\intercal}\phi^{*}+e_i \quad \text{with} \quad \E (e_i Z_{i\cdot})={\bf 0} \quad \text{and}\quad \E (e_i X_{i\cdot})={\bf 0},
\label{eq: outcome}
\end{equation}
where $\betap\in \R$ denotes the treatment effect, $\pi^*\in \R^{\pz}$, and $\phi^{*}\in \R^{\px}.$ We set the first element of $X_{i\cdot}$ as the constant $1$. If the IVs satisfy (A2) and (A3), this leads to $\pi^*={\bf 0}$ in  \eqref{eq: outcome}.  A non-zero vector $\pi^*$ indicates that the IVs violate the classical IV assumptions (A2) and (A3); see the following Figure \ref{fig: robust IV} for an illustration.

We consider the association model for the treatment $D_i$, 
\begin{equation}
D_i=Z_{i\cdot}^{\intercal}\gamma^*+X^{\intercal}_{i\cdot}\psi^*+\delta_i \quad \text{with}\quad \E (\delta_i Z_{i\cdot})={\bf 0} \quad \text{and}\quad \E (\delta_i X_{i\cdot})={\bf 0}.
\label{eq: treatment}
\end{equation}
The model \eqref{eq: treatment} can be viewed as the best linear approximation of $D_i$ by $Z_{i\cdot}$ and $X_{i\cdot}$ instead of a casual model. In \eqref{eq: treatment}, $\gamma^*_j\neq 0$ indicates that the $j$-th IV satisfies assumption (A1). Due to unmeasured confounders, $e_i$ and $\delta_i$ can be correlated, and  the treatment $D_i$ is endogeneous with $\E(D_i e_i)\neq 0$. 

Following \citet{Kang16}, we discuss the causal interpretation of the model \eqref{eq: outcome} and explain why $\pi^*\neq 0$ represents the violation of assumptions (A2) and (A3).  
For two treatment values $d, d'\in \R$ and two realizations of IVs ${\bz}, \bz'\in \R^{\pz}$, define the following potential outcome model:
\begin{equation*}
Y_i^{(d',{\bf z}')}-Y_i^{(d,{\bf z})}=(d'-d)\betap+({\bf z}'-{\bf z})^{\intercal}\kappa^{*} \quad \text{and}\quad \E(Y_i^{(0,{\bf 0})}\mid Z_{i\cdot}, X_{i\cdot})=Z_{i\cdot}^{\intercal}\eta^{*}+ X_{i\cdot}^{\intercal}\phi^{*}
\end{equation*}
where $\betap\in \R$ is the treatment effect, $\kappa^{*},\eta^*\in \R^{\pz}$ and $\phi^{*}\in \R^{\px}.$ As illustrated in Figure \ref{fig: robust IV}, $\kappa^*_j\neq 0$ indicates that the $j$-th candidate IV has a direct effect on the outcome, which violates the assumption (A3); $\eta_j\neq 0$ indicates that the $j$-th candidate IV is associated with the unmeasured confounder, which violates the assumption (A2). Under the consistency condition $Y_i=Y_i^{(D_i, Z_{i\cdot})},$ the above potential outcome model implies \eqref{eq: outcome} with the invalidity vector $\pi^{*}=\kappa^{*}+\eta^{*}$ and $e_i=Y_i^{(0,{\bf 0})}-\E(Y_i^{(0,{\bf 0})}\mid Z_{i\cdot}, X_{i\cdot}).$ 

\tikzstyle{format} = [draw, thin,minimum height=1.35em, minimum width=1.35cm]
\tikzstyle{format1} = [draw, thin,minimum height=1.35em, minimum width=2.35cm]
\tikzstyle{medium} = [draw, thin,minimum height=1.35em]

\begin{figure}[htp!]
\centering
\begin{tikzpicture}
   \node[format] at (9.5, 5) (treatment2) {\scriptsize Treatment $D_i$};
    \node[format, right of=treatment2, node distance=4.15cm] (outcome2) {\scriptsize  Outcome $Y_i$};
    \draw[->]   (treatment2) -- node[above] {\tiny Treatment effect} (outcome2);
\node[format, above of=outcome2, dashed, left of=outcome2, node distance=2.2cm](unmeasured2){\scriptsize Unmeasured confounder};
\draw[->] (unmeasured2) edge[] node {} (treatment2);
\draw[->] (unmeasured2) edge[]  node{} (outcome2);
\node[format,left of=treatment2, node distance=2.6cm] (IV) {\scriptsize $Z_{i\cdot}$};
\draw[->] (IV) edge node[swap,above]{\small $\gamma^*$}(treatment2);
\draw[->] (IV) edge[bend right=25] node [swap, below=1mm]{\small $\kappa^* \neq 0$ }(outcome2);
\draw[->] (IV) edge[bend left=20] node [swap, above=2mm]{\small $\eta^* \neq  0$}(unmeasured2);
\end{tikzpicture}
\caption{Illustration of (A2) and (A3) being violated in the model \eqref{eq: outcome} with $\pi^{*}=\kappa^{*}+\eta^{*}$.}
\label{fig: robust IV}
\end{figure}

We define the set $\mathcal{S}$ of relevant instruments and the set $\mathcal{V}$ of valid instruments as, 
\begin{equation}
\mathcal{S}=\{1\leq j\leq \pz: \gammap_j\neq 0\} \quad \text{and}\quad \mathcal{V}=\{j\in \mathcal{S}: \pip_j=0\}.
\label{eq: relevant valid}
\end{equation}
The IVs belonging to $\mathcal{S}$ satisfy the IV assumption {\rm (A1)}. The set $\mathcal{V}$ is a subset of $\mathcal{S}$ and the IVs belonging to $\mathcal{V}$ satisfy the classical IV assumptions {\rm (A1)}-{\rm (A3)} simultaneously.

We now review the identification strategy under models \eqref{eq: outcome} and \eqref{eq: treatment}.
We plug in the treatment model \eqref{eq: treatment} into the outcome model \eqref{eq: outcome} and  obtain the reduced-form model,
\begin{equation}
\begin{aligned}
Y_i&=& Z_{i\cdot}^{\intercal}\Gammap+X^{\intercal}_{i\cdot}\Psi^*+\epsilon_i  \quad \text{with}\quad  \E (Z_{i\cdot} \epsilon_i)={\bf 0},\;\E (X_{i\cdot} \epsilon_i)={\bf 0},\\
D_i&=&Z_{i\cdot}^{\intercal}\gamma^*+X^{\intercal}_{i\cdot}\psi^*+\delta_i \quad \text{with}\quad \E (Z_{i\cdot}\delta_i)={\bf 0},\;\E (X_{i\cdot} \delta_i)={\bf 0},\\
\end{aligned}
\label{eq: reduced-form}
\end{equation}
where $\Gammap=\betap \gammap+\pip \in \R^{\pz}$, $\Psi^*=\betap\psi^*+\phi^* \in \R^{\px},$ and $\epsilon_i=\betap \delta_i+e_i.$  We shall devise our methods under the model \eqref{eq: reduced-form}, which is induced by the models \eqref{eq: outcome} and  \eqref{eq: treatment}. 

Since $Z_{i\cdot}$ and $X_{i\cdot}$ are uncorrelated with $\epsilon_i$ and $\delta_i,$ we can identify the reduced-form parameters $\Gammap $ and $\gammap$ in \eqref{eq: reduced-form}. However, since $\pip\neq {\bf 0},$ the identification of $\betap$ through solving the equation $\Gammap=\betap \gammap+\pip \in \R^{\pz}$ requires extra assumptions. { \citet{Kang16,Bowden16} proposed the following majority rule to identify $\beta$ for $\pip\neq {\bf 0}$.}

\begin{Condition}[Population Majority Rule]
More than half of the relevant IVs are valid; that is, $
|\mathcal{V}|>|\mathcal{S}|/2,$
where $\mathcal{V}$ and $\mathcal{S}$ are defined in  \eqref{eq: relevant valid}.
\label{cond: rule 1}
\end{Condition}

Condition \ref{cond: rule 1} requires that more than half of the relevant IVs are valid but does not directly require the knowledge of $\mathcal{V}$. For the $j$-th IV, we may identify the effect as $\beta^{[j]}=\Gammap_j/\gammap_j.$ If the $j$-th IV is valid, $\beta^{[j]}=\betap$; otherwise, $\beta^{[j]}\neq\betap.$ Under Condition \ref{cond: rule 1}, $\betap$ can be identified using the majority of $\{\beta^{[j]}\}_{j\in \mathcal{S}}.$ { \citet{guo2018confidence,hartwig2017robust} proposed the following plurality rule as a weaker identification condition.} 

\begin{Condition}[Population Plurality Rule] The number of valid IVs is larger than the number of invalid IVs with any given invalidity level $\nu\neq 0$, that is, 
\begin{equation*}
|\mathcal{V}|>\max_{\nu \neq 0} \left|\mathcal{I}_{\nu}\right| \quad \text{with}\quad \mathcal{I}_{\nu}=\left\{j\in \mathcal{S}: {\pip_j}/{\gammap_j}=\nu\right\},
\label{eq: Plurality}
\end{equation*}
where the set $\mathcal{V}$ of valid IVs is defined in  \eqref{eq: relevant valid}.
\label{cond: rule 2}
\end{Condition}
 The term ${\pip_j}/{\gammap_j}$ represents the invalidity level: 
${\pip_j}/{\gammap_j}\neq 0$ indicates that the $j$-th IV violates assumptions (A2) and (A3).  $\mathcal{I}_{\nu}$ denotes the set of all IVs with the same invalidity level $\nu.$ Note that $\mathcal{V}=\mathcal{I}_0.$
Under the plurality rule, $\betap$ can be identified using the largest cluster of $\{\beta^{[j]}\}_{j\in \mathcal{S}}.$
We refer to Conditions \ref{cond: rule 1} and \ref{cond: rule 2} as {\it population} identification conditions since they are used to identify $\betap$ with an infinite amount of data. We will propose finite-sample versions of Conditions \ref{cond: rule 1} and \ref{cond: rule 2} in the following Conditions \ref{cond: majority-finite} and \ref{cond: plurality-finite}, respectively.

We now present the data-dependent estimators of $\gammap$ and $\Gammap$ in the model \eqref{eq: reduced-form}. 
Define $p=\px+\pz,$ $W=(Z,X)\in \R^{n\times p},$ and $\widehat{\Sigma}=\frac{1}{n}\sum_{i=1}^{n}W_{i\cdot}(W_{i\cdot})^{\intercal}.$ We estimate $\gammap$ and $\Gammap$ by  the Ordinary Least Squares (OLS) estimators, defined as 
\begin{equation}
(\widehat{\Gamma}^{\intercal}, \widehat{\Psi}^{\intercal})^{\intercal}= (W^{\intercal} W)^{-1} W^{\intercal} Y \quad \text{and}\quad  (\widehat{\gamma}^{\intercal}, \widehat{\psi}^{\intercal})^{\intercal}= (W^{\intercal} W)^{-1} W^{\intercal} D.
\label{eq: OLS}
\end{equation}
Under regularity conditions, { as $n\rightarrow \infty$}, the OLS estimators satisfy
\begin{equation} 
\sqrt{n}\begin{pmatrix}\widehat{\Gamma}-\Gammap\\ \widehat{\gamma}-\gammap\end{pmatrix}
\cid N\left({\bf 0},\Cov\right) \quad \text{with}\quad \Cov=\begin{pmatrix}\V^{\Gamma}& {\C}\\ {\C}^{\intercal}& \V^{\gamma}\end{pmatrix},
\label{eq: reduced-form limiting} 
\end{equation}
where $\V^{\Gamma}\in \R^{\pz\times \pz}$, $\V^{\gamma}\in \R^{\pz\times \pz}$, and ${\C}\in \R^{\pz\times \pz}$ are explicitly defined in Lemma 1 in the supplement. Define the residues $\widehat{\epsilon}_i=Y_i-Z_{i\cdot}^{\intercal}\widehat{\Gamma}-X_{i\cdot}^{\intercal}\widehat{\Psi}$ and $\widehat{\delta}_i=D_i-Z_{i\cdot}^{\intercal}\widehat{\gamma}-X_{i\cdot}^{\intercal}\widehat{\psi}$ for $1\leq i\leq n.$
We estimate $\V^{\Gamma}$, $\V^{\gamma}$, and ${\C}$ in \eqref{eq: reduced-form limiting} by 
\begin{equation}
\begin{aligned}
\widehat{\V}^{\Gamma}&= \left[\widehat{\Sigma}^{-1}\left(\frac{1}{n}\sum_{i=1}^{n}\widehat{\epsilon}_i^2 W_{i\cdot} W_{i\cdot}^{\intercal}\right)\widehat{\Sigma}^{-1}\right]_{1:\pz,1:\pz}, \widehat{\V}^{\gamma}=\left[\widehat{\Sigma}^{-1}\left(\frac{1}{n}\sum_{i=1}^{n}\widehat{\delta}_i^2 W_{i\cdot} W_{i\cdot}^{\intercal}\right)\widehat{\Sigma}^{-1}\right]_{1:\pz,1:\pz},\\
\widehat{\C}&=\left[\widehat{\Sigma}^{-1}\left(\frac{1}{n}\sum_{i=1}^{n}\widehat{\epsilon}_i\widehat{\delta}_i W_{i\cdot} W_{i\cdot}^{\intercal}\right)\widehat{\Sigma}^{-1}\right]_{1:\pz,1:\pz},
\end{aligned}
\label{eq: hetero var}
\end{equation}
{ where for a matrix ${\bf A}\in \R^{p\times p}$, we use ${\bf A}_{1:\pz,1:\pz}$ to denote the $p_z\times p_z$ submatrix containing the first $p_z$ rows and columns of ${\bf A}$.}
{The variance estimators in \eqref{eq: hetero var} are robust to the heteroskedastic errors and referred to as the sandwich estimators \citep{eicker1967limit,huber1967under};
see Chapter 4.2.3 of \citet{wooldridge2010econometric} for details.} We use the OLS estimators in \eqref{eq: OLS} and the sandwich estimators in \eqref{eq: hetero var} as a prototype to discuss our proposed methods. Our proposed methods are effective for any reduced-form estimators satisfying \eqref{eq: reduced-form limiting}. In Section A.3 in the supplement, we consider the high-dimensional setting with $p>n$ and construct the debiased Lasso estimators $\widehat{\Gamma}$ and $\widehat{\gamma}$ satisfying \eqref{eq: reduced-form limiting}.

\section{IV Selection Errors and Non-uniform Inference}
\label{sec: post-selection}
In this section, we demonstrate that even if the majority rule (Condition \ref{cond: rule 1}) or plurality rule (Condition \ref{cond: rule 2}) holds, the CIs by $\TSHT$ \citep{guo2018confidence} and $\CIIV$ \citep{windmeijer2019confidence} may be unreliable. The main idea of \citet{guo2018confidence} and \citet{windmeijer2019confidence} is to estimate $\mathcal{V}$ by a set estimator $\Vhat$ and then identify $\betap$ through the following expression or its weighted version,
\begin{equation}
\beta(\Vhat)=\sum_{j\in \Vhat}\Gammap_{j}\gammap_{j} /\sum_{j\in \Vhat}(\gammap_{j})^2 .
\label{eq: identification}
\end{equation}
When there are no selection errors (i.e. $\Vhat=\mathcal{V}$), we have $\beta(\Vhat)=\betap$. The validity of the CIs by $\TSHT$  and $\CIIV$ requires $\Vhat$ to recover $\mathcal{V}$ correctly. However, there are chances to make mistakes in estimating $\mathcal{V}$ in finite samples, and the CIs by \texttt{TSHT} and \texttt{CIIV} are unreliable when invalid IVs are included in $\Vhat$. The IV selection error leads to a bias in identifying $\betap$ with $\beta(\Vhat)$. { Even if $\Vhat$ used in \eqref{eq: identification} is selected in a data-dependent way, the target causal effect $\betap$ is fixed, which is a main difference from the post-selection inference literature \citep[e.g.]{berk2013valid,lee2016exact,leeb2005model}. We provide more detailed discussions in Section A.1 in the supplement.} 

In the following, we define ``locally invalid IVs" as invalid IVs that are hard to be separated from valid IVs with a given amount of data. 
For $j,k\in \mathcal{S}$, define 
\begin{equation}
{\Diff}_{j,k}\coloneqq \min\{{\Diff}^0_{j,k},{\Diff}^0_{k,j}\}\quad \text{with}\quad {\Diff}^0_{j,k}=\sqrt{\frac{1}{n}\left(\RR^{[j]}_{k,k}/[\gammap_k]^2+\RR^{[j]}_{j,j}/[\gammap_j]^2-2\RR^{[j]}_{j,k}/[\gammap_k\gammap_j]\right)},
\label{eq: exact difference}
\end{equation}
where $\RR^{[j]}=\V^{\Gamma}+\left(\beta^{[j]}\right)^2\V^{\gamma}-2\beta^{[j]}\C$ and $\beta^{[j]}=\Gammap_j/\gammap_j.$ { As shown in the following Proposition \ref{prop: voting}, if the absolute difference between $\pip_j/\gammap_j$ and $\pip_k/\gammap_k$ for $j,k\in \mathcal{S}$ is above $2\sqrt{\log n} \cdot{\Diff}_{j,k}$, then we can tell that the $j$-th and $k$-th IVs have different invalidity levels. Particularly, the term $\gammap_k {\Diff}^0_{j,k}$ represents the standard error of estimating $\pip_k$ by $\widehat{\Gamma}_k-\widehat{\gamma}_k\widehat{\beta}^{[j]}$ where $\widehat{\beta}^{[j]}={\widehat{\Gamma}_j}/{\widehat{\gamma}_j}$ denotes the causal effect estimator by assuming the $j$-th IV to be valid. By symmetry, $\gammap_j {\Diff}^0_{k,j}$ denotes the standard error of estimating $\pip_j$ by $\widehat{\Gamma}_j-\widehat{\gamma}_j\widehat{\beta}^{[k]}.$ } We now formally define locally invalid IVs. 

\begin{Definition}[Locally invalid IV] For $j \in \mathcal{S}$, the $j$-th IV is locally invalid if \begin{equation*}
0<|{\pip_j}/{\gammap_j}|< s_j(n)\quad \text{with}\quad s_j(n)\coloneqq 2\sqrt{\log n} \cdot \max_{k\in \mathcal{V}}|\Diff_{j,k}|.
\end{equation*}
where $\Diff_{j,k}$ is defined in \eqref{eq: exact difference}.
\label{def: locally invalid}
\end{Definition}

The definition of locally invalid IVs depends on the invalidity level ${\pip_j}/{\gammap_j}$ and the separation level $s_j(n)$, which stands for the uncertainty level in separating the $j$-th IV and valid IVs. We show in the following Proposition \ref{prop: voting} that the $j$-th IV, if invalid, can be separated from valid IVs if $|{\pip_j}/{\gammap_j}|\geq s_j(n).$ The separation level $s_j(n)$ is of the order $\sqrt{\log n/n}$ if all of $\{\gammap_j\}_{j\in \mathcal{S}}$ are constants. The large sample size enhances the power of detecting invalid IVs. For a sufficiently large $n$, the set of locally invalid IVs becomes empty since $s_j(n)\rightarrow 0$. Definition \ref{def: locally invalid} is related to the local violation of IV exogeneity assumptions and valid moment conditions studied in \citet{caner2018adaptive,berkowitz2012validity,guggenberger2012asymptotic,hahn2005estimation}.

The theoretical results of $\TSHT$
\citep{guo2018confidence} or $\CIIV$ \citep{windmeijer2019confidence} essentially assume that there are no locally invalid IVs; see Assumption 8 in \citet{guo2018confidence}. However, the absence of locally invalid IVs may not properly accommodate real data analysis with a finite sample. For a given sample size $n$, there are chances that an invalid IV with  $|{\pip_j}/{\gammap_j}|<s_j(n)$ is taken as valid by mistake. We now consider a numerical example and demonstrate that the coverage levels of CIs by $\TSHT$ and $\CIIV$ may be below the nominal level when locally invalid IVs exist. 

\vspace{-3mm}

\begin{Example}[Setting {\bf S2} in Section \ref{sec: simulation}]
For the models \eqref{eq: outcome} and \eqref{eq: treatment}, set 
 $\gammap=0.5 \cdot {\bf 1}_{10}$ and $\pip=({\bf 0}^{\intercal}_4,\tau/2,\tau/2,-1/3,-2/3,-1,-4/3)^{\intercal}$. The plurality rule is satisfied with $|\mathcal{V}|=4>\max_{v\neq 0}|\mathcal{I}_{v}|=2.$ We vary $\tau$ across $\{0.025,0.05,0.075,0.1,0.2,0.3,0.4,0.5\}.$ $s_j(n)$ for $j=5,6$ in Definition \ref{def: locally invalid} takes the value  $0.96$ for $n=500$ and $0.53$ for $n=2000.$
 \label{example 1}
\end{Example}
\vspace{-7mm}

\begin{figure}[H]
\includegraphics[scale=0.8]{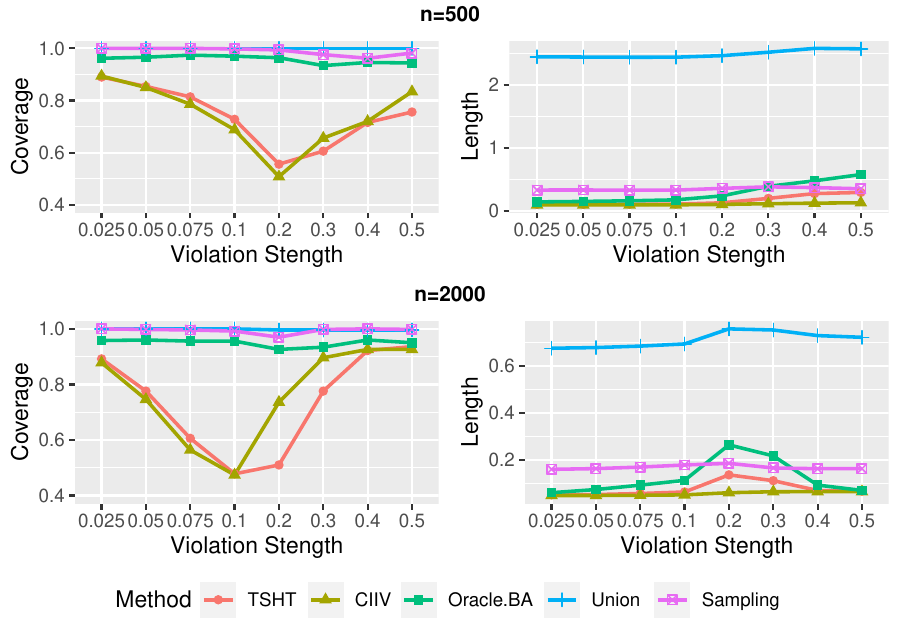}
\vspace{-2.5mm}
 \caption{\small Empirical coverage and average lengths for {\bf Example 1} with $\tau$ varying across $\{0.025,0.05,0.075,0.1,0.2,0.3,0.4,0.5\}$. $\TSHT$,  $\CIIV$, and \texttt{Union} stand for the CIs by \citet{guo2018confidence}, \citet{windmeijer2019confidence}, and \citet{kang2020two}, respectively. \texttt{Oracle BA} stands for the oracle bias-aware CI in \eqref{eq: bias aware}. \texttt{Sampling} represents our proposed sampling CI in Algorithm \ref{algo: USS-plurality}.}
 \label{fig: coverage}
\end{figure}

In this example, the fifth and sixth IVs (with a small $\tau$) are locally invalid IVs, which are hard to detect for $n=500, 2000.$ The empirical coverage is reported in Figure \ref{fig: coverage}. 
For a small $n$ and $\tau$, the estimated sets $\Vhat$ by $\TSHT$ and $\CIIV$ often contain locally invalid IVs, and the coverage levels of $\TSHT$ and $\CIIV$ are below $95\%$. There are nearly no locally invalid IVs for $n=2000$ and $\tau= 0.5$, and the CIs by $\TSHT$ and $\CIIV$ achieve the nominal level $95\%$. For $\tau=0.1,$ with $n$ increasing from $500$ to $2000$, the fifth and sixth IVs may still be included in $\Vhat$. Consequently, the bias due to the locally invalid IVs remains unchanged, and the empirical coverage levels of $\TSHT$ and $\CIIV$ decrease since the standard errors get smaller with a larger $n$. In contrast, our proposed sampling CI in the following Algorithm \ref{algo: USS-plurality} and the  \texttt{Union} interval in \citet{kang2020two} achieve the desired coverage at the expense of wider intervals. Our proposed sampling CIs are significantly shorter than the \texttt{Union} intervals.

We introduce an oracle bias-aware CI as the benchmark when IV selection errors exist. The oracle bias-aware CI serves as a better benchmark than the oracle CI assuming the knowledge of $\mathcal{V}$ since it accounts for the IV selection error. For the $\TSHT$ estimator $\widehat{\beta}$ and its standard error ${\rm SE}(\widehat{\beta})$, we assume $(\widehat{\beta}-\betap)/{\rm SE}(\widehat{\beta})\cid N(b,1)$ with $b$ denoting the asymptotic bias. Following (7) in \citet{armstrong2020bias}, we leverage the oracle knowledge of $|\E\widehat{\beta}-\betap|$ and form the oracle bias-aware CI as 
\begin{equation}
\left(\widehat{\beta}-\chi,\widehat{\beta}+\chi\right) \quad \text{with}\quad \chi=\widehat{\rm SE}(\widehat{\beta})\cdot \sqrt{{\rm cv}_{\alpha}(|\E\widehat{\beta}-\betap|^2/\widehat{\rm SE}^2(\widehat{\beta}))},
\label{eq: bias aware}
\end{equation}
where $\widehat{\rm SE}(\widehat{\beta})$ is the empirical standard error computed over 500 simulations and ${\rm cv}_{\alpha}(B^2)$ is the $1-\alpha$ quantile of the $\chi^2$ distribution with 1 degree of freedom and non-centrality parameter $B^2.$ In Figure \ref{fig: coverage}, the oracle bias-aware CI achieves the desired coverage, and the sampling CI has a comparable length to the oracle bias-aware CI. 

\vspace{-5mm}

\section{Searching and Sampling: Robust Inference Methods under Majority Rule}
\label{sec: majority}
In this section, we focus on the majority rule setting and generalize the proposed methods to the plurality rule in Section \ref{sec: plurality}. 
Our proposed procedure consists of two stages as an analogy to \texttt{TSLS}. We fit the treatment model in the first stage and select strong IVs; see Section \ref{sec: first stage}. In the second stage, we propose novel searching and sampling CIs in Sections \ref{sec: searching} to \ref{sec: sampling}, which are robust to the mistakes in separating valid and invalid IVs.
\vspace{-2.5mm}

\subsection{First-stage selection and the finite-sample majority rule}
\label{sec: first stage}
Following \citet{guo2018confidence}, we estimate $\mathcal{S}$ by applying the first-stage hard thresholding \citep{donoho1994ideal} to the reduced-form estimator $\widehat{\gamma}$ defined in \eqref{eq: OLS},
\begin{equation}
\widehat{\mathcal{S}}=\left\{1\leq j\leq \pz: |\widehat{\gamma}_j|\geq \sqrt{\log n} \cdot \sqrt{\widehat{\V}^{\gamma}_{jj}/n}\right\},
\label{eq: relevant est}
\end{equation}
with $\widehat{\V}^{\gamma}$ defined in \eqref{eq: hetero var}. The main purpose of \eqref{eq: relevant est} is to estimate the set of relevant IVs and screen out IVs weakly associated with the treatment. The term $\sqrt{\log n}$ is introduced to adjust for the multiplicity of testing $\pz$ hypothesis;  see more discussions in the following Remark \ref{rem: thresholding}.   The screening step in \eqref{eq: relevant est} guarantees the robustness of our proposal when some IVs are weakly associated with the treatment. With a high probability, our estimated set $\Shat$ in \eqref{eq: relevant est} belongs to $\mathcal{S}$ and contains the set $\mathcal{S}_{\rm str}$ of strongly relevant IVs, defined as
\begin{equation}
\mathcal{S}_{\rm str}=\left\{1\leq j\leq \pz: \left|\gamma^*_j\right|\geq 2\sqrt{\log n} \cdot \sqrt{{\V}^{\gamma}_{jj}/n}\right\} \quad \text{with}\;{\V}^{\gamma}\;\text{defined in \eqref{eq: reduced-form limiting}}.
\label{eq: str relevant}
\end{equation}
The set $\mathcal{S}_{\rm str}$ contains the $j$-th IV if its individual strength $|\gammap_j|$ is well above its estimation accuracy $\sqrt{{\V}^{\gamma}_{jj}/n}.$  The factor $2$ in \eqref{eq: str relevant} ensures that the individual IV strength is sufficiently large such that the $j$-th IV can be included in the set $\Shat$ defined in \eqref{eq: relevant est}; { see more detailed discussion in Section A.2 in the supplement.}

We now modify Condition \ref{cond: rule 1} to accommodate the finite-sample uncertainty.
\begin{Condition}{\rm \bf (Finite-sample Majority Rule)} More than half of the relevant IVs are strongly relevant and valid, that is, 
$|\mathcal{V}\cap\mathcal{S}_{\rm str}|> |\mathcal{S}|/2,$
where $\mathcal{S}$ and $\mathcal{V}$ are defined in \eqref{eq: relevant valid} and $\mathcal{S}_{\rm str}$ is defined in \eqref{eq: str relevant}.
\label{cond: majority-finite}
\end{Condition}

When the sample size $n$ is small, Condition \ref{cond: majority-finite} is slightly stronger than Condition \ref{cond: rule 1}. When $n\rightarrow \infty$ and the IV strengths $\{\gammap_j\}_{1\leq j\leq \pz}$ do not change with $n$, the set $\mathcal{S}_{\rm str}$ converges to $\mathcal{S}$ and Condition \ref{cond: majority-finite} is asymptotically the same as Condition \ref{cond: rule 1}.

\begin{Remark}\rm 
\label{rem: thresholding} {
There are other choices of the threshold level in \eqref{eq: relevant est}. We can replace $\sqrt{\log n}$ in \eqref{eq: relevant est} and \eqref{eq: str relevant} by any $f(n)$ satisfying $f(n)\rightarrow \infty$ with $n\rightarrow \infty$ and $f(n)> \sqrt{2 \log \pz}.$  \citet{guo2018confidence} used $f(n)=\sqrt{2.01 \log \max\{n,p_z\}},$ which is applicable in both low and  high dimensions. In Section B.1 in the supplement, we consider the settings with weak IVs and demonstrate that our proposed methods have nearly the same performance with both the thresholds in \eqref{eq: relevant est} and that in \citet{guo2018confidence}. }   
\end{Remark} 
 
 \vspace{-7mm}
\subsection{The searching confidence interval}
\label{sec: searching}
In the following, we restrict our attention to the estimated set $\Shat$ of relevant IVs and leverage Condition \ref{cond: majority-finite} to devise the searching CI for $\betap$. The main idea is to search for $\beta$ values that lead to the majority of the instruments being detected as valid. The searching idea is highly relevant to the Anderson-Rubin test \citep{anderson1949estimation} for the weak IV problem, which searches for a range of $\beta$ values leading to a sufficiently small $\chi^2$ test statistic; see the following Remark \ref{rem: AR} for more discussions. 

For any given $\beta\in \R,$ we apply the relation $\Gammap_{j}=\betap \gammap_{j}+\pip_j$ and construct the initial estimator of $\pip_j$ as $\widehat{\Gamma}_j-\beta\widehat{\gamma}_j$. The estimation error of $\widehat{\Gamma}_j-\beta\widehat{\gamma}_j$ is decomposed as,
\begin{equation}
\begin{aligned}
(\widehat{\Gamma}_j-\beta\widehat{\gamma}_j)-\pip_j
=\widehat{\Gamma}_j-\Gammap_j-\beta(\widehat{\gamma}_j-\gammap_j)+(\betap-\beta) \gammap_j.
\end{aligned}
\label{eq: error decomposition}
\end{equation}
When $\beta=\betap,$ the above estimation error is further simplied as $\widehat{\Gamma}_j-\Gammap_j-\beta(\widehat{\gamma}_j-\gammap_j).$ We quantify the uncertainty of $\{\widehat{\Gamma}_j-\Gammap_j-\beta(\widehat{\gamma}_j-\gammap_j)\}_{j \in \Shat}$ by the following union bound,
\begin{equation}
\liminf_{n\rightarrow\infty}\PP\left[\max_{j\in \Shat} \frac{|\widehat{\Gamma}_j-\Gammap_j-\beta(\widehat{\gamma}_j-\gammap_j)|}{\sqrt{(\widehat{\V}^{\Gamma}_{jj}+\beta^2\widehat{\V}^{\gamma}_{jj}-2\beta \widehat{\C}_{jj})/n}}\leq \Phi^{-1}\left(1-\frac{\alpha}{2|\Shat|}\right)\right]\geq 1- \alpha,
\label{eq: quantile}
\end{equation}
where $\alpha\in (0,1)$ and $\Phi^{-1}$ is the inverse CDF of the standard normal distribution.  By rescaling \eqref{eq: quantile}, we construct the following threshold for $\widehat{\Gamma}_j-\Gammap_j-\beta(\widehat{\gamma}_j-\gammap_j)$ with $j \in \Shat$,
\begin{equation}
 \widehat{\rho}_j(\beta)\coloneqq \Phi^{-1}\left(1-\frac{\alpha}{2|\Shat|}\right) \cdot \sqrt{(\widehat{\V}^{\Gamma}_{jj}+\beta^2\widehat{\V}^{\gamma}_{jj}-2\beta \widehat{\C}_{jj})/n}.
 \label{eq: key threshold}
\end{equation}
We further apply the hard thresholding to $\{\widehat{\Gamma}_j-\beta\widehat{\gamma}_j\}_{j\in \Shat}$ and estimate $\pip_j$ by 
\begin{equation}
\widehat{\pi}_j(\beta)=\left(\widehat{\Gamma}_{j}-\beta\widehat{\gamma}_j\right)\cdot \mathbf{1}\left(\left|\widehat{\Gamma}_{j}-\beta\widehat{\gamma}_j\right|\geq  \widehat{\rho}_j(\beta)\right) \quad \text{for} \quad j \in \Shat.
\label{eq: thresholding}
\end{equation} 
If $\beta=\betap,$ the hard thresholding in \eqref{eq: thresholding} guarantees $\widehat{\pi}_j(\betap)=0$ for $j\in \mathcal{V}$ with probability larger than $1-\alpha.$
For any $\beta\in \R$, we can construct the vector $\widehat{\pi}_{\Shat}(\beta)=(\widehat{\pi}_{j}(\beta))_{j\in \Shat}$ and calculate the number of non-zero entries in $\widehat{\pi}_{\Shat}(\beta)$, denoted as $\|\widehat{\pi}_{\Shat}(\beta)\|_0.$ Due to the hard thresholding, $\widehat{\pi}_{\Shat}(\beta)$ can be a sparse vector and $\|\widehat{\pi}_{\Shat}(\beta)\|_0$ stands for the number of invalid IVs corresponding to the given $\beta$ value. 

\begin{figure}[htp!]
\centering
\includegraphics[scale=0.55]{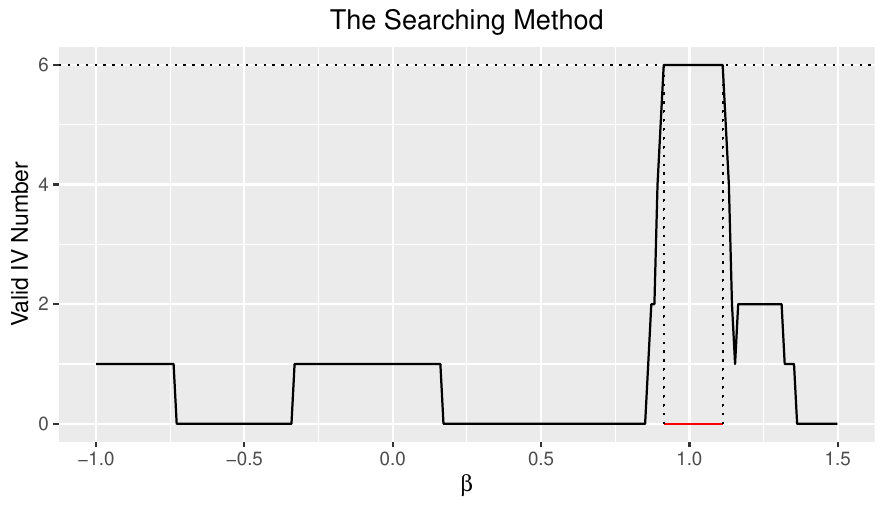}
\vspace{-3mm}
\caption{\small The x-axis plots a range of $\beta$ values, and the y-axis plots the number of valid IVs (i.e., $|\Shat|-\|\widehat{\pi}_{\Shat}(\beta)\|_0$) for every given $\beta$. The red interval $(0.914, 1.112)$ denotes $\CIsearchdef$ in \eqref{eq: searching CI def}. }
\label{fig: searching}
\end{figure}

We search for a range of $\beta$ such that the number of invalid IVs is below $|\Shat|/2$, 
\begin{equation}
\CIsearchdef \coloneqq \left\{\beta\in \R: \|\widehat{\pi}_{\Shat}(\beta)\|_0<{|\Shat|}/{2}\right\}.
\label{eq: searching CI def}
\end{equation}
The above searching CI is valid since Condition \ref{cond: majority-finite} implies that $\|\widehat{\pi}_{\Shat}(\betap)\|_0<{|\Shat|}/{2}$ with probability $1-\alpha$; see equation (56) in the supplement. We consider the following example and illustrate the construction of $\CIsearchdef$ in Figure \ref{fig: searching}.

\begin{Example}
Generate the models \eqref{eq: outcome} and \eqref{eq: treatment} with no baseline covariates, set $\betap=1,$ $n=2000,$
$\gammap_j=0.5\cdot {\bf 1}_{10}$ and  $\pip=({\bf 0}_6^{\intercal},0.05,0.05,-0.5,-1)^{\intercal}$.  In Figure \ref{fig: searching}, we plot $|\Shat|-\|\widehat{\pi}_{\Shat}(\beta)\|_0$ over different $\beta$ and the red interval $(0.914, 1.112)$ covers $\betap=1.$
\label{ex: illustration}
\end{Example}

\begin{Remark}\rm
{ The proposed searching idea is related to but different from the Anderson-Rubin test  \citep{anderson1949estimation} for the weak IV problem.} In \eqref{eq: searching CI def}, we use the sparsity as the test statistic and invert the sparsity level to determine the confidence region for $\betap$. Our inverted test statistics $\|\widehat{\pi}_{\Shat}(\beta)\|_0$ is a discrete function of $\beta$, and its limiting distribution is not standard, which is fundamentally different from the $\chi^2$ statistics used in AR test; see Figure \ref{fig: searching}. Due to the discreteness of the test statistics, the searching interval in \eqref{eq: searching CI def} can be a union of disjoint intervals.  
\label{rem: AR}
\end{Remark}

\vspace{-5mm}

\subsection{Efficient implementation of the searching CI}
\label{sec: num implemenation}
We propose a computationally efficient implementation of $\CIsearchdef$ in \eqref{eq: searching CI def}. To implement \eqref{eq: searching CI def}, we can  enumerate all values of $\beta$ and calculate $\|\widehat{\pi}_{\Shat}(\beta)\|_0.$ However, the enumeration method for constructing $\CIsearchdef$ can be time-consuming if there is a huge amount of $\beta$ values. To improve the computational efficiency, we construct an initial set $[L,U]$ such that $\PP(\betap\in [L,U])\rightarrow 1$ and construct a grid set with the grid size $n^{-a}$ for $a>1/2$. Then we discretize $[L,U]$ into the following grid set, 
\begin{equation}
\mathcal{B}=\{\beta_{1},\beta_{2},\cdots, \beta_{K}\}\quad \text{with}\;\beta_1=L,\;\beta_{K}=U,\;|\beta_{j+1}-\beta_j|=n^{-a} \; \text{for} \;1\leq j\leq K-2,
\label{eq: grid construction}
\end{equation} 
and  $|\beta_{K}-\beta_{K-1}|\leq n^{-a}.$ The reason for requiring $a>1/2$ is to ensure that the approximation error due to interval discretization is smaller than the parametric rate $n^{-1/2}$. { Importantly, the constructed confidence interval is almost invariant to the choices of $[L,U]$ and the grid size $n^{-a}$ as long as $[L,U]$ contains $\betap$ and $a>1/2.$} 

Throughout the paper, we choose the default grid size $n^{-0.6}$ and present the following default construction of the initial range $[L,U]$. For $j\in \Shat,$ we estimate $\betap$ by the ratio $\widehat{\Gamma}_j/\widehat{\gamma}_j$ and estimate its variance by 
$\widehat{\rm Var}\left(\widehat{\Gamma}_j/\widehat{\gamma}_j\right)= \frac{1}{n}\left({\widehat{\V}^\Gamma_{jj}}/{\widehat{\gamma}_j^2} + {\widehat{\V}^\gamma_{jj} \widehat{\Gamma}_j^2 }/{\widehat{\gamma}_j^4} - 2 {\widehat{\C}_{jj} \widehat{\Gamma}_j}/{\widehat{\gamma}_j^3}\right).$
We then construct $L$ and $U$ as
{\small
\begin{equation}
L=\min_{j\in \widehat{\mathcal{S}}} \left\{\widehat{\Gamma}_j/\widehat{\gamma}_j -\sqrt{\log n\cdot \widehat{\rm Var}\left(\widehat{\Gamma}_j/\widehat{\gamma}_j\right)}\right\},\;  U=\max_{j\in \widehat{\mathcal{S}}} \left\{\widehat{\Gamma}_j/\widehat{\gamma}_j+\sqrt{\log n\cdot\widehat{\rm Var}\left(\widehat{\Gamma}_j/\widehat{\gamma}_j\right)}\right\},
\label{eq: rough bound}
\end{equation}
}
where $\sqrt{\log n}$ is used to adjust for multiplicity. This initial range has been discussed in Section 3 of \citet{windmeijer2019confidence}. 


With the grid set defined in \eqref{eq: grid construction}, we modify \eqref{eq: searching CI def} and propose the following computationally efficient implementation, 
\begin{equation}
\CIsearch=\left[\min_{\{\beta\in \mathcal{B}: \|\widehat{\pi}_{\Shat}(\beta)\|_0<{|\Shat|}/{2}\}} \beta, \quad\max_{\{\beta\in \mathcal{B}: \|\widehat{\pi}_{\Shat}(\beta)\|_0<{|\Shat|}/{2}\}} \beta\right].
\label{eq: searching CI}
\end{equation}
Instead of searching over the real line as in \eqref{eq: searching CI def}, we restrict to the grid set $\mathcal{B}$ defined in \eqref{eq: grid construction} and \eqref{eq: rough bound} and search for the smallest and largest grid value such that more than half of the relevant IVs are valid. 
In contrast to $\CIsearchdef$ in \eqref{eq: searching CI def}, $\CIsearch$ is guaranteed to be an interval. { In Section A.6 in the supplement, we compare $\CIsearch$ and $\CIsearchdef$ and find that both intervals guarantee the coverage properties and achieve similar lengths.}

We summarize the construction of searching CI in Algorithm \ref{algo: USS-majority-searching}.

\begin{algorithm}[H]
\caption{Searching CI (Uniform Inference under Majority Rule)}
\begin{flushleft}
\textbf{Input:} Outcome $Y\in \R^{n}$; Treatment $D\in \R^{n}$; IVs $Z\in \R^{n\times \pz}$; Covariates $X\in \R^{n\times \px}$; Significance level $\alpha\in (0,1).$\\
\textbf{Output:} Confidence interval $\CIsearch$; Majority rule check $R\in \{0,1\}$.\\
\end{flushleft}
\vspace{-3mm}
\begin{algorithmic}[1]
    \State Construct $\widehat{\Gamma}\in \R^{\pz},\widehat{\gamma}\in \R^{\pz}$ as in \eqref{eq: OLS} and $\widehat{\V}^{\Gamma},\widehat{\V}^{\gamma}$ and $\widehat{\C}$ as in \eqref{eq: hetero var};
 \State Construct $\widehat{\mathcal{S}}$ as in \eqref{eq: relevant est}; \Comment{First-stage selection}
 \vspace{1mm}
 \State Construct $L$ and $U$ as in \eqref{eq: rough bound};
 \State Construct the grid set $\mathcal{B}$ as in \eqref{eq: grid construction} with $a=0.6$;
\State Construct  $\{\widehat{\pi}_j(\beta)\}_{j\in \Shat, \beta\in \mathcal{B}}$ as in \eqref{eq: thresholding};
 \State Construct $\CIsearch$ as in \eqref{eq: searching CI} and set $R={\bf 1}(\CIsearch\neq \emptyset)$. \Comment{Searching CI} 
\end{algorithmic}
\label{algo: USS-majority-searching}
\end{algorithm}

When the majority rule is violated, there is no $\beta$ such that $\|\widehat{\pi}_{\Shat}(\beta)\|_0<{|\widehat{\mathcal{S}}|}/{2}$ and  $\CIsearch$ in \eqref{eq: searching CI} is empty. This indicates that the majority rule is violated, which can be used as a partial check of the majority rule. Moreover, Algorithm \ref{algo: USS-majority-searching} can be implemented with the summary statistics $\widehat{\Gamma},\widehat{\gamma}$ and  $\widehat{\V}^{\Gamma},\widehat{\V}^{\gamma},\widehat{\C}$, which are the inputs for steps 2 to 6 of Algorithm \ref{algo: USS-majority-searching}.

\subsection{The Sampling CI}
\label{sec: sampling}
In this subsection, we devise a novel sampling method to improve the precision of the searching CI. Conditioning on the observed data, we resample $\{\widehat{\Gamma}^{\m},\widehat{\gamma}^{\m}\}_{1\leq m\leq M}$ as \begin{equation}
\begin{pmatrix}
\widehat{\Gamma}^{\m}\\
\widehat{\gamma}^{\m}
\end{pmatrix} \stackrel{\rm i.i.d.}{\sim} N\left[\begin{pmatrix}
\widehat{\Gamma}\\
\widehat{\gamma}
\end{pmatrix} ,\begin{pmatrix}\widehat{\V}^{\Gamma}/n& \widehat{\C}/n\\ \widehat{\C}^{\intercal}/n& \widehat{\V}^{\gamma}/n\end{pmatrix}\right]  \quad \text{for}\quad 1\leq m\leq M,
\label{eq: sampling}
\end{equation}
where $M$ denotes the resampling size (with the default value $1000$),  $\widehat{\gamma}$ and $\widehat{\Gamma}$ are defined in \eqref{eq: OLS}, and  $\widehat{\V}^{\Gamma}$, $\widehat{\C}$ and $\widehat{\V}^{\gamma}$ are defined in \eqref{eq: hetero var}.

For $1\leq m\leq M,$ we  use $\{\widehat{\Gamma}^{\m},\widehat{\gamma}^{\m}\}$ to replace $\{\widehat{\Gamma},\widehat{\gamma}\}$ and implement Algorithm \ref{algo: USS-majority-searching} to construct a searching CI. We refer to this CI as the $m$-th sampled searching CI. In implementing Algorithm \ref{algo: USS-majority-searching}, we decrease the thresholding level used in the hard thresholding step of estimating $\pip$. Consequently, each sampled searching CI can be much shorter than the original searching CI; see the following Figure \ref{fig: sampling}. Our proposal of decreasing the threshold relies on the following observation: there exists {$1\leq m^*\leq M$} such that 
\begin{equation}
\max_{j\in \Shat} \frac{|\widehat{\Gamma}^{[m^*]}_j-\Gammap_j-\beta(\widehat{\gamma}^{[m^*]}_j-\gammap_j)|}{\sqrt{(\widehat{\V}^{\Gamma}_{jj}+\beta^2\widehat{\V}^{\gamma}_{jj}-2\beta \widehat{\C}_{jj})/n}}\leq \lambda\cdot \Phi^{-1}\left(1-\frac{\alpha}{2|\Shat|}\right) \quad \text{with} \quad \lambda\asymp \left(\frac{\log n}{M}\right)^{\frac{1}{2|\Shat|}}.
\label{eq: sampling property}
\end{equation}
The rigorous statement is provided in the following Proposition \ref{prop: sampling}. Compared with \eqref{eq: quantile}, the upper bound in \eqref{eq: sampling property} is shrunk by a factor $\lambda,$ which is close to zero with a large $M$ (e.g., $M=1000$). If we had access to $\{\widehat{\Gamma}^{[m^*]}, 
\widehat{\gamma}^{[m^*]}\},$ we could use a much smaller threshold level in constructing the searching CI. 

We now provide the complete details about constructing the sampling CI. 
We firstly consider that the tuning parameter $\lambda$ in \eqref{eq: sampling property} is given and will present a data-dependent way of choosing $\lambda$ in the following Remark \ref{rem: tuning}. 
Motivated by \eqref{eq: sampling property}, we multiply the threshold level by $\lambda$ and implement Algorithm \ref{algo: USS-majority-searching} for each of  $\{\widehat{\Gamma}^{[m]},
\widehat{\gamma}^{[m]}\}_{1\leq m\leq M}.$ That is, for $1\leq m\leq M$, we modify \eqref{eq: thresholding} and estimate $\pib_{\widehat{\mathcal{S}}}$ by 
\vspace{1mm}
\begin{equation}
\widehat{\pi}_j^{\m}(\beta,\lambda)=\left(\widehat{\Gamma}^{\m}_{j}-\beta\widehat{\gamma}^{\m}_j\right)\cdot \mathbf{1}\left(\left|\widehat{\Gamma}^{\m}_{j}-\beta\widehat{\gamma}^{\m}_j\right|\geq \lambda\cdot \widehat{\rho}_j(\beta)\right) \quad \text{for} \quad j \in\widehat{\mathcal{S}},
\label{eq: thresholding sample}
\end{equation}
with $\widehat{\rho}_j(\beta)$ defined in \eqref{eq: key threshold}. We apply \eqref{eq: searching CI} and construct the $m$-th sampled searching CI as  $[\beta^{\m}_{\min}(\lambda),\beta^{\m}_{\max}(\lambda)]$ with
\begin{equation}
\beta^{\m}_{\min}(\lambda)=\min_{\left\{\beta\in \mathcal{B}: \|\widehat{\pi}^{\m}_{\Shat}(\beta,\lambda)\|_0<|\Shat|/2\right\}}\beta \quad \text{and}\quad \beta^{\m}_{\max}(\lambda)=\max_{\left\{\beta\in \mathcal{B}: \|\widehat{\pi}^{\m}_{\Shat}(\beta,\lambda)\|_0<|\Shat|/2\right\}}\beta.
\label{eq: sampled interval}
\end{equation}
Similarly to \eqref{eq: searching CI}, we set $[\beta^{\m}_{\min}(\lambda),\beta^{\m}_{\max}(\lambda)]=\emptyset$ if there is no $\beta$ such that 
$\|\widehat{\pi}^{\m}(\beta,\lambda)\|_0<|\Shat|/2.$ 
We aggregate the $M$ searching CIs and propose the following sampling CI: 
\begin{equation}
\CIsample=\left[\min_{m\in \mathcal{M}}\beta^{\m}_{\min}(\lambda),\max_{m\in \mathcal{M}}\beta^{\m}_{\max}(\lambda)\right],
\label{eq: sampling CI}
\end{equation}
where the index set 
$
\mathcal{M}=\{1\leq m\leq M: [\beta^{\m}_{\min}(\lambda),\beta^{\m}_{\max}(\lambda)]\neq \emptyset \}
$ contains all indexes $m$ corresponding to non-empty $[\beta^{\m}_{\min}(\lambda),\beta^{\m}_{\max}(\lambda)]$. In Figure \ref{fig: sampling}, we demonstrate the sampling CI in \eqref{eq: sampling CI} using {Example \ref{ex: illustration}}. Many of the $M$ sampled searching CIs are empty, and the non-empty ones can be much shorter than the searching CI. { Consequently, the sampling CI (in red) is much shorter than the searching CI (in blue).} 

\vspace{-2mm}
\begin{figure}[H]
    \centering
    \includegraphics[scale=0.6]{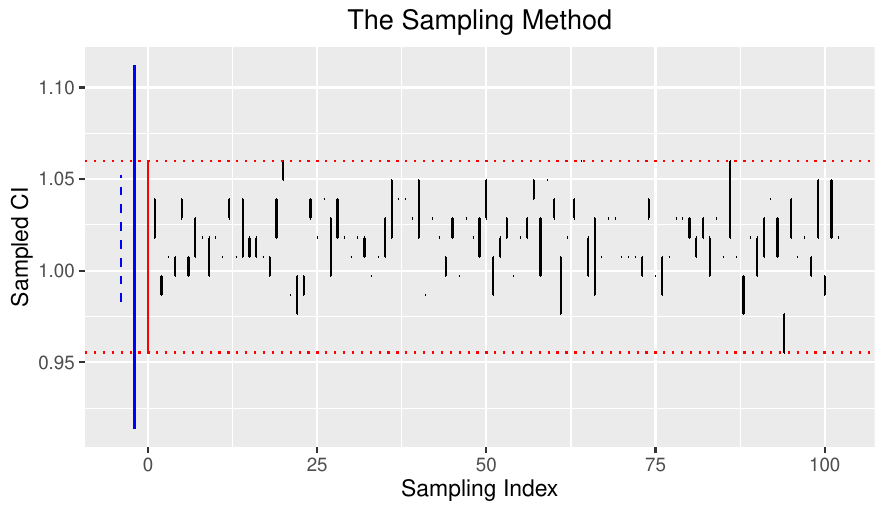}
    \vspace{-1mm}
    \caption{\small The axis corresponds to sampling indexes $\{1,2,\cdots,102\}$ (after re-ordering), and the y-axis reports the sampled CIs. The red interval  is $\CIsample=(0.955,1.060),$ the blue interval is $\CIsearch=(0.914, 1.112),$ and the blue dashed interval is the oracle CI $(0.983, 1.052)$ with prior information on valid IVs.}
    \label{fig: sampling}
\end{figure}
\vspace{-3mm}

An important reason for the sampling method improving the precision is that the test statistics $\|\widehat{\pi}_{\Shat}(\beta)\|_0$ 
used for constructing the searching CI in \eqref{eq: searching CI} is a discrete function of $\beta$, and the limiting distribution of the test statistics is not standard.

We summarize the sampling CI in Algorithm \ref{algo: USS-majority-sampling} and will provide two remarks about the implementation of Algorithm \ref{algo: USS-majority-sampling}.

\begin{algorithm}[htp!]
\caption{Sampling CI (Uniform Inference under Majority Rule)}
\begin{flushleft}
\textbf{Input:} Outcome $Y\in \R^{n}$; Treatment $D\in \R^{n}$; IVs $Z\in \R^{n\times \pz}$; Covariates $X\in \R^{n\times \px}$; Sampling number $M$=1000; $\lambda=c_*\left({\log n}/{M}\right)^{{1}/(2|\Shat|)}$; Significance level $\alpha\in (0,1)$\\
\textbf{Output:} Confidence interval $\CIsample$\\
\end{flushleft}
\vspace{-3mm}
\begin{algorithmic}[1]
\State Implement steps 1 to 4 as in Algorithm \ref{algo: USS-majority-searching};  
\For{$m \gets 1$ to $M$}                      \State Sample $\widehat{\Gamma}^{\m}$ and $\widehat{\gamma}^{\m}$ as in \eqref{eq: sampling};
   \State Compute $\{\widehat{\pi}_j^{\m}(\beta,\lambda)\}_{j\in \widehat{\mathcal{S}},\beta\in \mathcal{B}}$ as in \eqref{eq: thresholding sample};
   \State Compute $\beta^{\m}_{\min}(\lambda)$ and $\beta^{\m}_{\max}(\lambda)$  in \eqref{eq: sampled interval};
  \EndFor 
\State Construct $\CIsample$ as in \eqref{eq: sampling CI} \Comment{Sampling CI}
\end{algorithmic}
\label{algo: USS-majority-sampling}
\end{algorithm}

\begin{Remark}[Tuning parameter selection for sampling]\rm We demonstrate in Section B.2 in the supplement that the sampling CIs in Algorithm \ref{algo: USS-majority-sampling} do not materially change with different choices of $L,U,a$ and the resampling number $M$. The  following Theorem \ref{thm: inference-sampling} suggests the form of the tuning parameter $\lambda$ as $c_*\left({\log n}/{M}\right)^{{1}/(2|\Shat|)}.$ We present a data-dependent way to specify the constant $c_*$. If the $\lambda$ value is too small,  very few of the resampled reduced-form estimators will pass the majority rule and most of the $M=1000$ sampled intervals will be empty. Hence, the proportion of the non-empty intervals indicates whether $\lambda$ is large enough. We start with a small value $\lambda=1/6\cdot \left({\log n}/{M}\right)^{{1}/(2|\Shat|)}$ and increase the value of $\lambda$ by a factor of $1.25$ until more than $\texttt{prop}=10\%$ of the $M=1000$ intervals are non-empty. We choose the smallest $\lambda$ value achieving this and use this $\lambda$ value to implement Algorithm \ref{algo: USS-majority-sampling}. 
\label{rem: tuning}
\end{Remark}

\begin{Remark}[Alternative aggregation]\rm We may combine the sampled CIs as
$
\CIsamplenull=\cup_{m=1}^{M}[\beta^{\m}_{\min}(\lambda),\beta^{\m}_{\max}(\lambda)].$ Since $\CIsamplenull$ may not be an interval due to its definition, we focus on $\CIsample$ defined in \eqref{eq: sampling CI}. In addition, we may filter out  $\{\widehat{\gamma}^{\m},\widehat{\Gamma}^{\m}\}_{1\leq m\leq M}$ near the boundary of the sampling distribution in \eqref{eq: sampling}. Particularly, define
{\small
\begin{equation*}
\mathcal{M}_0=\left\{1\leq m\leq M: \max_{j\in \widehat{\mathcal{S}}}\max\left\{{\left|\widehat{\gamma}^{\m}_j-\widehat{\gamma}_j\right|}/{\sqrt{\widehat{\V}^{\gamma}_{jj}/n}},{\left|\widehat{\Gamma}^{\m}_j-\widehat{\Gamma}_j\right|}/{\sqrt{\widehat{\V}^{\Gamma}_{jj}/n}}\right\}\leq 1.1 \Phi^{-1}\left(1-\frac{\alpha_0}{4|\Shat|}\right)\right\}
\end{equation*}}
with $\alpha_0=0.05.$ 
We modify \eqref{eq: sampling CI} and construct the sampling CI as 
\begin{equation}
\CIsample(\mathcal{M}_0)=\left[\min_{m\in \mathcal{M}'}\beta^{\m}_{\min}(\lambda),\max_{m\in \mathcal{M}'}\beta^{\m}_{\max}(\lambda)\right],
\label{eq: sampling CI theory}
\end{equation}
with the index set $\mathcal{M}'=\{ m\in \mathcal{M}_0: [\beta^{\m}_{\min}(\lambda),\beta^{\m}_{\max}(\lambda)]\neq \emptyset \}.$
In Table B.6 in the supplement, we show that $\CIsample(\mathcal{M}_0)$ is nearly the same as $\CIsample$ in 
\eqref{eq: sampling CI}.
\label{rem: alter aggre}
\end{Remark}

\subsection{Robustness to the existence of weak IVs}
\label{sec: weak IV}
The validity of our proposed CIs in Sections \ref{sec: searching} to \ref{sec: sampling} relies on Condition \ref{cond: majority-finite}, which requires more than half of the IVs to be strongly relevant and valid but allows the remaining IVs to be arbitrarily weak and invalid. In the following, we make two important remarks about weak IVs. Firstly, our focused setting is completely different from the classical weak IV framework \citep[e.g.]{staiger1997instrumental}, where all IVs as a total are weak. The F test for the treatment model and the concentration parameters can provide evidence on whether the IVs are weak \citep[e.g.]{stock2002survey}. { Similarly, the estimated set $\Shat$ in \eqref{eq: relevant est} can also be used to check whether the given data set falls into the classical weak IV framework. This paper focuses on the regime with non-empty $\Shat$, indicating the existence of strong IVs.} Secondly, the challenge due to weak IVs is fundamentally different from locally invalid IVs. In the first-stage selection, the set of relatively weak IVs $\mathcal{S}\backslash \mathcal{S}_{\rm str}$ are not guaranteed to be selected by $\Shat$ in \eqref{eq: relevant est}. The uncertainty of selecting IVs inside $\mathcal{S}\backslash \mathcal{S}_{\rm str}$ is of a different nature from the uncertainty of detecting locally invalid IVs. In Section B.1 in the supplement, we demonstrate that both $\TSHT$ and $\CIIV$ perform well even if there exists uncertainty of selecting IVs belonging to $\mathcal{S}\backslash \mathcal{S}_{\rm str}$.


\section{Uniform Inference Methods under Plurality Rule}
\label{sec: plurality}

This section considers the more challenging setting with the plurality rule. We introduce the finite-sample plurality rule and then generalize the method proposed in Section \ref{sec: majority}.
For a given invalidity level $v\in \R$, we define the set of IVs, 
\begin{equation}
\mathcal{I}(v,\tau)=\left\{j\in \mathcal{S}: \left|{\pip_j}/{\gammap_j}-v\right|\leq \tau\right\} \quad \text{with}\quad \tau\in \R.
\label{eq: set notation}
\end{equation}
When $\tau$ is small,  $\mathcal{I}(v,\tau)$ denotes the set of IVs with the invalidity level around $v.$ With $\Diff_{j,k}$ defined in \eqref{eq: exact difference}, we define the separation level as 
\begin{equation}
\sep \coloneqq 2\sqrt{\log n}\max_{j,k \in \Shat}\Diff_{j,k}. 
\label{eq: separation}
\end{equation}
{Note that  $\beta^{[j]}=\Gammap_j/\gammap_j=\betap+\pip_j/\gammap_j.$ If the pairwise difference $\beta^{[j]}-\beta^{[k]}$ for ${j,k\in \widehat{\mathcal{S}}}$ is above $\sep$, the $j$-th and $k$-th IVs can be separated based on their invalidity levels. Particularly, the term $\max_{j,k \in \Shat}\Diff_{j,k}$ denotes the largest error of estimating the pairwise difference $\{\beta^{[j]}-\beta^{[k]}\}_{j,k\in \Shat}$ and $\sqrt{\log n}$ is used to adjust for the multiplicity of hypothesis testing;} see more discussion after the following Proposition \ref{prop: voting}. When $\{\gammap_j\}_{j\in \Shat}$ are non-zero constants, $\sep$ is of order $\sqrt{\log n/n}$.

With the separation level $\sep$, we introduce the finite-sample plurality rule.

\begin{Condition}{\rm \bf (Finite-sample Plurality Rule)} For $\tau_n= 3\sep,$ 
\begin{equation}
|\mathcal{V}\cap\mathcal{S}_{\rm str}|>\max_{v\in \R} |\mathcal{I}(v,\tau_n)\backslash \mathcal{V}| 
\label{eq: plurality-finite}
\end{equation}
where $\sep$, $\mathcal{V},$ $\mathcal{S}_{\rm str}$ and $\mathcal{I}(v,\tau_n)$ are defined in \eqref{eq: separation}, \eqref{eq: relevant valid}, \eqref{eq: str relevant} and \eqref{eq: set notation}, respectively.
 \label{cond: plurality-finite}
 \end{Condition}
The set $\mathcal{V}\cap\mathcal{S}_{\rm str}$ consists of the strongly relevant and valid IVs, which we rely on to make inferences for $\betap.$ Since $\tau_n\rightarrow 0$ { as $n\rightarrow \infty$}, the set $\mathcal{I}(v,\tau_n)\backslash \mathcal{V}$ contains all invalid IVs with invalidity levels $\pip_j/\gammap_j\approx v.$ Condition \ref{cond: plurality-finite} requires that the cardinality of $\mathcal{V}\cap\mathcal{S}_{\rm str}$ is larger than that of invalid IVs with $\pip_j/\gammap_j\approx v$. When $v=0$, the set $\mathcal{I}(0,\tau_n)\backslash \mathcal{V}$ is the set of invalid IVs with invalidity levels below $\tau_n.$ Condition \ref{cond: plurality-finite} also requires that there are more IVs in $\mathcal{V}\cap\mathcal{S}_{\rm str}$
than the locally invalid IVs with invalidity levels below $\tau_n.$

In comparison to Condition \ref{cond: rule 2}, Condition \ref{cond: plurality-finite} accommodates for the finite-sample approximation error by grouping together invalid IVs with similar invalidity levels. For a large sample size, Condition \ref{cond: plurality-finite} is the same as Condition \ref{cond: rule 2}. Specifically, if $n\rightarrow \infty$ and $\{\pip_j\}_{1\leq j\leq \pz}$ and $\{\gammap_j\}_{1\leq j\leq \pz}$ do not change with  $n,$ then
\begin{equation*}
\lim_{n\rightarrow\infty} |\mathcal{V}\cap\mathcal{S}_{\rm str}|=|\mathcal{V}|, \quad \lim_{n\rightarrow\infty}|\mathcal{I}(v,\tau_n)\backslash \mathcal{V}|=|\mathcal{I}_{\nu}\backslash \mathcal{V}|=\begin{cases}0 &\text{if} \; \nu=0\\ |\mathcal{I}_{\nu}|&\text{if}\; \nu\neq 0\end{cases} 
\end{equation*}
{with $\mathcal{I}_{\nu}$ defined in Condition \ref{cond: rule 2}. We shall emphasize that $\sep$ defined in \eqref{eq: separation} is not required in the following confidence interval construction.}

%
%

We propose a two-step inference procedure for $\betap$ under Condition \ref{cond: plurality-finite}. 
In the first step, we construct an initial set $\widehat{\mathcal{V}}$ satisfying
\begin{equation}
\mathcal{V}\cap \mathcal{S}_{\rm str}\subset \widehat{\mathcal{V}}\subset \mathcal{I}(0,\tau_n) \quad \text{with}\quad \tau_n= 3\sep.
\label{eq: initial condition}
\end{equation}
Since Condition \ref{cond: plurality-finite} implies that  $\mathcal{V}\cap\mathcal{S}_{\rm str}$ is the majority of the set $\mathcal{I}(0,\tau_n)$, we apply \eqref{eq: initial condition} and establish that the set $\mathcal{V}\cap\mathcal{S}_{\rm str}$ becomes the majority of the initial set $\Vhat$. In the second step, we restrict our attention to $\Vhat$ and generalize the methods in Section \ref{sec: majority}. Importantly, the initial set $\widehat{\mathcal{V}}$ is allowed to contain locally invalid IVs.

We now construct an initial set $\Vhat$ satisfying \eqref{eq: initial condition} by modifying  $\TSHT$ \citep{guo2018confidence}. Without loss of generality, we set $\widehat{\mathcal{S}}=\{1,2,\cdots,|\widehat{\mathcal{S}}|\}.$ 
For any $j \in \widehat{\mathcal{S}}$, we  construct an estimator of $\betap$ and ${\pi}^*$ as 
\begin{equation} \label{eq:pi_est}
\widehat{\beta}^{[j]}={\widehat{\Gamma}_j}/{ \widehat{\gamma}_j} \quad \text{and}\quad
\widehat{{\pi}}^{[j]}_k = \widehat{{\Gamma}}_k -\widehat{\beta}^{[j]}\widehat{{\gamma}}_k \quad \text{for}\quad k\in {\Shat},
 \end{equation}
 where the super index $j$ stands for the model identification by assuming the $j$-th IV to be valid. 
We define $\widehat{\RR}^{[j]} = \widehat{\V}^{\Gamma} + (\widehat{\beta}^{[j]})^2 \widehat{\V}^{\gamma} - 2 \widehat{\beta}^{[j]} \widehat{\C},$ and further estimate the standard error of $\widehat{\pi}_{k}^{[j]}$ with $k\in \widehat{\mathcal{S}}$ by
\begin{equation}
    \widehat{\rm SE}(\widehat{\pi}_{k}^{[j]})=\sqrt{
   \left(\widehat{\RR}^{[j]}_{k,k} + \left({\widehat{\gamma}_k}/{\widehat{\gamma}_j}\right)^2 \widehat{\RR}^{[j]}_{j,j} - 2 {\widehat{\gamma}_k}/{\widehat{\gamma}_j} \widehat{\RR}^{[j]}_{k,j}
    \right)/n}.
\label{eq: SE pi}
\end{equation}
For $1\leq k,j\leq |\Shat|,$ we apply the following hard thresholding and construct the $(k,j)$ entry of the voting matrix $\widehat{\Pi}\in \R^{|\Shat|\times|\Shat|}$ as
\begin{equation}
\widehat{\Pi}_{k,j}={\bf 1}\left(|\widehat{\pi}_{k}^{[j]}| \leq \widehat{\rm SE}(\widehat{\pi}_{k}^{[j]}) \cdot \sqrt{\log n} \quad \text{and}\quad |\widehat{\pi}_{j}^{[k]}| \leq \widehat{\rm SE}(\widehat{\pi}_{j}^{[k]}) \cdot \sqrt{\log n}\right),
\label{eq: col def}
\end{equation}
with $\widehat{\pi}_{k}^{[j]}$ and $\widehat{\pi}_{j}^{[k]}$ defined in \eqref{eq:pi_est}, $\widehat{\rm SE}(\widehat{\pi}_{k}^{[j]})$ and $\widehat{\rm SE}(\widehat{\pi}_{j}^{[k]})$ defined in \eqref{eq: SE pi}, and $\sqrt{\log n}$ used to adjust for multiplicity. In \eqref{eq: col def}, $\widehat{\Pi}_{k,j}=1$ represents that the $k$-th and $j$-th IVs support each other to be valid while $\widehat{\Pi}_{k,j}=0$ represents that they do not. The voting matrix in \eqref{eq: col def} is a symmetric version of the voting matrix proposed in \citet{guo2018confidence}.

We now construct the initial set by leveraging the voting matrix in \eqref{eq: col def}. Define $
\widehat{\mathcal{W}}=\argmax_{1\leq j\leq |\widehat{\mathcal{S}}|} \|\widehat{\Pi}_{j\cdot}\|_0
$ as the set of IVs receiving the largest number of votes.
We construct the following initial set,
\begin{equation}
\VTSHT\coloneqq \{1\leq l\leq |\widehat{\mathcal{S}}|: \text{there exist} \; 1\leq k\leq |\Shat| \; \text{and}\; j\in \widehat{\mathcal{W}}\; \text{such that}\; \widehat{\Pi}_{j,k}\widehat{\Pi}_{k,l}=1 \}.
\label{eq: initial valid}
\end{equation}
In words, if the $l$-th IV from $\Shat$ and the $j$-th IV from $\widehat{\mathcal{W}}$ are claimed to be valid by any IV from $\Shat$, then the $l$-th IV is also included in $\VTSHT.$ We show in the following Proposition \ref{prop: voting} that $\VTSHT$ is guaranteed to satisfy \eqref{eq: initial condition}. Together with Condition \ref{cond: plurality-finite}, $\mathcal{V}\cap\mathcal{S}_{\rm str}$ becomes the majority of the initial set $\VTSHT$. Then we restrict to the set $\VTSHT$ and apply Algorithms \ref{algo: USS-majority-searching} and \ref{algo: USS-majority-sampling} with $\Shat$ replaced by $\VTSHT.$ 

\begin{algorithm}[htp!]
\caption{Uniform Inference with Searching and Sampling (Plurality Rule)}
\begin{flushleft}
\textbf{Input:} Outcome $Y\in \R^{n}$; Treatment $D\in \R^{n}$; IVs $Z\in \R^{n\times \pz}$; Covariates $X\in \R^{n\times \px}$; Sampling number $M$=1000; $\lambda=c_*\left({\log n}/{M}\right)^{{1}/(2|\Shat|)}$; Significance level $\alpha\in (0,1).$\\
\textbf{Output:} Confidence intervals $\CIsearch$ and $\CIsample$;  Plurality rule check $R.$\\
\end{flushleft}
\vspace{-3mm}
\begin{algorithmic}[1]
    \State Construct $\widehat{\Gamma}\in \R^{\pz},\widehat{\gamma}\in \R^{\pz}$ as in \eqref{eq: OLS} and $\widehat{\V}^{\Gamma},\widehat{\V}^{\gamma}$ and $\widehat{\C}$ as in \eqref{eq: hetero var};
 \State Construct $\widehat{\mathcal{S}}$ as in \eqref{eq: relevant est};
\State Construct the voting matrix $\widehat{\Pi}\in \R^{|\widehat{\mathcal{S}}|\times |\widehat{\mathcal{S}}|}$ as in \eqref{eq: col def};
\State Construct $\widehat{\mathcal{V}}=\VTSHT$ as in \eqref{eq: initial valid};
\Comment{Construction of $\widehat{\mathcal{V}}$}
 \State Construct $L$ and $U$ as in \eqref{eq: rough bound} with $\Shat=\Vhat$;
 \State Construct the grid set $\mathcal{B}\subset [L, U]$ as in \eqref{eq: grid construction} with the grid size $n^{-0.6}$; 
 \State Compute  $\{\widehat{\rho}_j(\beta)\}_{j\in \Vhat,\beta\in \mathcal{B}}$ as in \eqref{eq: key threshold} with $\Shat=\Vhat$;
\State Compute $\{\widehat{\pi}_j(\beta)\}_{j\in \Vhat,\beta\in \mathcal{B}}$ as in \eqref{eq: thresholding} with $\Shat=\Vhat$;
   \State Construct $\CIsearch$ as in \eqref{eq: searching CI} with $\Shat=\Vhat$ and set $R={\bf 1}(\CIsearch\neq \emptyset)$; \Comment{Searching CI} 
\vspace{2mm}
\For{$m \gets 1$ to $M$}                      \State Sample $\widehat{\Gamma}^{\m}$ and $\widehat{\gamma}^{\m}$ as in \eqref{eq: sampling};
   \State Compute $\{\widehat{\pi}_j^{\m}(\beta,\lambda)\}_{j\in \widehat{\mathcal{V}},\beta\in \mathcal{B}}$ as in \eqref{eq: thresholding sample} with $\Shat=\Vhat$;
   \State Construct  $\beta^{\m}_{\min}(\lambda)$ and $\beta^{\m}_{\max}(\lambda)$ as in \eqref{eq: sampled interval}  with $\Shat=\Vhat$;
  \EndFor
\State Construct $\CIsample$ as in \eqref{eq: sampling CI}. \Comment{Sampling CI}
\end{algorithmic}
\label{algo: USS-plurality}
\end{algorithm}

{We summarize our proposed searching and sampling CIs in Algorithm \ref{algo: USS-plurality}, with the tuning parameters selected in the same way as that in Remark \ref{rem: tuning}. Algorithm \ref{algo: USS-plurality} can be implemented without requiring the raw data, but with $\widehat{\Gamma},\widehat{\gamma}$ and $\widehat{\V}^{\Gamma},\widehat{\V}^{\gamma},\widehat{\C}$ as the inputs. We have demonstrated our method by constructing $\widehat{\mathcal{V}}=\VTSHT$ as in \eqref{eq: initial valid}, but Algorithm \ref{algo: USS-plurality} can be applied with any $\widehat{\mathcal{V}}$ satisfying \eqref{eq: initial condition}.} 



\vspace{2mm}
\noindent {\bf Comparison with the CIIV method.}
The idea of searching has been developed in \citet{windmeijer2019confidence} to select valid IVs. We now follow \citet{windmeijer2019confidence} and sketch the intuitive idea of the $\CIIV$ method. For any grid value $\delta_g\in [L,U],$ define the set $$\widehat{\mathcal{V}}(\delta_g)=\left\{j\in \mathcal{S}: {\Gammap_j}/{\gammap_j}=\delta_{g} \;\; \text{is not rejected}\right\}.$$ 
Here, $\widehat{\mathcal{V}}(\delta_g)$ denotes a subset of IVs such that the corresponding hypothesis ${\Gammap_j}/{\gammap_j}=\delta_{g}$ is not rejected.  As explained in Section 3 of \citet{windmeijer2019confidence}, the \texttt{CIIV} method examines all values of $\delta_{g}$ and selects the largest set $\widehat{\mathcal{V}}(\delta_g)$  as the set of valid IVs, that is, 
\begin{equation}
\VCIIV= \widehat{\mathcal{V}}({\widehat{\delta_{g}}}) \quad \text{with}\quad \widehat{\delta_{g}}=\argmax_{\delta_{g}\in [L,U]} |\widehat{\mathcal{V}}(\delta_g)|.
\label{eq: CIIV}
\end{equation}

Our proposed searching CI differs from \citet{windmeijer2019confidence} since we directly construct CIs by searching for a range of suitable $\beta$ values, while the $\CIIV$ method applies the searching idea to select the set of valid IVs. 

We provide some intuitions on why our proposal is more robust to the IV selection error. We first screen out the strongly invalid IVs and construct an initial set estimator $\Vhat=\VTSHT$; then, we restrict to the set $\Vhat$ and apply searching and sampling CIs developed under the majority rule. The majority rule in the second step explains the robustness: we compare the number of votes to $|\Vhat|/2,$, which is fixed after computing $\Vhat.$ However, the optimization in \eqref{eq: CIIV} chooses $\delta_g$, giving the largest number of votes, which can be more vulnerable to selection/testing errors. The validity of the $\CIIV$ method requires that $\VCIIV$ does not contain any invalid IVs; in contrast, our method is still effective even if the initial set $\Vhat$ contains the invalid IVs but satisfies \eqref{eq: initial condition}.

\section{Theoretical Justification}
\label{sec: theory}
 We focus on the low-dimensional setting with heteroscedastic errors and introduce the following regularity conditions. { We always consider the asymptotic expressions in the limit with $n\rightarrow \infty.$ } 
\begin{enumerate}
\item[(C1)] 
 For $1\leq i\leq n,$ $W_{i\cdot}=(X_{i\cdot}^{\intercal},Z_{i\cdot}^{\intercal})^{\intercal}\in \R^{p}$ are i.i.d. Sub-gaussian random vectors with $\Sigma=\E (W_{i\cdot}(W_{i\cdot})^{\intercal})$ satisfying $c_0\leq \lambda_{\min}(\Sigma)\leq \lambda_{\max}(\Sigma)\leq C_0$ for some positive constants $C_0\geq c_0>0$.
\item[(C2)]For $1\leq i\leq n,$ the errors $(\epsilon_i,\delta_i)^{\intercal}$ in \eqref{eq: reduced-form} are i.i.d Sub-gaussian random vectors; the conditional  covariance matrix satisfying $c_1\leq \lambda_{\min}\left(\E\left[(\epsilon_i,\delta_i)^{\intercal}(\epsilon_i,\delta_i)\mid W_{i\cdot}\right]\right)\leq \lambda_{\max}\left(\E\left[(\epsilon_i,\delta_i)^{\intercal}(\epsilon_i,\delta_i)\mid W_{i\cdot}\right]\right)\leq C_1$ for some positive constants $C_1\geq c_1>0.$
 \end{enumerate}
 
Conditions {\rm (C1)} and {\rm (C2)} are imposed on the reduced-form model \eqref{eq: reduced-form}, which includes the outcome model \eqref{eq: outcome} and the treatment model \eqref{eq: treatment} as a special case. We assume that the covariance matrix of $W_{i\cdot}$ is well conditioned and also the covariance matrix of the errors is well conditioned. Condition {\rm (C2)} in general holds if $e_i$ in \eqref{eq: outcome} and $\delta_i$ in \eqref{eq: treatment} are not perfectly correlated. Our setting allows for the heteroscedastic errors. If we further assume $(\epsilon_i,\delta_i)^{\intercal}$ to be independent of $W_{i\cdot}$, Condition {\rm (C2)} assumes the covariance matrix of $(\epsilon_i,\delta_i)^{\intercal}$ to be well conditioned. As a remark, the Sub-gaussian conditions on both $W_{i\cdot}$ and the errors might be relaxed to the moment conditions in low dimensions.

{We start with the majority rule setting and will move to the plurality rule.}
The following theorem justifies the searching CI under the majority rule.
\begin{Theorem} Consider the model \eqref{eq: reduced-form}. 
 Suppose that Condition \ref{cond: majority-finite}, Conditions {\rm (C1)} and {\rm (C2)} hold, and $\alpha\in (0,1/4)$ is the significance level. Then $\CIsearchdef$ defined in \eqref{eq: searching CI def} and $\CIsearch$ defined in \eqref{eq: searching CI} satisfy
$$
\liminf_{n\rightarrow \infty} \PP\left(\betap\in \CIsearchdef\right)\geq 1-\alpha \quad \text{and} \quad \liminf_{n\rightarrow \infty} \PP\left(\betap\in \CIsearch\right)\geq 1-\alpha.
$$
There exists a positive constant $C>0$ such that
\begin{equation*}
\liminf_{n\rightarrow \infty}\PP\left(\max\{\mathbf{L}(\CIsearchdef),\mathbf{L}(\CIsearch)\}\leq  \frac{C}{\min_{j\in\Shat\cap \mathcal{V}}|\gammap_j|\cdot \sqrt{n}}\right)\geq 1-\alpha,
\label{eq: length}
\end{equation*}
where $\mathbf{L}(\cdot)$ denotes the interval length.
\label{thm: inference-searching}
\end{Theorem}

 If $\min_{j\in\Shat\cap \mathcal{V}}|\gammap_j|$ is of a constant order, Theorem \ref{thm: inference-searching} implies that the length of the searching CI is of the parametric rate $1/\sqrt{n}.$ The searching CI achieves the desired coverage level without relying on a perfection separation of valid and invalid IVs, which brings in a sharp contrast to the well-separation conditions required in $\TSHT$ \citep{guo2018confidence} and $\CIIV$ \citep{windmeijer2019confidence}. 



We now turn to the sampling CI. Before presenting the theory for the sampling CI, we justify in
the following Proposition \ref{prop: sampling} why we are able to decrease the thresholding level in constructing the sampling CI. For $\alpha_0\in (0,1/4),$
define the positive constant 
\begin{equation}
c^{*}(\alpha_0)= \frac{1}{\left[{3}\pi\cdot\lambda_{\min}(\Cov)\right]^{{|\Shat|}}}\exp\left(-\frac{|\Shat|\cdot 3\lambda_{\max}(\Cov)}{\lambda_{\min}(\Cov)}\cdot \left[\Phi^{-1}\left(1-\frac{\alpha_0}{4|\Shat|}\right)\right]^2\right),
\label{eq: key constant 2}
\end{equation}
with $\Cov$ defined in \eqref{eq: reduced-form limiting} and $\Phi^{-1}$ denoting the inverse CDF of the standard normal distribution. For a fixed $\pz$ and $\alpha_0\in (0,1),$ we have $c_1/C_0\leq \lambda_{\min}(\Cov)\leq \lambda_{\max}(\Cov)\leq C_1/c_0$ and $c^{*}(\alpha_0)$ is a positive constant independent of $n$. { With $c^{*}(\alpha_0)$ defined in \eqref{eq: key constant 2}, we introduce the following term quantifying the resampling property,
\begin{equation}
\err_n(M,\alpha_0) =\left[\frac{2 \log n}{c^*(\alpha_0)M}\right]^{\frac{1}{2|\Shat|}},
\label{eq: sampling constant}
\end{equation}
where $\err_n(M,\alpha_0)$ converges to $0$ with $M\rightarrow \infty$ and $M$ being much larger than $\log n.$}

\begin{Proposition}
Suppose Conditions {\rm (C1)} and {\rm (C2)} hold and $\alpha_0\in(0,1/4)$. If $\err_n(M,\alpha_0) \leq c$ for a small positive constant $c>0$, then there exists a positive constant $C>0$ such that
\begin{equation}
\liminf_{n\rightarrow\infty}\PP\left(\min_{1\leq m\leq M}\left[\max_{\beta\in \mathcal{U}(a)}\max_{j\in \Shat} \frac{|\widehat{\Gamma}^{[m]}_j-\Gammap_j-\beta(\widehat{\gamma}^{[m]}_j-\gammap_j)|}{\sqrt{(\widehat{\V}^{\Gamma}_{jj}+\beta^2\widehat{\V}^{\gamma}_{jj}-2\beta \widehat{\C}_{jj})/n}}\right]\leq C \err_n(M,\alpha_0)\right)\geq 1- \alpha_0,
\label{eq: sample quantile}
\end{equation}
with $\err_n(M,\alpha_0)$ defined in \eqref{eq: sampling constant} and $\mathcal{U}(a)\coloneqq\{\beta\in \R:|\beta-\betap|\leq n^{-a}\}$ for any $a>1/2.$
\label{prop: sampling}
\end{Proposition}
Since $c^*(\alpha_0)$ is of a constant order, the condition $\err_n(M,\alpha_0) \leq c$ holds for a sufficiently large resampling size $M$. The above proposition states that, with a high probability, there exists $1\leq m^*\leq M$ such that $$\max_{\beta\in \mathcal{U}(a)}\max_{j\in \Shat} \frac{|\widehat{\Gamma}^{[m^*]}_j-\Gammap_j-\beta(\widehat{\gamma}^{[m^*]}_j-\gammap_j)|}{\sqrt{(\widehat{\V}^{\Gamma}_{jj}+\beta^2\widehat{\V}^{\gamma}_{jj}-2\beta \widehat{\C}_{jj})/n}}\leq C \err_n(M,\alpha_0).$$ 
In comparison to \eqref{eq: quantile}, the threshold decreases from $\Phi^{-1}\left(1-\frac{\alpha}{2|\Shat|}\right)$ to $C \err_n(M,\alpha_0).$
A related sampling property was established in \citet{guo2020inference} to address a different nonstandard inference problem.  

We now apply Proposition \ref{prop: sampling} to justify the sampling CI under the majority rule.

\begin{Theorem}
Suppose that the conditions of Proposition \ref{prop: sampling} hold, $\alpha_0\in(0,1/4)$, and
$\lambda$ used in \eqref{eq: sampled interval} satisfies 
 $\lambda \geq 2C {\err_n(M,\alpha_0)}/\Phi^{-1}[1-{\alpha}/(2|\Shat|)]$ and $\lambda\gg n^{1/2-a}$
 with the constant $C$ used in \eqref{eq: sample quantile}, $\alpha\in (0,1/4)$ and $a>1/2$.
Then $\CIsample$ defined in \eqref{eq: sampling CI} and $\CIsample(\mathcal{M}_0)$ defined in \eqref{eq: sampling CI theory} satisfy 
$$
\liminf_{n\rightarrow \infty}\PP\left(\betap\in \CIsample\right)\geq  \liminf_{n\rightarrow \infty}\PP\left(\betap\in \CIsample(\mathcal{M}_0)\right)\geq 1-\alpha_0.$$
There exists a positive constant $C>0$ such that  
$$\liminf_{n\rightarrow \infty}\PP\left(\max\left\{\frac{\mathbf{L}(\CIsample)}{\sqrt{\log |\mathcal{M}|}}, \mathbf{L}( \CIsample(\mathcal{M}_0))\right\}\leq \frac{C}{\min_{j\in\Shat\cap \mathcal{V}}|\gammap_j|\cdot \sqrt{n}}\right)\geq 1-\alpha_0.
$$
\label{thm: inference-sampling}
\end{Theorem}

Note that $2C {\err_n(M,\alpha_0)}/\Phi^{-1}[1-{\alpha}/(2|\Shat|)]=c(\log n/M)^{1/(2|\Shat|)}$ for some positive constant $c>0.$
Motivated by the condition  $\lambda \geq 2C {\err_n(M,\alpha_0)}/\Phi^{-1}[1-{\alpha}/(2|\Shat|)],$ we choose the tuning parameter $\lambda$ in the form $\lambda=c_{*}(\log n/M)^{1/(2|\Shat|)}$ in Remark \ref{rem: tuning}. Similar to Theorem \ref{thm: inference-searching}, Theorem \ref{thm: inference-sampling} shows that our proposed searching CI does not require the well-separation condition on the invalidity levels. If the IV strengths $\{\gammap_j\}_{j\in \mathcal{S}}$ are assumed to be of a constant order, then the length of $\CIsample(\mathcal{M}_0)$ defined in \eqref{eq: sampling CI theory} is $1/\sqrt{n}$. We can only establish the upper bound $\sqrt{\log  |\mathcal{M}|/{n}}$ for the length of $\CIsample$ defined in \eqref{eq: sampling CI}. However, we believe that this is mainly a technical artifact since $\CIsample(\mathcal{M}_0)$ and $\CIsample$ are nearly the same in the numerical studies; see Section B.2 in the supplement. 

{We now switch to the more challenging setting only assuming the finite-sample plurality rule (Condition \ref{cond: plurality-finite}).} The main extra step is to show that our constructed initial set $\widehat{\mathcal{V}}=\VTSHT$ satisfies \eqref{eq: initial condition}. To establish this, we provide a careful finite-sample analysis of the voting scheme described in \eqref{eq: col def} and $\VTSHT$ defined in \eqref{eq: initial valid}.
\begin{Proposition} Suppose that Conditions {\rm (C1)} and {\rm (C2)} hold. Consider the indexes $j\in \widehat{\mathcal{S}}$ and $k\in \widehat{\mathcal{S}}.$ (a) If ${\pib_k}/{\gammab_k}={\pib_j}/{\gammab_j},$ then 
$
\liminf_{n\rightarrow\infty}\PP\left(\widehat{\Pi}_{k,j}=\widehat{\Pi}_{j,k}=1\right)=1.
$ (b)If $
\left|{\pib_k}/{\gammab_k}-{\pib_j}/{\gammab_j}\right| \geq 2\sqrt{\log n}\cdot {\Diff}_{j,k},$ then $\liminf_{n\rightarrow\infty}\PP\left(\widehat{\Pi}_{k,j}=\widehat{\Pi}_{j,k}=0\right)=1,$ where ${\Diff}_{j,k}$ is defined in \eqref{eq: exact difference}. 
Under the additional Condition \ref{cond: plurality-finite}, 
the constructed $\widehat{\mathcal{V}}=\VTSHT$ in \eqref{eq: initial valid} satisfies $\liminf_{n\rightarrow \infty}\PP\left(\mathcal{V}\cap \mathcal{S}_{\rm str}\subset \widehat{\mathcal{V}}\subset \mathcal{I}(0,3 \sep)\right)=1,$ with $\sep$ defined in \eqref{eq: separation}.
\label{prop: voting}
\end{Proposition}
The above proposition shows that if two IVs are of the same invalidity level, then they vote for each other with a high probability. If two IVs are well-separated (i.e. $|{\pib_k}/{\gammab_k}-{\pib_j}/{\gammab_j}| \geq 2\sqrt{\log n}\cdot {\Diff}_{j,k}$), then they vote against each other with a high probability. If $0<|{\pib_k}/{\gammab_k}-{\pib_j}/{\gammab_j}|<2\sqrt{\log n}\cdot {\Diff}_{j,k},$ there is no theoretical guarantee on how the two IVs will vote. The above proposition explains why we are likely to make a mistake in detecting some locally invalid IVs defined in Definition \ref{def: locally invalid}.

With Proposition \ref{prop: voting}, we connect the finite-sample plurality rule to the finite-sample majority rule. We apply the voting algorithm to remove all strongly invalid IVs and the set $\VTSHT$ only consists of valid IVs and the locally invalid IVs. Under Condition \ref{cond: plurality-finite}, $\mathcal{V}\cap \mathcal{S}_{\rm str}$ becomes the majority of $\VTSHT.$ We then establish the following theorem by applying the theoretical analysis of the majority rule by replacing $\Shat$ with $\VTSHT$. 

\begin{Theorem} Consider the model \eqref{eq: reduced-form}. Suppose that Condition \ref{cond: plurality-finite}, Conditions {\rm (C1)} and {\rm (C2)} hold. Then with $\Shat$ replaced by $\VTSHT$, the coverage and precision properties in Theorem \ref{thm: inference-searching} hold for the searching interval in Algorithm \ref{algo: USS-plurality} and the coverage and precision properties in Theorem \ref{thm: inference-sampling} hold for the sampling interval in Algorithm \ref{algo: USS-plurality}. 
\label{thm: inference-searching plurality}
\end{Theorem}


\section{Simulation Studies}
\label{sec: simulation}


Throughout the numerical studies, we implement our proposed $\CIsearch$ and $\CIsample$  detailed in Algorithm \ref{algo: USS-plurality}. We illustrate the robustness of Algorithm \ref{algo: USS-plurality} to different initial estimators of ${\mathcal{V}}$ by considering two choices of $\widehat{\mathcal{V}}$: $\VTSHT$ defined in \eqref{eq: initial valid} and the set of valid IVs $\VCIIV$ outputted by \texttt{CIIV} \citep{windmeijer2019confidence}.  { We focus on the low-dimensional setting in the current section and will present the high-dimensional results in Section A.3 in the supplement.} The code for replicating all numerical results in the current paper is available at \url{https://github.com/zijguo/Searching-Sampling-Replication}.

 We compare with three existing CIs allowing for invalid IVs: \texttt{TSHT} \citep{guo2018confidence}, \texttt{CIIV} \citep{windmeijer2019confidence}, and the  \texttt{Union} method \citep{kang2020two}. \texttt{TSHT} and \texttt{CIIV} are implemented with the codes on the Github websites \footnote{The code for \texttt{TSHT} is obtained from
\texttt{https://github.com/hyunseungkang/invalidIV} and for \texttt{CIIV} is obtained from \texttt{https://github.com/xlbristol/CIIV}.} 
while the \texttt{Union} method is implemented with the code shared by the authors of \citet{kang2020two}.
The \texttt{Union} method takes a union of intervals that are constructed by a given number of candidate IVs and are not rejected by the Sargan test. An upper bound $\bar{s}$ for the number of invalid IVs is required for the implementation. We consider two specific upper bounds: $\bar{s}=\pz-1$ corresponds to the existence of two valid IVs, and $\bar{s}=\lceil \pz/2\rceil$ corresponds to the majority rule being satisfied. We conduct $500$ replications of simulations and compare different CIs in terms of empirical coverage and average lengths. 

We implement two oracle methods as the benchmark. Firstly, we implement the {oracle} TSLS assuming the prior knowledge of $\mathcal{V}$. This method serves as a benchmark when the set $\mathcal{V}$ of valid IVs is correctly recovered. Secondly, we implement the oracle bias-aware confidence interval in \eqref{eq: bias aware} assuming the oracle knowledge of the bias of \texttt{TSHT} estimator. We argue that the oracle bias-aware confidence interval serves as a better benchmark, especially when $\mathcal{V}$ might not be correctly recovered in finite samples.


We generate the i.i.d. data $\{Y_i,D_i, Z_{i\cdot}, X_{i\cdot}\}_{1\leq i\leq n}$ using the outcome model \eqref{eq: outcome} and the treatment model \eqref{eq: treatment}.  We generate $\gammap\in \R^{\pz}$ and $\pip \in \R^{\pz}$ as follows,

\vspace{-2mm}
\begin{enumerate}
\item[\bf S1] (Majority rule): set 
 $\gammap= \gamma_0\cdot{\bf 1}_{10}$ and  $\pip=({\bf 0}_{6},\tau\cdot \gamma_0,\tau\cdot\gamma_0,-0.5,-1)^{\intercal}$;  
\item[\bf S2] (Plurality rule): set 
 $\gammap= \gamma_0\cdot{\bf 1}_{10}$ and  $\pip=({\bf 0}_{4},\tau\cdot \gamma_0,\tau\cdot\gamma_0,-\frac{1}{3},-\frac{2}{3},-1,-\frac{4}{3})^{\intercal}$;
\item[\bf S3] (Plurality rule): set 
 $\gammap= \gamma_0\cdot{\bf 1}_{10}$ and  $\pip=({\bf 0}_{4},\tau\cdot \gamma_0,\tau\cdot\gamma_0,-\frac{1}{6},-\frac{1}{3},-\frac{1}{2},-\frac{2}{3})^{\intercal}$;
 \item[\bf S4] (Plurality rule): set 
 $\gammap= \gamma_0\cdot{\bf 1}_{6}$ and  $\pip=({\bf 0}_{2}, -0.8,-0.4, \tau\cdot \gamma_0, 0.6)^{\intercal}$;
\item[\bf S5] (Plurality rule): set 
 $\gammap= \gamma_0\cdot{\bf 1}_{6}$ and  $\pip=({\bf 0}_{2},-0.8,-0.4,\tau\cdot \gamma_0,\tau\cdot \gamma_0+0.1)^{\intercal}$.
 \end{enumerate}
 \vspace{-2mm}
The parameter $\gamma_0$ denotes the IV strength and is set as $0.5$. 
The parameter $\tau$ denotes the invalidity level and is varied across $\{0.2,0.4\}$.
The setting {\bf S1} satisfies the population majority rule while the settings {\bf S2}  to {\bf S5} only satisfy the population plurality rule. Settings {\bf S4} and {\bf S5} represent the challenging settings where there are only two valid IVs. We introduce settings {\bf S3} and {\bf S5} to test the robustness of our proposed method when the finite-sample plurality rule might be violated. For example, 
for the setting {\bf S5} with small $n$ (e.g. $n=500$), the invalid IVs with $\pip_j$ values $\tau\cdot \gamma_0,\tau\cdot \gamma_0+0.1$ have similar invalidity levels and  may violate Condition \ref{cond: plurality-finite}.

We now specify the remaining details for the generating models \eqref{eq: outcome} and \eqref{eq: treatment}. 
 Set $\px=10$, $\phi^*=(0.6,0.7,\cdots,1.5)^{\intercal}\in \R^{10}$ in \eqref{eq: treatment} and $\psi^*=(1.1,1.2,\cdots,2)^{\intercal}\in \R^{10}$ in \eqref{eq: outcome}. 
We vary $n$ across $\{500, 1000, 2000, 5000\}.$  
 For $1\leq i\leq n,$ generate the covariates $W_{i\cdot}=(Z_{i\cdot}^{\intercal}, X^{\intercal}_{i\cdot})^{\intercal}\in \R^{p}$ following a multivariate normal distribution with zero mean and covariance $\Sigma\in \R^{p\times p}$ where $\Sigma_{jl}=0.5^{|j-l|}$ for $1\leq j,l\leq p$; generate the errors $(e_i,\delta_i)^{\intercal}$ following bivariate normal with zero mean, unit variance and ${\rm Cov}(e_i,\delta_i)=0.8.$

\begin{table}[htp!]
\centering
\resizebox{\linewidth}{!}{
\begin{tabular}[t]{|c|c|c c|c c|c c|c c|c|c c|}
\multicolumn{13}{c}{Empirical Coverage of Confidence Intervals for $\tau=0.2$} \\
\hline
\multicolumn{2}{|c|}{ } & \multicolumn{2}{c|}{Oracle} & \multicolumn{2}{c|}{ } & \multicolumn{2}{c|}{Searching} & \multicolumn{2}{c|}{Sampling} & \multicolumn{1}{c|}{ } & \multicolumn{2}{c|}{\texttt{Union}} \\
\hline
Set & n & \texttt{TSLS} & \texttt{BA} & \texttt{TSHT} & \texttt{CIIV} & $\VTSHT$ & $\VCIIV$ & $\VTSHT$ & $\VCIIV$ & Check & $p_z-1$ & $\lceil p_z/2 \rceil$\\
\hline
 & 500 & 0.95 & 0.97 & 0.53 & 0.61 & 1.00 & 1.00 & 1.00 & 1.00 & 1.00 & 1.00 & 0.99\\
 & 1000 & 0.94 & 0.97 & 0.45 & 0.69 & 1.00 & 1.00 & 1.00 & 1.00 & 1.00 & 1.00 & 0.98\\
 & 2000 & 0.96 & 0.95 & 0.63 & 0.79 & 1.00 & 1.00 & 1.00 & 1.00 & 1.00 & 1.00 & 0.96\\
\multirow{-4}{*}{\centering\arraybackslash {\bf S1}} & 5000 & 0.94 & 0.95 & 0.89 & 0.92 & 1.00 & 1.00 & 1.00 & 1.00 & 1.00 & 1.00 & 0.96\\
\cline{1-13}
 & 500 & 0.95 & 0.96 & 0.56 & 0.51 & 1.00 & 1.00 & 1.00 & 0.99 & 1.00 & 1.00 & 0.27\\
 & 1000 & 0.94 & 0.95 & 0.45 & 0.58 & 0.99 & 0.99 & 1.00 & 0.97 & 1.00 & 1.00 & 0.03\\
 & 2000 & 0.94 & 0.93 & 0.51 & 0.74 & 0.98 & 0.98 & 0.98 & 0.97 & 1.00 & 1.00 & 0.00\\
\multirow{-4}{*}{\centering\arraybackslash {\bf S2}} & 5000 & 0.96 & 0.93 & 0.85 & 0.95 & 0.99 & 1.00 & 1.00 & 1.00 & 1.00 & 1.00 & 0.00\\
\cline{1-13}
 & 500 & 0.95 & 0.95 & 0.63 & 0.63 & 0.99 & 0.99 & 0.99 & 0.99 & 1.00 & 1.00 & 0.62\\
 & 1000 & 0.94 & 0.97 & 0.62 & 0.60 & 0.99 & 0.99 & 0.99 & 0.97 & 1.00 & 1.00 & 0.18\\
 & 2000 & 0.94 & 0.93 & 0.62 & 0.73 & 0.97 & 0.98 & 0.97 & 0.96 & 1.00 & 1.00 & 0.01\\
\multirow{-4}{*}{\centering\arraybackslash {\bf S3}} & 5000 & 0.96 & 0.93 & 0.85 & 0.95 & 0.99 & 1.00 & 0.99 & 1.00 & 1.00 & 1.00 & 0.00\\
\cline{1-13}
 & 500 & 0.95 & 0.96 & 0.72 & 0.66 & 0.94 & 0.95 & 0.94 & 0.93 & 0.98 & 0.98 & 0.00\\
 & 1000 & 0.94 & 0.95 & 0.65 & 0.58 & 1.00 & 0.97 & 0.99 & 0.96 & 0.95 & 0.98 & 0.00\\
 & 2000 & 0.93 & 0.99 & 0.68 & 0.58 & 0.98 & 0.97 & 0.97 & 0.93 & 0.99 & 0.95 & 0.00\\
\multirow{-4}{*}{\centering\arraybackslash {\bf S4}} & 5000 & 0.95 & 0.95 & 0.91 & 0.88 & 0.98 & 0.95 & 0.98 & 0.95 & 1.00 & 0.94 & 0.00\\
\cline{1-13}
 & 500 & 0.95 & 0.97 & 0.49 & 0.51 & 0.81 & 0.89 & 0.88 & 0.86 & 0.98 & 0.97 & 0.14\\
 & 1000 & 0.94 & 0.96 & 0.31 & 0.50 & 0.68 & 0.89 & 0.76 & 0.86 & 0.91 & 0.98 & 0.00\\
 & 2000 & 0.93 & 0.98 & 0.46 & 0.57 & 0.86 & 0.96 & 0.86 & 0.92 & 0.84 & 0.95 & 0.00\\
\multirow{-4}{*}{\centering\arraybackslash {\bf S5}} & 5000 & 0.95 & 0.95 & 0.90 & 0.88 & 0.98 & 0.95 & 0.97 & 0.94 & 0.98 & 0.94 & 0.00\\
\hline
\multicolumn{13}{c}{Average Lengths of Confidence Intervals for $\tau=0.2$} \\
\hline
\multicolumn{2}{|c|}{ } & \multicolumn{2}{c|}{Oracle} & \multicolumn{2}{c|}{ } & \multicolumn{2}{c|}{Searching} & \multicolumn{2}{c|}{Sampling} & \multicolumn{1}{c|}{ } & \multicolumn{2}{c|}{\texttt{Union}} \\
\hline
Set & n & \texttt{TSLS} & \texttt{BA} & \texttt{TSHT} & \texttt{CIIV} & $\VTSHT$ & $\VCIIV$ & $\VTSHT$ & $\VCIIV$ & Check & $p_z-1$ & $\lceil p_z/2 \rceil$\\

\cline{1-13}
 & 500 & 0.10 & 0.16 & 0.08 & 0.08 & 0.59 & 0.61 & 0.34 & 0.34 & - & 1.16 & 0.25\\
 & 1000 & 0.07 & 0.14 & 0.06 & 0.06 & 0.39 & 0.43 & 0.24 & 0.24 & - & 0.63 & 0.16\\
 & 2000 & 0.05 & 0.10 & 0.05 & 0.05 & 0.27 & 0.30 & 0.17 & 0.17 & - & 0.42 & 0.09\\
\multirow{-4}{*}{\centering\arraybackslash {\bf S1}} & 5000 & 0.03 & 0.04 & 0.03 & 0.03 & 0.17 & 0.18 & 0.10 & 0.10 & - & 0.27 & 0.04\\
\cline{1-13}
 & 500 & 0.13 & 0.24 & 0.13 & 0.10 & 0.58 & 0.59 & 0.37 & 0.36 & - & 2.46 & 0.07\\
 & 1000 & 0.09 & 0.24 & 0.13 & 0.08 & 0.37 & 0.41 & 0.26 & 0.26 & - & 1.45 & 0.02\\
 & 2000 & 0.06 & 0.26 & 0.14 & 0.06 & 0.25 & 0.29 & 0.19 & 0.18 & - & 0.76 & 0.00\\
\multirow{-4}{*}{\centering\arraybackslash {\bf S2}} & 5000 & 0.04 & 0.13 & 0.08 & 0.04 & 0.16 & 0.17 & 0.10 & 0.10 & - & 0.28 & 0.00\\
\cline{1-13}
 & 500 & 0.13 & 0.22 & 0.10 & 0.10 & 0.62 & 0.62 & 0.45 & 0.38 & - & 1.77 & 0.13\\
 & 1000 & 0.09 & 0.21 & 0.10 & 0.08 & 0.38 & 0.41 & 0.29 & 0.27 & - & 1.36 & 0.03\\
 & 2000 & 0.06 & 0.25 & 0.13 & 0.06 & 0.26 & 0.29 & 0.19 & 0.18 & - & 0.86 & 0.00\\
\multirow{-4}{*}{\centering\arraybackslash {\bf S3}} & 5000 & 0.04 & 0.13 & 0.08 & 0.04 & 0.16 & 0.17 & 0.10 & 0.10 & - & 0.35 & 0.00\\
\cline{1-13}
 & 500 & 0.23 & 0.62 & 0.24 & 0.18 & 0.56 & 0.56 & 0.48 & 0.44 & - & 0.87 & 0.00\\
 & 1000 & 0.16 & 0.56 & 0.17 & 0.13 & 0.44 & 0.36 & 0.38 & 0.29 & - & 0.42 & 0.00\\
 & 2000 & 0.11 & 0.32 & 0.14 & 0.10 & 0.27 & 0.24 & 0.22 & 0.18 & - & 0.20 & 0.00\\
\multirow{-4}{*}{\centering\arraybackslash {\bf S4}} & 5000 & 0.07 & 0.13 & 0.08 & 0.07 & 0.14 & 0.13 & 0.11 & 0.10 & - & 0.09 & 0.00\\
\cline{1-13}
 & 500 & 0.23 & 0.63 & 0.27 & 0.17 & 0.42 & 0.51 & 0.41 & 0.42 & - & 1.01 & 0.05\\
 & 1000 & 0.16 & 0.61 & 0.18 & 0.13 & 0.32 & 0.36 & 0.30 & 0.29 & - & 0.50 & 0.00\\
 & 2000 & 0.11 & 0.38 & 0.12 & 0.10 & 0.28 & 0.24 & 0.25 & 0.18 & - & 0.22 & 0.00\\
\multirow{-4}{*}{\centering\arraybackslash {\bf S5}} & 5000 & 0.07 & 0.15 & 0.08 & 0.07 & 0.15 & 0.13 & 0.12 & 0.10 & - & 0.09 & 0.00\\
\hline
\end{tabular}}
\caption{Empirical coverage and average lengths for {\bf S1} to {\bf S5} with $\tau=0.2$. The columns indexed with \texttt{TSLS}, \texttt{BA}, \texttt{TSHT} and \texttt{CIIV} represent the oracle TSLS CI with the knowledge of $\mathcal{V}$, the oracle bias-aware CI in \eqref{eq: bias aware}, the CI by  \citet{guo2018confidence}, and the CI by \citet{windmeijer2019confidence}, respectively. Under the columns indexed with ``Searching'' (or ``Sampling''), the columns indexed with $\VTSHT$ and $\VCIIV$ represent our proposed searching (or sampling) CI in Algorithm \ref{algo: USS-plurality} with $\VTSHT$ and $\VCIIV$, respectively. The column indexed with ``Check" reports the proportion of simulations passing the plurality rule check in Algorithm \ref{algo: USS-plurality}. The columns indexed with \texttt{Union} represent the method by \citet{kang2020two}. The columns indexed with $\pz-1$ and $\lceil \pz/2\rceil$ correspond to the \texttt{Union} methods assuming two valid IVs and the majority rule, respectively.}
\label{tab: CI coverage length Main}
\end{table}

In Table \ref{tab: CI coverage length Main}, we report the empirical coverage and interval length for $\tau=0.2.$ The CIs by \texttt{TSHT} and \texttt{CIIV}  undercover for $n=500, 1000$ and $2000$ and only achieve the 95\% coverage level for a large sample size   $n=5000$.  Our proposed searching and sampling CIs achieve the desired coverage levels in most settings. For settings {\bf S1} to {\bf S4}, both initial estimates of set of valid IVs $\VTSHT$ and $\VCIIV$ lead to CIs achieving the 95\% coverage level.
For the more challenging settings {\bf S5}, the empirical coverage level of our proposed searching and sampling CIs achieve the desired coverage with sample sizes above $2000.$ For $n=500$ and $n=1000$, our proposed searching and sampling methods improve the coverage of \texttt{TSHT} and \texttt{CIIV}.
The undercoverage happens mainly due to the fact that the finite-sample plurality rule might fail for setting {\bf S5} with a relatively small sample size. The CIs by the \texttt{Union} method \citep{kang2020two} with $\bar{s}=\pz-1$ (assuming there are two valid IVs) achieve the desired coverage level while those with $\bar{s}=\lceil \pz/2\rceil$ (assuming the majority rule) do not achieve the desired coverage level for settings {\bf S2} to {\bf S5}.

We now compare the CI lengths. When the CIs by \texttt{TSHT} and \texttt{CIIV} are valid, their lengths are similar to the length of the CI by {oracle} TSLS, which match with the theory in \citet{guo2018confidence,windmeijer2019confidence}. For the CIs achieving valid coverage, our proposed sampling CI is in general shorter than the searching CI and the \texttt{Union} CI with $\pz-1$.  We shall point out that our proposed sampling CI can be even shorter than the oracle bias-aware CI (the benchmark).  As an important remark, the oracle bias-aware CI, the sampling CI, the searching CI, and the \texttt{Union} CI are in general longer than the CI by the \texttt{oracle} TSLS, which is a price to pay for constructing uniformly valid CIs. In Section E.1 in the supplement, 
we consider the settings with $\tau=0.4$ and heteroscedastic errors. In Section E.3 in the supplement, we further explore the settings considered in  \citet{windmeijer2019confidence}. The results for these settings are similar to those in Table \ref{tab: CI coverage length Main}.  

\vspace{2mm}

\begin{figure}[H]
    \centering
    \includegraphics[scale=0.7]{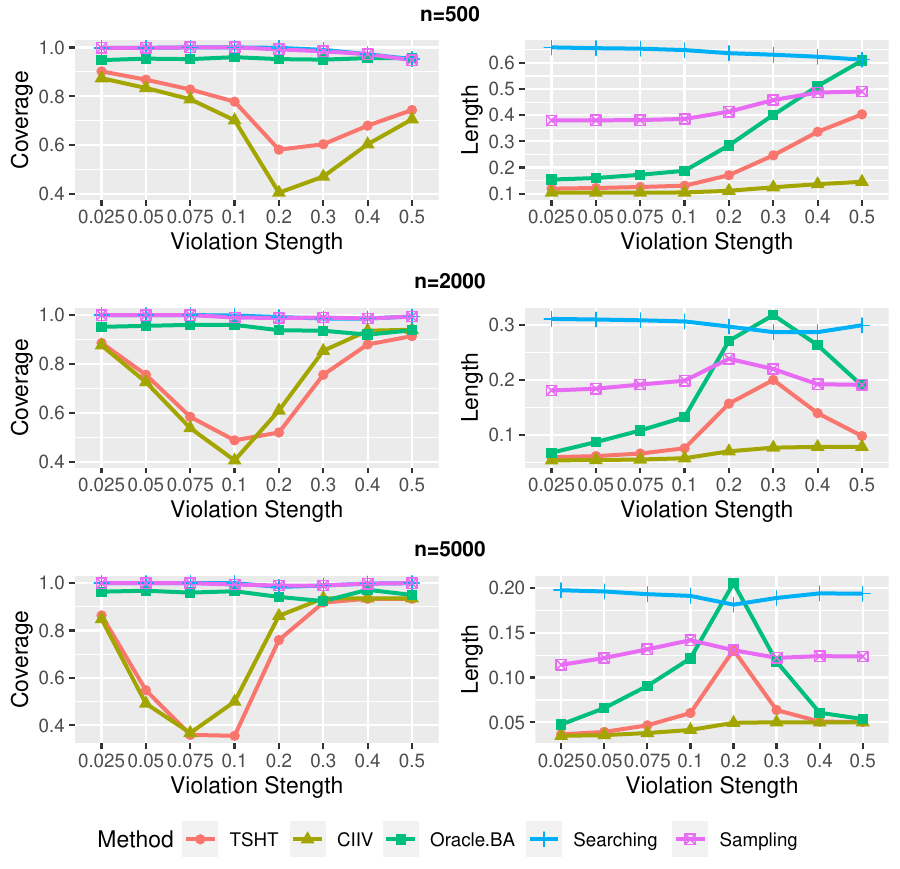}
    \caption{Empirical coverage and average lengths for the setting {\bf S2} with $\tau\in\{0.025,0.05,0.075,0.1,0.2,0.3,0.4,0.5\}$ and heteroscedastic errors. \texttt{Oracle-BA}, \texttt{TSHT}, and \texttt{CIIV} represent the oracle bias-aware CI in \eqref{eq: bias aware}, the CI by  \citet{guo2018confidence}, and the CI by \citet{windmeijer2019confidence},  respectively. The searching and sampling CIs are implemented as in Algorithm \ref{algo: USS-plurality}.} 
    \label{fig: varying tau hetero}
\end{figure}

\noindent {{\bf Varying violation strength.} In Figure \ref{fig: varying tau hetero}, we focus on the setting {\bf S2} and vary $\tau$ across $\{0.025,0.05,0.075,0.1,0.2,0.3,0.4,0.5\}$.} We follow \citet{bekker2015jackknife} and generate heteroscedastic errors as follows: for $1\leq i\leq n,$ generate $\delta_i \sim N(0,1)$ and  $
e_i=0.3 \delta_i+\sqrt{{[1-0.3^2]}/[0.86^4+1.38^2]}(1.38\cdot \tau_{1,i}+0.86^2\cdot \tau_{2,i}),
$
where conditioning on $Z_i$, $\tau_{1,i}\sim N(0, [0.5\cdot Z_{i,1}^2+0.25]^2),$ $\tau_{2,i} \sim N(0,1),$ and $\tau_{1,i}$ and $\tau_{2,i}$ are independent of $\delta_i.$ 

\vspace{-3mm}

As reported in Figure \ref{fig: varying tau hetero}, our proposed searching and sampling CIs achieve the desired coverage (95\%) while \texttt{TSHT} and \texttt{CIIV} only achieve the desire coverage for $\tau= 0.5$ with $n=2000$ and $\tau\geq 0.3$ with $n=5000$. In terms of length, we observe that the sampling CI is shorter than the searching CI. The sampling CI can be even shorter than the oracle bias-aware CI (the benchmark). We do not plot the length of the \texttt{Union} CI, which is three to six times longer than our proposed sampling CI; see Figure \ref{fig: coverage} for details. An interesting observation is that the empirical coverage of \texttt{TSHT} and \texttt{CIIV} is around 90\% for $\tau=0.025.$ This happens since the invalidity levels of the IVs are small, and even the inclusion of such invalid IVs does not significantly worsen the empirical coverage.

We present the results for homoscedastic errors in Figure E.1 in the supplement. 
{ In Section E.2 in the supplement, we explore the performance of different methods for settings with the locally invalid IVs where the violation levels are scaled to $\sqrt{\log n/n}.$}

\vspace{2mm}

{\noindent{\bf Tuning parameter selection.}
We investigate the robustness of Algorithm \ref{algo: USS-plurality} to different choices of tuning parameters.} For the searching CI, we observe in Table \ref{tab: searching tuning} that the empirical coverage and the average lengths are almost invariant to different choices of $L$, $U$ and $a$. For two intervals $(a_1,b_1)$ and $(a_2,b_2)$, we define its difference as $|a_1-a_2|+|b_1-b_2|.$ In Table \ref{tab: searching tuning}, we report the difference between the searching CI constructed with the default choice of $L,U,a$ in Algorithm \ref{algo: USS-plurality} and searching CIs with other choices of $L,U,a.$ The average interval difference is smaller than twice the default grid size $n^{-0.6}$.

\begin{table}[H]
\centering
\resizebox{0.95\linewidth}{!}{
\begin{tabular}[t]{|c|c|ccc|ccc|ccc|ccc|}
\hline

\multicolumn{2}{|c|}{ } & \multicolumn{3}{c|}{{\bf S1} ($n=2000$)} & \multicolumn{3}{c|}{{\bf S2} ($n=2000$)} & \multicolumn{3}{c|}{{\bf S3} ($n=2000$)} & \multicolumn{3}{c|}{{\bf S4} ($n=2000$)} \\
\hline
[L,U] & a & Cov & Len & Diff & Cov & Len & Diff & Cov & Len & Diff & Cov & Len & Diff\\
\hline
\multirow{3}{*}{\centering\arraybackslash \text{As in} \eqref{eq: rough bound}} & 0.6 & 1 & 0.287 & 0.000 & 1 & 0.268 & 0.000 & 0.994 & 0.275 & 0.000 & 0.980 & 0.274 & 0.000\\
 & 0.8 & 1 & 0.295 & 0.008 & 1 & 0.276 & 0.008 & 0.996 & 0.283 & 0.008 & 0.982 & 0.282 & 0.008\\
 & 1.0 & 1 & 0.297 & 0.010 & 1 & 0.278 & 0.010 & 0.996 & 0.284 & 0.010 & 0.984 & 0.284 & 0.010\\
\hline
\multirow{3}{*}{\centering\arraybackslash [-10,10]} & 0.6 & 1 & 0.287 & 0.007 & 1 & 0.268 & 0.008 & 0.994 & 0.275 & 0.008 & 0.978 & 0.274 & 0.007\\
& 0.8 & 1 & 0.295 & 0.008 & 1 & 0.277 & 0.009 & 0.996 & 0.283 & 0.009 & 0.984 & 0.282 & 0.009\\
 & 1.0 & 1 & 0.297 & 0.010 & 1 & 0.278 & 0.010 & 0.996 & 0.285 & 0.010 & 0.984 & 0.284 & 0.010\\

\hline
\multirow{3}{*}{\centering\arraybackslash [-20,20]} & 0.6 & 1 & 0.287 & 0.007 & 1 & 0.268 & 0.007 & 0.992 & 0.275 & 0.007 & 0.982 & 0.274 & 0.007\\

 & 0.8 & 1 & 0.295 & 0.008 & 1 & 0.277 & 0.009 & 0.996 & 0.283 & 0.009 & 0.982 & 0.282 & 0.009\\
 
 & 1.0 & 1 & 0.297 & 0.010 & 1 & 0.278 & 0.010 & 0.996 & 0.285 & 0.010 & 0.984 & 0.284 & 0.010\\
\hline
\end{tabular}}
\caption{\small Searching CIs with different choices of $L,U,a$ for settings {\bf S1} to {\bf S4} with $\tau=0.4$ and $n=2000$. The columns indexed with ``Cov" and ``Len" denote empirical coverage and average lengths, respectively. The column indexed with ``Diff" represents the average length difference between the searching CI in Algorithm \ref{algo: USS-plurality} and the searching CIs with other choices of $L$, $U$, and $n^{-a}$. }
\label{tab: searching tuning}
\end{table}

\vspace{-2mm}

\begin{figure}[H]
    \centering
    \includegraphics[scale=0.7]{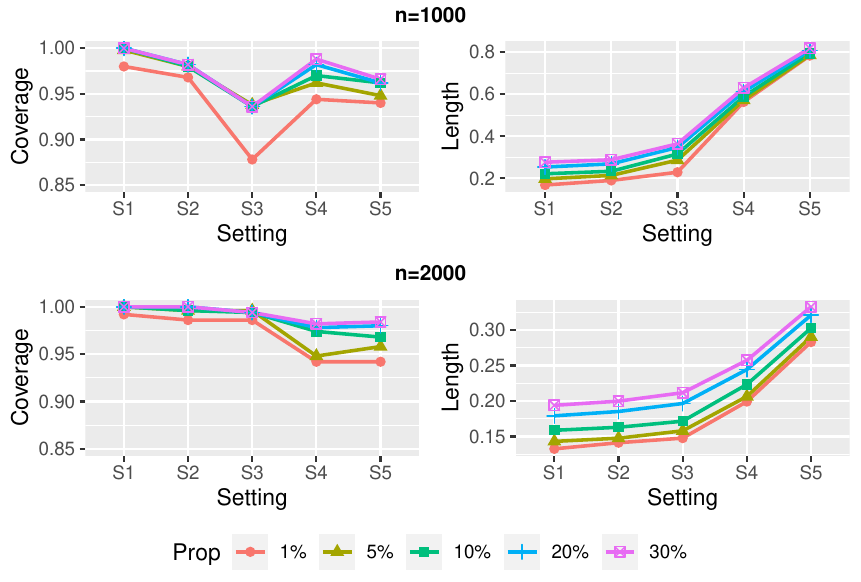}
    \caption{\small Comparison of the sampling CIs with different $\lambda$ for $\tau=0.4$ and $M=1000$. We choose the smallest value of $\lambda$ such that $1000\cdot \texttt{prop}$ sampled intervals are non-empty. For example, for $\texttt{prop}=10\%$, we choose the smallest $\lambda$ leading to 100 non-empty sampled intervals.}
    \label{fig: different lambda}
\end{figure}

In Section B.2 in the supplement, we demonstrate that the sampling CIs have nearly the same empirical coverage and length for different choices of $L, U,a$ and the resampling size $M$. The choice of the shrinkage parameter $\lambda>0$ has a more obvious effect on the sampling CI. As explained in Remark \ref{rem: tuning}, we choose the smallest $\lambda$ such that more than a pre-specified proportion (denoted as $\texttt{prop}$) of the $M=1000$ intervals are non-empty. In Figure \ref{fig: different lambda}, we compare the coverage and length properties of the sampling CIs by varying $\texttt{prop}$ across $\{1\%,5\%,10\%,20\%,30\%\}.$ If $\texttt{prop}\geq 5\%$, the empirical coverage reaches the nominal level and vary slightly with $\texttt{prop}.$ In terms of length, the intervals get slightly longer with a larger value of \texttt{prop}. The empirical coverage and average lengths of the sampling CIs are robust to a wide range of $\lambda$ values, as long as the corresponding $\lambda$ guarantees a sufficient proportion of non-empty sampled searching CIs.

\section{Real Data Analysis}
\label{sec: real data}
We study the effect of education on earnings by analyzing the 2018 survey of China Family Panel Studies \citep{xie2012china}.  
The outcome is the logarithm transformation of salary in the past 12 months, and the treatment is the years of education. We include three binary baseline covariates: gender, urban (whether the subject lives in the urban area), and hukou (whether the subject's hukou is agricultural or non-agricultural). { Following the literature, we have induced the following nine IVs:} \\

\vspace{-3.5mm}
(a) Family background variables \citep[e.g.]{trostel2002estimates,blackburn1993omitted,behrman2012financial}: the father's education level, the mother's education level, the spouse's education level, the family size, and the log transformation of the education expenditure in the past 12 months;\\

\vspace{-3.5mm}
 (b) The group-level years of education where the groups are formulated by age, gender, and region; \\
 
 \vspace{-3.5mm}
 (c) Personal traits \citep{behrman2012financial}: the statement on the fair competition, the statement on the talent pay-off, and whether the subject reads some books or not in the past 12 months.\\
The statement on fair competition measures an individual's willingness to compete through ability. The statement on the talent pay-off is about an individual's viewpoint on whether their educational endeavors will pay off. After removing the missing values, the data consists of $3758$ observations. In the supplement, we report the summary statistics of all IVs and baseline covariates in Table E.8. 

In Table \ref{tab: CI comparison}, we compare our proposed searching and sampling CIs with existing methods. By applying $\texttt{TSHT},$ we identified six relevant instruments: the father's education level, the mother's education level, the spouse's education level, the group-level years of education, the family size, whether to read some books or not in the past 12 months. Out of these IVs, the family size is detected as the invalid IV. \texttt{CIIV} outputs the same set of valid IVs. When we include all 9 IVs, the concentration parameter is 2906.36. If we only include the 5 IVs selected by \texttt{TSHT} and \texttt{CIIV}, the concentration parameter is 2850.57. This indicates that the whole set of IVs are strongly associated with the treatment, but some IVs (e.g., the family size) are possibly invalid.

\begin{table}[htp!]
    \centering
    \begin{tabular}{|r|r|r|r|}
    \hline
    Method & CI & Method & CI \\
    \hline
    OLS & (0.0305, 0.0503) & Searching CI & (0.0409, 0.1698)  \\
    \hline
    TSLS & (0.0959, 0.1190) & Sampling CI & (0.0552, 0.1268) \\
    \hline
    TSHT & (0.0946, 0.1178) & Union ($\bar{s}=p_z-1$) & (-0.4915, 1.6043) \\ 
    \hline
    CIIV & (0.0948, 0.1175) & Union($\bar{s}=\lceil p_z/2 \rceil$) & (0.0409, 0.1342)\\
    \hline
    \end{tabular}
    \caption{Confidence intervals for the effect of education on earnings.}
    \label{tab: CI comparison}
\end{table}

As reported in Table \ref{tab: CI comparison}, CIs by $\TSHT$ and $\CIIV$ are relatively short, but they may undercover due to the IV selection error. We compare the lengths of \texttt{Union} method and our proposed searching and sampling CIs, which are all robust to the IV selection error. The sampling CI is the shortest among these CIs. 
{We further plot the searching and sampling CIs in Figure E.4 in the supplement.} The validity of the \texttt{Union} CI with $\bar{s}=\lceil \pz/2\rceil$ requires half of the candidate IVs to be valid; if we can only assume that two of candidate IVs are valid, then the CI by \texttt{Union}  with $\bar{s}=\lceil \pz/2\rceil$ may not be valid but the CI by \texttt{Union}  with $\bar{s}=\pz-1$  is valid.

\section{Conclusion and Discussion}
\label{sec: discussion}
Causal inference from observational studies is a challenging task. Typically, stringent identification conditions are required to facilitate various causal inference approaches. The valid IV assumption is one of such assumptions to handle unmeasured confounders. In the current paper, we devise uniformly valid confidence intervals for the causal effect when the candidate IVs are possibly invalid. Our proposed searching and sampling confidence intervals add to the fast-growing literature on robust inference with possibly invalid IVs. The proposed method has the advantage of being more robust to the mistakes in separating the valid and invalid IVs at the expense of a wider confidence interval. The proposed intervals are computationally efficient and less conservative than existing uniformly valid confidence intervals. 

\vspace{-2mm}
 
 \section*{Data Availability Statement}
 The data underlying this article is the 2018 survey of China Family Panel Studies \citep{xie2012china}, available at \url{https://www.isss.pku.edu.cn/cfps/en/data/public/index.htm}.
 
\section*{Acknowledgement}
The research of Z. Guo was partly supported by the NSF grants DMS 1811857 and 2015373 and NIH grants R01GM140463 and R01LM013614. Z. Guo is grateful to the participants at Penn Causal Reading Group and CUHK econometrics seminar for their helpful discussion, to Dr. Frank Windmeijer for pointing out the connection to the CIIV method, and to Dr. Hyunseung Kang for sharing the code for replicating the \texttt{Union} method. Z. Guo thanks the editors, two anonymous referees, and Drs. Hyunseung Kang, Zhonghua Liu, and  Molei Liu for constructive comments on a previous draft. 
Z. Guo thanks Mr. Zhenyu Wang for the help with the numerical implementation and Dr. Junhui Yang and Mr. Zhenyu Wang for cleaning the CFHS data set. 
\bibliographystyle{chicago.bst}
\bibliography{IVRef}

\newpage 
\appendix
\counterwithin{figure}{section}
\counterwithin{table}{section}
\setcounter{page}{1}

{ 
\section{Additional Discussions}
\subsection{Relationship with Post-selection Inference}
\label{sec: diff post-selection}
Throughout the paper, the causal effect $\beta^*$ is fixed since it represents the effect after receiving the treatment and does not depend on which IVs are used. This focused regime is fundamentally different from the selective inference literature \citep[e.g.]{berk2013valid,lee2016exact,leeb2005model}, where the targeted parameter depends on the selected model. 

For our focused setting with the IV selection, the set $\mathcal{V}$ of valid IVs are treated as the nuisance parameters, which are useful in identifying the treatment effect $\beta^*$. \citet{guo2018confidence} and \citet{windmeijer2019confidence} proposed to estimate $\mathcal{V}$ by data-dependent set estimators $\Vhat$. \citet{guo2018confidence} and \citet{windmeijer2019confidence} identified the target parameter $\betap$ through the following expression,
$$
\beta(\Vhat)=\sum_{j\in \Vhat}\Gammap_{j}\gammap_{j} /\sum_{j\in \Vhat}(\gammap_{j})^2,
$$
with $\Vhat$ estimated in a data-dependent way.
The validity of \citet{guo2018confidence} and \citet{windmeijer2019confidence} required $\Vhat=\mathcal{V},$ which implies $\beta(\Vhat)=\betap.$ The error in constructing the set $\mathcal{V}$ leads to the bias of identifying the targeted causal effect and the following confidence intervals being unreliable. Our inference problem is framed as the post-selection problem because the main challenge arises from the possible IV selection error.

Even though the parameter of interest (e.g., the targeted causal effect $\beta^*$) is fixed, the nuisance parameters (e.g., the set $\mathcal{V}$ of valid IVs) are randomly selected by the data. 
 The setting with a fixed target parameter but randomly selected nuisance parameters can be viewed as a special case of the general post-selection problem. We also would like to mention that this kind of setting has been extensively investigated in the high-dimensional inference literature 
\citep[e.g.]{zhang2014confidence,javanmard2014confidence,van2014asymptotically,chernozhukov2015post}. Particularly, these works considered the high-dimensional linear models and made inference for a fixed one-dimensional parameter (e.g., a given regression coefficient). Such an inference problem in high dimensions is also termed as ``post-selection" \citep[e.g.]{belloni2014inference} since the main challenge arises from the randomness in selecting high-dimensional nuisance variables/parameters.

Our proposed searching and sampling methods are similar to other post-selection inference methods: the constructed confidence intervals are uniformly valid regardless of whether selection errors exist.


\subsection{Reasoning behind the definition of $\mathcal{S}^{\rm str}$}
\label{sec: factor 2}
The definitions of $\widehat{\mathcal{S}}$ in \eqref{eq: relevant est} and $\mathcal{S}_{\rm str}$ in \eqref{eq: str relevant} are motivated from the following finite-sample upper bound: with probability larger than $1-\exp(-c\sqrt{\log n})$,
\begin{equation}
\max_{1\leq j\leq \pz}\frac{|\widehat{\gamma}_j-\gammap_j|}{\sqrt{\widehat{\V}^{\gamma}_{jj}/n}}\leq \sqrt{\log n}.
\label{eq: key bound}
\end{equation}
The above result is implied by the definition of $\mathcal{G}_2$ in \eqref{eq: events} and Lemma \ref{lem: good event 1}.
If \eqref{eq: key bound} holds, then $|\widehat{\gamma}_j|\geq {\sqrt{\widehat{\V}^{\gamma}_{jj}/n}}\sqrt{\log n}$ implies $\gammap_j \neq 0$, which explains that $\widehat{\mathcal{S}}$ excludes all indexes corresponding to $\gammap_j=0$.
We now discuss what values of $\gammap_j$ will lead to the $j$-th IV being included into $\widehat{\mathcal{S}}.$ If $\left|\gamma^*_j\right|\geq 2\sqrt{\log n} \cdot \sqrt{{\V}^{\gamma}_{jj}/n}$ and $\widehat{\V}^{\gamma}_{jj}\cip {\V}^{\gamma}_{jj}$, then we apply \eqref{eq: key bound} and obtain that 
$$|\widehat{\gamma}_j|\geq {\sqrt{\widehat{\V}^{\gamma}_{jj}/n}}\sqrt{\log n},$$
that is $j\in \widehat{\mathcal{S}}.$ The above derivation shows that $\mathcal{S}_{\rm str} \subset \widehat{\mathcal{S}},$ with probability larger than $1-\exp(-c\sqrt{\log n}).$ The factor $2$ in \eqref{eq: str relevant} ensures that the individual IV strength is sufficiently large such that the $j$-th IV can be included in the set $\Shat$ defined in \eqref{eq: relevant est}. 

\subsection{High dimensional IVs and covariates}
\label{sec: high dim}
In this subsection, we extend our proposed searching and sampling method to the settings with high-dimensional IVs and covariates. We modify Algorithm \ref{algo: USS-plurality} to accommodate for the high-dimensional setting. Particularly, for step 1 of Algorithm \ref{algo: USS-plurality}, we construct $\widehat{\Gamma}\in \R^{\pz},\widehat{\gamma}\in \R^{\pz}$ by applying the debiased lasso estimators \citep{zhang2014confidence,javanmard2014confidence,van2014asymptotically}. The detailed discussions about the reduced-form estimators and its covariance estimators can be found in Section 4.1 of \citet{guo2018testing}. For step 2 of Algorithm \ref{algo: USS-plurality},  we modify the definition of $\widehat{\mathcal{S}}$ in \eqref{eq: relevant est} by replacing $\sqrt{\log n}$ with $\sqrt{2.01 \cdot \log \max\{\pz,n\}};$ for step 3 of Algorithm \ref{algo: USS-plurality}, we modify the construction of $\widehat{\Pi}$ in \eqref{eq: col def} by replacing $\sqrt{\log n}$ with $2.01\cdot\sqrt{\log \max\{\pz,n\}}$; see Remark \ref{rem: thresholding} and \citet{guo2018confidence} for more discussions. We report the numerical performance of our proposed high-dimensional method in Section \ref{sec: high dim sim}.

\subsection{Homoscadastic setting}
We assume the homoscedastic regression errors: $\E(e_i^2\mid Z_{i\cdot},X_{i\cdot})=\sigma_{e}^2$ and $\E(\delta_i^2\mid Z_{i\cdot},X_{i\cdot})=\sigma_{\delta}^2.$
We estimate the matrices $\V^{\Gamma}$, $\V^{\gamma}$ and $\C$ in \eqref{eq: reduced-form limiting} by 
\begin{equation*}
\widehat{\V}^{\Gamma}=\widehat{\sigma}_{\epsilon}^2 \widehat{\Omega}, \; \widehat{\V}^{\gamma}=\widehat{\sigma}_{\delta}^2 \widehat{\Omega}, \; \widehat{\C}=\widehat{\sigma}_{\epsilon,\delta} \widehat{\Omega} \quad \text{with}\quad \widehat{\Omega}=[(W^{\intercal} W/n)^{-1}]_{1:\pz,1:\pz},
\end{equation*}
where $
\widehat{\sigma}_{\epsilon}^2= {\|Y - Z \widehat{\Gamma} - X \widehat{\Psi}\|_2^2}/{(n-1)},$  $\widehat{\sigma}_{\delta}^2={\|D - Z \widehat{\gamma} - X \widehat{\psi}\|_2^2}/{(n-1)},$ and
$$\widehat{\sigma}_{\epsilon,\delta}= {(Y - Z \widehat{\Gamma} - X \widehat{\Psi})^{\intercal}(D - Z \widehat{\gamma} - X \widehat{\psi})}/{(n-1)}.$$
In the homoscedastic setting, we can simplify $\widehat{\rm SE}(\widehat{\pi}_{k}^{[j]})$ defined in \eqref{eq: SE pi} as
\begin{equation*}
\widehat{\rm SE}(\widehat{\pi}_{k}^{[j]})=\sqrt{(\widehat{\sigma}_{\epsilon}^2+(\widehat{\beta}^{[j]})^2\widehat{\sigma}^2_{\delta}-2\widehat{\beta}^{[j]}\widehat{\sigma}_{\epsilon,\delta})/{n}}\cdot \sqrt{\widehat{\Omega}_{kk}-2{\widehat{\gamma}_k}/{\widehat{\gamma}_j}\cdot\widehat{\Omega}_{jk}+({\widehat{\gamma}_k}/{\widehat{\gamma}_j})^2\widehat{\Omega}_{jj}}.
\end{equation*}

\subsection{Equivalent Definition of $\VTSHT$ in \eqref{eq: initial valid}}
\label{sec: equi definition}
Recall 
$
\widehat{\mathcal{W}}=\argmax_{1\leq j\leq |\widehat{\mathcal{S}}|} \|\widehat{\Pi}_{j\cdot}\|_0.
\label{eq: winner set}
$
With $\widehat{\mathcal{W}}$, we further define the index set $\widetilde{\mathcal{V}}$ as 
\begin{equation}
\widetilde{\mathcal{V}}=\cup_{j\in \widehat{\mathcal{W}}}\left\{1\leq k\leq |\widehat{\mathcal{S}}|:\widehat{\Pi}_{j,k}=1\right\}.
\label{eq: middle}
\end{equation}
The set $\widetilde{\mathcal{V}}$ denotes the set of IVs who support (and are also supported by) at least one element in $\widehat{\mathcal{W}}.$ We finally construct the index set $\widehat{\mathcal{V}}\subset \{1,2,\cdots,|\widehat{\mathcal{S}}|\}$ as
\begin{equation}
\VTSHT=\cup_{k\in \widetilde{\mathcal{V}}}\left\{1\leq l\leq |\widehat{\mathcal{S}}|:\widehat{\Pi}_{k,l}=1\right\}.
\label{eq: initial V construction equi}
\end{equation}
This set $\VTSHT$ contains all candidate IVs that are claimed to be valid by at least one element from $\widetilde{\mathcal{V}}.$ The set defined in \eqref{eq: initial V construction equi} is equivalent to that in \eqref{eq: initial valid}. We illustrate the definitions of the voting matrix, $\widehat{\Pi}$, $\widehat{\mathcal{W}}$ and $\VTSHT$ using the following example.
\begin{Example}
We consider an example with $\pz=8$ IVs, where $\{Z_1,Z_2,Z_3, Z_4\}$ are valid IVs, $\{Z_{5},Z_6,Z_7\}$ are invalid IVs sharing the same invalidity level and $Z_8$ is invalid with a different invalidity level. The left panel of Table \ref{tab: voting table} corresponds to a favorable scenario where the valid IVs $\{Z_1,Z_2,Z_3, Z_4\}$ only vote for each other. On the right panel of Table \ref{tab: voting table}, the candidate IV $Z_5$ receives the votes (by mistake) from three valid IVs $\{Z_2,Z_3, Z_4\}.$
This might happen when the IV $Z_{5}$ is a locally invalid IV. 
\end{Example}

\begin{table}[htp!]
    \begin{minipage}{.458\linewidth}
      \centering
      \resizebox{\columnwidth}{!}{
        \begin{tabular}{r|llllllllll}
        & $Z_1$ & $Z_2$& $Z_3$&$Z_4$&$Z_5$&$Z_6$&$Z_7$&$Z_8$\\            \hline
         $Z_1$&$\yes$&$\yes$&$\yes$&$\yes$&$\no$&$\no$&$\no$&$\no$\\
         $Z_2$&$\yes$&$\yes$&$\yes$&$\yes$&$\no$&$\no$&$\no$&$\no$\\
         $Z_3$&$\yes$&$\yes$&$\yes$&$\yes$&$\no$&$\no$&$\no$&$\no$\\
         $Z_4$&$\yes$&$\yes$&$\yes$&$\yes$&$\no$&$\no$&$\no$&$\no$\\
         $Z_5$&$\no$&$\no$&$\no$&$\no$&$\yes$&$\yes$&$\yes$&$\no$\\
         $Z_6$&$\no$&$\no$&$\no$&$\no$&$\yes$&$\yes$&$\yes$&$\no$\\
         $Z_7$&$\no$&$\no$&$\no$&$\no$&$\yes$&$\yes$&$\yes$&$\no$\\
         $Z_8$&$\no$&$\no$&$\no$&$\no$&$\no$&$\no$&$\no$&$\yes$\\
         \hline 
         Votes&4&4&4&4&3&3&3&1\\
        \end{tabular}}
    \end{minipage}%
    \quad\quad\quad 
    \begin{minipage}{.458\linewidth}
      \centering
\resizebox{\columnwidth}{!}{
 \begin{tabular}{r|llllllllll}
           & $Z_1$ & $Z_2$& $Z_3$&$Z_4$&$Z_5$&$Z_6$&$Z_7$&$Z_8$\\
            \hline
        $Z_1$ &$\yes$&$\yes$&$\yes$&$\yes$&$\no$&$\no$&$\no$&$\no$\\
         \cline{6-6}
         $Z_2$&$\yes$&$\yes$&$\yes$&$\yes$&\multicolumn{1}{|c|}{$\yes$}&$\no$&$\no$&$\no$\\
         $Z_3$&$\yes$&$\yes$&$\yes$&$\yes$&\multicolumn{1}{|c|}{$\yes$}&$\no$&$\no$&$\no$\\
         $Z_4$&$\yes$&$\yes$&$\yes$&$\yes$&\multicolumn{1}{|c|}{$\yes$}&$\no$&$\no$&$\no$\\
         \cline{6-6}
         \cline{3-5}
         $Z_5$&$\no$&\multicolumn{1}{|c}{$\yes$}&$\yes$&\multicolumn{1}{c|}{$\yes$}&$\yes$&$\yes$&$\yes$&$\no$\\
         \cline{3-5}
         $Z_6$&$\no$&$\no$&$\no$&$\no$&$\yes$&$\yes$&$\yes$&$\no$\\
         $Z_7$&$\no$&$\no$&$\no$&$\no$&$\yes$&$\yes$&$\yes$&$\no$\\
         $Z_8$&$\no$&$\no$&$\no$&$\no$&$\no$&$\no$&$\no$&$\yes$\\ 
         \hline 
         Votes&4&5&5&5&6&3&3&1\\
        \end{tabular}
        }
    \end{minipage} 
    \caption{The left voting matrix $\widehat{\Pi}$ denotes that all valid IVs $\{Z_1,Z_2,Z_3,Z_4\}$ support each other but not any invalid IV. The right voting matrix $\widehat{\Pi}$ denotes that the (weakly) invalid IV $Z_5$ receives support from valid IVs $\{Z_2,Z_3,Z_4\}$ and invalid IVs $\{Z_6,Z_7\}.$} 
    \label{tab: voting table}
\end{table}

On the left panel of Table \ref{tab: voting table}, we have $\VTSHT=\widehat{\mathcal{W}}=\{1,2,3,4\}=\mathcal{V}$ and the property \eqref{eq: initial condition} is satisfied. On the right panel of Table \ref{tab: voting table}, we have 
$\widehat{\mathcal{W}}=\{5\}$ and $\VTSHT=\{1,2,3,4,5,6,7\}.$ Only $\VTSHT$ satisfies \eqref{eq: initial condition} but not $\widehat{\mathcal{W}}$.

\subsection{Union of disjoint intervals for the searching method}
\label{sec: discreteness}
We consider the setting {\bf S1} with homoscadastic errors and $\tau=({\rm VF}/\gamma_0)\cdot \sqrt{\log n/n}$ where $\text{Violation Factor} ({\rm VF})$ is varied across $\{0.5,0.75,1,1.25,1.5,1.75,2,2.25,2.5,2.75,3\}.$ We compare the two searching CIs: $\CIsearchdef$ in \eqref{eq: searching CI def} and $\CIsearch$ in \eqref{eq: searching CI}. We observe that both intervals guarantee the valid coverage property in the presence of locally invalid IVs. The lengths of these two intervals are similar in the sense that the length of $\CIsearch$ is at most $1.1$ times that of $\CIsearchdef$ when $\CIsearchdef$ is a union of disjoint intervals. When the $\text{Violation Factor} ({\rm VF})$ falls into the range $[1,2.25]$, $\CIsearchdef$ will be a union of disjoint intervals for a proportion of the 500 simulations. The proportion of disjoint intervals is at most 30\% and most times below 10\%. However, when VF is below 1 or above 2.25, the chances of observing disjointness in $\CIsearchdef$ are low.

The existence of locally invalid IVs is the main reason for the disjointness in $\CIsearchdef$. For a relatively small $\text{Violation Factor} ({\rm VF})$, some locally invalid IVs cannot be correctly separated from the valid IVs. Together with a subset of valid IVs, these locally invalid IVs may identify values of $\beta$ deviating away from the true value, creating disjoint intervals. In contrast, when the value of ${\rm VF}$ is above 2.25, the invalid IVs can be correctly identified in most simulations. Hence, the chances of having disjointness in $\CIsearchdef$ are low. On the other side, when the value of ${\rm VF}$ is very small (e.g., smaller than 1), the range of $\beta$ values supported by locally invalid IVs are not very separated from the true values, which explains why the proportion of having disjoint confidence intervals is low.

\begin{figure}[H]
    \centering
    \includegraphics[scale=0.75]{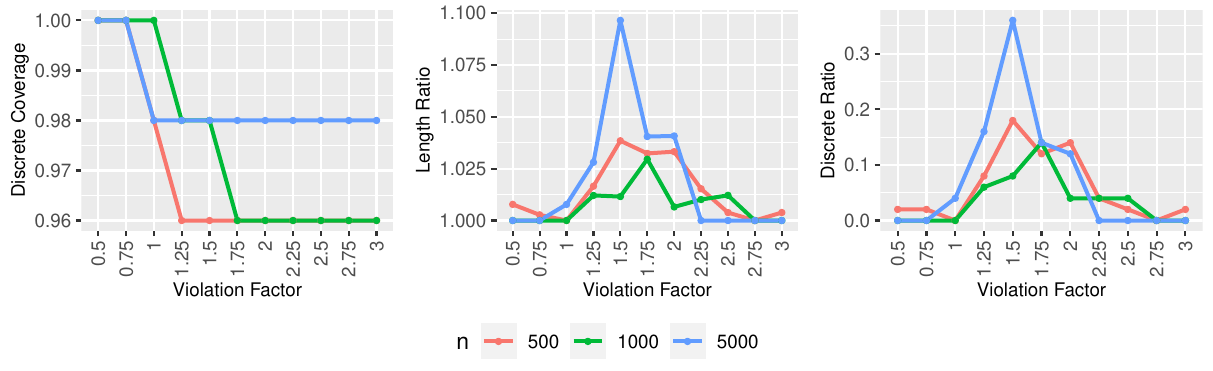}
     \caption{Comparison of $\CIsearchdef$ in \eqref{eq: searching CI def} and $\CIsearch$ in \eqref{eq: searching CI} for the setting {\bf S1} with homoscadastic errors and $\tau=({\rm VF}/\gamma_0)\cdot \sqrt{\log n/n}$ where $\text{Violation Factor} ({\rm VF})$  is varied across $\{0.5,0.75,1,1.25,1.5,1.75,2,2.25,2.5,2.75,3\}.$ The leftmost figure reports the empirical coverage of $\CIsearchdef$ in \eqref{eq: searching CI def}; the middle figure reports the ratio of the length of $\CIsearch$ in \eqref{eq: searching CI} to that of $\CIsearchdef$ in \eqref{eq: searching CI def}; the rightmost figure reports the proportion of $\CIsearchdef$ being the union of disjoint intervals.} 
    \label{fig: non_interval}
\end{figure}

}
\section{Extra Discussions on Tuning Parameter Selection}
\label{sec: extra tuning selection}
We demonstrate in the following that our proposed searching and sampling methods are  invariant to different choices of tuning parameters. 
\subsection{First-stage selection}
\label{sec: 1st stage supp}
In selecting the valid instruments, the first stage is to construct an initial set $\Shat$ of relevant IVs. Such a first-stage selection is required for \texttt{TSHT} \citet{guo2018confidence}, \texttt{CIIV} \citet{windmeijer2019confidence}, and our current proposal. We shall first show that the existing \texttt{TSHT} and \texttt{CIIV} can still achieve the desired coverage level even in the presence of individually weak IVs. Particularly, we consider the following simulation setting, 
\begin{enumerate}
    \item[\bf S0]: set $\gammap= (\gamma_1 \cdot {\bf 1}_2, 0.5 \cdot {\bf 1}_4, \gamma_1, 0.5, 0.5, 0.5, 0.5)$ and  $\pip=({\bf 0}_{2},{\bf 0}_{4},\tau/2,\tau/2,-0.5,-1)^{\intercal}$.
\end{enumerate}
We first set $\tau=0.5$ such that the invalid IVs have a large invalidity level. The first, second, and seventh IVs are generated as individually weak IVs, where the individual IV strength is controlled by $\gamma_1.$ We vary 
$\gamma_1$ across $\{0.05, 0.075, 0.10\}$.

As reported in Table \ref{tab: weak IV}, the three individually weak IVs are only included into $\Shat$ for a proportion of the 500 simulations. However, even if there are uncertainties of including these relatively weak IVs, the existing CIs by  \texttt{TSHT} and \texttt{CIIV} in general achieve the desire coverage. By comparing the results in Table \ref{tab: weak IV} and Figure \ref{fig: coverage} in the main paper, we reach the conclusion that the challenge of locally invalid IVs is fundamentally different from that of the weak IVs. Even if there are individually weak IVs, the existing CIs by \texttt{TSHT} and \texttt{CIIV} can still achieve the desired coverage level as long as there are enough strongly associated IVs. In contrast, as demonstrated in Figure \ref{fig: coverage} in the main paper, the existence of locally invalid IVs poses great challenges for inference with \texttt{TSHT} and \texttt{CIIV}.

\begin{table}[htp!]
\centering
\resizebox{0.9\linewidth}{!}{
\begin{tabular}[t]{|c|c|c c|c c|c c c|}
\hline
\multicolumn{2}{|c|}{ } & \multicolumn{2}{c|}{\TSHT} & \multicolumn{2}{c|}{\CIIV} & \multicolumn{3}{c|}{Prop of Belonging to $\Shat$} \\
\hline
$\gamma_1$ & n & Cov & Len & Cov & Len & First IV & Second IV & Seventh IV\\
\hline
 & 500 & 0.902 & 0.156 & 0.912 & 0.136 & 0.144 & 0.088 & 0.116\\
 & 1000 & 0.928 & 0.105 & 0.932 & 0.096 & 0.188 & 0.180 & 0.174\\
\multirow{-3}{*}{\centering\arraybackslash 0.050} & 2000 & 0.932 & 0.079 & 0.934 & 0.067 & 0.404 & 0.310 & 0.352\\
\cline{1-9}
 & 500 & 0.908 & 0.160 & 0.900 & 0.134 & 0.226 & 0.190 & 0.190\\
 & 1000 & 0.934 & 0.107 & 0.930 & 0.094 & 0.454 & 0.398 & 0.364\\
\multirow{-3}{*}{\centering\arraybackslash 0.075} & 2000 & 0.936 & 0.077 & 0.936 & 0.065 & 0.794 & 0.672 & 0.698\\
\cline{1-9}
 & 500 & 0.902 & 0.156 & 0.898 & 0.132 & 0.396 & 0.334 & 0.340\\
 & 1000 & 0.930 & 0.113 & 0.928 & 0.091 & 0.710 & 0.624 & 0.594\\
\multirow{-3}{*}{\centering\arraybackslash 0.100} & 2000 & 0.936 & 0.081 & 0.940 & 0.062 & 0.944 & 0.902 & 0.916\\
\hline
\end{tabular}
}
\caption{Empirical coverage (Cov) and average lengths (Len)  of CIs for the setting {\bf S0}. The columns indexed with $\TSHT$ and $\CIIV$ represent the confidence intervals by \citet{guo2018confidence} and \citet{windmeijer2019confidence}, respectively. The columns indexed with ``first IV",  ``second IV",  and ``seventh IV" report the proportion of the first IV, the second IV, and the seventh IV being included into the set $\Shat$.}
\label{tab: weak IV}
\end{table}

\begin{table}[H]
\centering
\resizebox{0.8\linewidth}{!}{
\begin{tabular}[t]{|c|c|c|c|c|c|c|c|c|c|c|}
\hline
\multicolumn{3}{|c|}{ } & \multicolumn{4}{c|}{$\sqrt{\log n}$} & \multicolumn{4}{c|}{$\sqrt{2.01\log \max\{n,\pz\}}$} \\
\cline{4-7} \cline{8-11}
\multicolumn{3}{|c|}{ } & \multicolumn{2}{c|}{Searching} & \multicolumn{2}{c|}{Sampling} & \multicolumn{2}{c|}{Searching} & \multicolumn{2}{c|}{Sampling} \\
\hline
$\tau$ & $\gamma_1$ & n & Cov & Len & Cov & Len & Cov & Len & Cov & Len\\
\hline
 &  & 500 & 0.998 & 0.640 & 0.996 & 0.418 & 0.998 & 0.637 & 0.996 & 0.417\\
 &  & 1000 & 1.000 & 0.442 & 0.996 & 0.292 & 1.000 & 0.441 & 0.996 & 0.291\\
 & \multirow{-3}{*}{\centering\arraybackslash 0.050} & 2000 & 1.000 & 0.309 & 0.996 & 0.203 & 1.000 & 0.306 & 0.998 & 0.199\\
\cline{2-11}
 &  & 500 & 1.000 & 0.648 & 0.998 & 0.424 & 0.998 & 0.636 & 0.996 & 0.418\\
 &  & 1000 & 1.000 & 0.446 & 1.000 & 0.295 & 1.000 & 0.441 & 0.996 & 0.291\\
 & \multirow{-3}{*}{\centering\arraybackslash 0.075} & 2000 & 1.000 & 0.320 & 0.998 & 0.216 & 1.000 & 0.307 & 0.998 & 0.201\\
\cline{2-11}
 &  & 500 & 1.000 & 0.660 & 0.996 & 0.426 & 0.998 & 0.636 & 0.994 & 0.420\\
 &  & 1000 & 1.000 & 0.458 & 1.000 & 0.303 & 1.000 & 0.441 & 1.000 & 0.292\\
\multirow{-9}{*}{\centering\arraybackslash 0.2} & \multirow{-3}{*}{\centering\arraybackslash 0.100} & 2000 & 1.000 & 0.337 & 1.000 & 0.229 & 1.000 & 0.316 & 1.000 & 0.210\\
\cline{1-11}
 &  & 500 & 0.998 & 0.610 & 0.992 & 0.397 & 0.998 & 0.603 & 0.992 & 0.390\\
 &  & 1000 & 1.000 & 0.409 & 0.998 & 0.255 & 1.000 & 0.407 & 1.000 & 0.252\\
 & \multirow{-3}{*}{\centering\arraybackslash 0.050} & 2000 & 1.000 & 0.284 & 1.000 & 0.174 & 1.000 & 0.283 & 1.000 & 0.170\\
\cline{2-11}
 &  & 500 & 1.000 & 0.617 & 0.998 & 0.400 & 0.998 & 0.604 & 0.994 & 0.392\\
 &  & 1000 & 1.000 & 0.417 & 0.998 & 0.259 & 1.000 & 0.408 & 1.000 & 0.251\\
 & \multirow{-3}{*}{\centering\arraybackslash 0.075} & 2000 & 1.000 & 0.295 & 1.000 & 0.185 & 1.000 & 0.283 & 1.000 & 0.174\\
\cline{2-11}
 &  & 500 & 1.000 & 0.628 & 0.996 & 0.404 & 0.998 & 0.606 & 0.994 & 0.391\\
 &  & 1000 & 0.998 & 0.434 & 0.998 & 0.273 & 1.000 & 0.410 & 1.000 & 0.254\\
\multirow{-9}{*}{\centering\arraybackslash 0.4} & \multirow{-3}{*}{\centering\arraybackslash 0.100} & 2000 & 1.000 & 0.311 & 1.000 & 0.201 & 1.000 & 0.294 & 1.000 & 0.187\\
\cline{1-11}
 &  & 500 & 0.996 & 0.596 & 0.996 & 0.378 & 0.996 & 0.590 & 0.996 & 0.372\\
 &  & 1000 & 1.000 & 0.401 & 0.998 & 0.245 & 1.000 & 0.400 & 1.000 & 0.242\\
 & \multirow{-3}{*}{\centering\arraybackslash 0.050} & 2000 & 1.000 & 0.284 & 1.000 & 0.175 & 1.000 & 0.283 & 1.000 & 0.172\\
\cline{2-11}
 &  & 500 & 0.998 & 0.603 & 0.996 & 0.382 & 0.996 & 0.591 & 0.994 & 0.374\\
 &  & 1000 & 1.000 & 0.407 & 0.996 & 0.252 & 1.000 & 0.402 & 1.000 & 0.246\\
 & \multirow{-3}{*}{\centering\arraybackslash 0.075} & 2000 & 1.000 & 0.295 & 1.000 & 0.188 & 1.000 & 0.284 & 1.000 & 0.175\\
\cline{2-11}
 &  & 500 & 0.998 & 0.616 & 0.996 & 0.388 & 0.996 & 0.593 & 0.992 & 0.374\\
 &  & 1000 & 0.998 & 0.424 & 0.998 & 0.267 & 1.000 & 0.403 & 1.000 & 0.249\\
\multirow{-9}{*}{\centering\arraybackslash 0.5} & \multirow{-3}{*}{\centering\arraybackslash 0.100} & 2000 & 1.000 & 0.318 & 1.000 & 0.206 & 1.000 & 0.302 & 1.000 & 0.192\\
\hline

\end{tabular}}
\caption{Empirical coverage (Cov) and average lengths (Len) of searching and sampling CIs for the setting {\bf S0}. The columns indexed with ``Searching" and ``Sampling" represent our proposed searching and sampling CIs in Algorithm \ref{algo: USS-plurality}. The columns indexed with $\sqrt{\log n}$ and  $\sqrt{2.01\log \max\{n,\pz\}}$ stands for the construction of $\Shat$ in \eqref{eq: relevant est} and the construction of $\Shat$ in \eqref{eq: relevant est}  with replacing $\sqrt{\log n}$ by $\sqrt{2.01 \log \max\{n,p_z\}}$, respectively. }
\label{tab: change of threshold}
\end{table}

We shall construct $\Shat$ with different threshold levels and compare the performance of our proposed searching and sampling CIs with the possibly different initial sets $\Shat$. We demonstrate the performance using the setting {\bf S0} since the change of the thresholds in the construction of $\Shat$ does not affect the performance over the settings {\bf S1} to {\bf S5}. We vary the invalidity level $\tau$ across $\{0.2,0.4,0.5\}$ and the IV strength $\gamma_1$ across $\{0.05,0.075,0.1\}.$ 
Particularly, we consider the threshold $\sqrt{\log n}$ in construction of $\Shat$ as in \eqref{eq: relevant est}. In addition, we follow \citet{guo2018confidence} and replace $\sqrt{\log n}$ with $\sqrt{2.01 \log \max\{n,p_z\}}$ in constructing the  $\Shat.$ As reported in Table \ref{tab: change of threshold}, the empirical coverage and the average lengths vary slightly with different first-stage threshold levels. 

\subsection{Searching and sampling: dependence on $L,U,a,$ $M$ and $\lambda$}
\label{sec: sampling tuning}
We further explore the effect of tuning parameters on the construction of the sampling CI. In Section \ref{sec: simulation} in the main paper, we have shown that the searching CI is almost invariant to the choices of the initial range $[L,U]$ and the grid size $n^{-a}$. In the following Table \ref{tab: tuning sampling}, we show that the sampling CI only changes slightly with different choices of $[L,U]$, $n^{-a}$ and the resampling size $M$, in terms of the empirical coverage and the average lengths.

\begin{table}[htp!]
\centering
\resizebox{0.8\linewidth}{!}{
\begin{tabular}[t]{|c|c c|c c|c c|c c|c c|}
\hline
\multicolumn{3}{|c|}{ } & \multicolumn{2}{c|}{$M=500$} & \multicolumn{2}{c|}{$M=1000$} & \multicolumn{2}{c|}{$M=1500$} & \multicolumn{2}{c|}{$M=2000$} \\
\hline
set & n & LUa & Cov & Len & Cov & Len & Cov & Len & Cov & Len\\
\hline
 &  & 1 & 1.000 & 0.301 & 1.000 & 0.333 & 1.000 & 0.349 & 1.000 & 0.356\\
 &  & 2 & 1.000 & 0.309 & 1.000 & 0.338 & 1.000 & 0.357 & 1.000 & 0.363\\
 & \multirow{-3}{*}{\centering\arraybackslash 500} & 3 & 0.998 & 0.303 & 1.000 & 0.331 & 1.000 & 0.349 & 1.000 & 0.357\\
\cline{2-11}
 &  & 1 & 0.998 & 0.204 & 0.998 & 0.221 & 0.998 & 0.233 & 0.998 & 0.240\\
 &  & 2 & 0.998 & 0.209 & 0.998 & 0.227 & 0.998 & 0.237 & 0.998 & 0.244\\
 & \multirow{-3}{*}{\centering\arraybackslash 1000} & 3 & 0.996 & 0.204 & 0.998 & 0.222 & 0.998 & 0.233 & 0.998 & 0.241\\
\cline{2-11}
 &  & 1 & 1.000 & 0.147 & 1.000 & 0.159 & 1.000 & 0.166 & 1.000 & 0.171\\
 &  & 2 & 1.000 & 0.151 & 1.000 & 0.162 & 1.000 & 0.169 & 1.000 & 0.174\\
\multirow{-9}{*}{\centering\arraybackslash S1} & \multirow{-3}{*}{\centering\arraybackslash 2000} & 3 & 1.000 & 0.147 & 1.000 & 0.159 & 1.000 & 0.166 & 1.000 & 0.171\\
\cline{1-11}
 &  & 1 & 0.958 & 0.377 & 0.958 & 0.413 & 0.960 & 0.429 & 0.968 & 0.441\\
 &  & 2 & 0.964 & 0.389 & 0.956 & 0.418 & 0.960 & 0.436 & 0.968 & 0.447\\
 & \multirow{-3}{*}{\centering\arraybackslash 500} & 3 & 0.958 & 0.378 & 0.956 & 0.413 & 0.960 & 0.430 & 0.968 & 0.442\\
\cline{2-11}
 &  & 1 & 0.976 & 0.217 & 0.980 & 0.238 & 0.980 & 0.246 & 0.982 & 0.257\\
 &  & 2 & 0.976 & 0.224 & 0.982 & 0.244 & 0.982 & 0.251 & 0.982 & 0.259\\
 & \multirow{-3}{*}{\centering\arraybackslash 1000} & 3 & 0.976 & 0.216 & 0.982 & 0.237 & 0.982 & 0.246 & 0.982 & 0.255\\
\cline{2-11}
 &  & 1 & 0.996 & 0.149 & 1.000 & 0.162 & 1.000 & 0.171 & 1.000 & 0.176\\
 &  & 2 & 0.996 & 0.154 & 1.000 & 0.166 & 0.998 & 0.174 & 1.000 & 0.178\\
\multirow{-9}{*}{\centering\arraybackslash S2} & \multirow{-3}{*}{\centering\arraybackslash 2000} & 3 & 0.996 & 0.149 & 1.000 & 0.162 & 1.000 & 0.170 & 1.000 & 0.176\\
\hline
\end{tabular}
}
\caption{Comparison of the sampling CIs with different choices of $M$ and $L,U,a$ for settings {\bf S1} to {\bf S5} with $\tau=0.4$. For column indexed with ``LUa", 1 stands for construction of $[L,U]$ as in \eqref{eq: rough bound} with $a=0.6$; 2 stands for construction of $[L,U]$ as in \eqref{eq: rough bound} with $a=0.8$; 3 stands for setting $[L,U]=[-5,5]$ and $a=0.6$. The columns indexed with ``Cov" and ``Len" denote the empirical coverage and the average lengths, respectively.  We shall increase the value of $\lambda$ such that there are $10\% M$ sampled intervals are non-empty. For example, for $M=1000$, we increase $\lambda$ until 100 sampled intervals are non-empty. }
\label{tab: tuning sampling}
\end{table}

In the following Tables \ref{tab: dependence on lambda} and \ref{tab: prop tuning}, we consider $\tau=0.2$ and $\tau=0.4$ and report the extra results for the dependence of the sampling on the choice of $\lambda$.

In the following Table \ref{tab: filtering}, we compare $\CIsample$ defined in \eqref{eq: sampling CI} and $\CIsample(\mathcal{M}_0)$ defined in \eqref{eq: sampling CI theory}  for settings {\bf S1} to {\bf S5} with $\tau=0.2$ and $\tau=0.4$. We observe that these two sampling CIs are nearly the same and the extra filtering step used for constructing $\CIsample(\mathcal{M}_0)$  in \eqref{eq: sampling CI theory} only decreases the average lengths slightly.

\begin{table}[htp!]
\centering
\resizebox{0.9\linewidth}{!}{
\begin{tabular}[t]{|c|c|c c|c c|c c|c c|c c|}
\hline
\multicolumn{2}{|c|}{ } & \multicolumn{2}{c|}{$\texttt{prop}=1\%$} & \multicolumn{2}{c|}{$\texttt{prop}=5\%$} & \multicolumn{2}{c|}{$\texttt{prop}=10\%$} & \multicolumn{2}{c|}{$\texttt{prop}=20\%$} & \multicolumn{2}{c|}{$\texttt{prop}=30\%$} \\
\hline
Set & n & Cov & Len & Cov & Len & Cov & Len & Cov & Len & Cov & Len\\
\hline
 & 500 & 0.982 & 0.231 & 1.000 & 0.298 & 1.000 & 0.336 & 1.000 & 0.383 & 1.000 & 0.412\\
 & 1000 & 0.966 & 0.155 & 0.994 & 0.210 & 0.996 & 0.240 & 0.998 & 0.272 & 1.000 & 0.294\\
\multirow{-3}{*}{\centering\arraybackslash S1} & 2000 & 0.954 & 0.099 & 0.998 & 0.142 & 0.998 & 0.167 & 0.998 & 0.194 & 1.000 & 0.210\\
\cline{1-12}
 & 500 & 0.952 & 0.257 & 0.984 & 0.321 & 0.996 & 0.370 & 0.998 & 0.426 & 1.000 & 0.460\\
 & 1000 & 0.904 & 0.172 & 0.978 & 0.228 & 0.988 & 0.260 & 0.992 & 0.294 & 0.992 & 0.311\\
\multirow{-3}{*}{\centering\arraybackslash S2} & 2000 & 0.870 & 0.109 & 0.974 & 0.162 & 0.988 & 0.188 & 0.988 & 0.213 & 0.984 & 0.226\\
\cline{1-12}
 & 500 & 0.928 & 0.265 & 0.980 & 0.389 & 0.984 & 0.446 & 0.986 & 0.521 & 0.986 & 0.554\\
 & 1000 & 0.884 & 0.184 & 0.980 & 0.253 & 0.990 & 0.286 & 0.990 & 0.322 & 0.990 & 0.339\\
\multirow{-3}{*}{\centering\arraybackslash S3} & 2000 & 0.864 & 0.114 & 0.966 & 0.168 & 0.976 & 0.195 & 0.976 & 0.223 & 0.976 & 0.240\\
\cline{1-12}
 & 500 & 0.946 & 0.471 & 0.954 & 0.484 & 0.950 & 0.484 & 0.944 & 0.493 & 0.942 & 0.510\\
 & 1000 & 0.988 & 0.375 & 0.986 & 0.375 & 0.978 & 0.380 & 0.988 & 0.398 & 0.994 & 0.416\\
\multirow{-3}{*}{\centering\arraybackslash S4} & 2000 & 0.954 & 0.217 & 0.952 & 0.222 & 0.970 & 0.226 & 0.976 & 0.245 & 0.974 & 0.259\\
\cline{1-12}
 & 500 & 0.698 & 0.306 & 0.852 & 0.379 & 0.876 & 0.407 & 0.844 & 0.424 & 0.836 & 0.430\\
 & 1000 & 0.688 & 0.280 & 0.802 & 0.308 & 0.760 & 0.310 & 0.696 & 0.304 & 0.688 & 0.305\\
\multirow{-3}{*}{\centering\arraybackslash S5} & 2000 & 0.870 & 0.254 & 0.866 & 0.254 & 0.858 & 0.256 & 0.856 & 0.266 & 0.856 & 0.277\\
\hline
\end{tabular}}
\caption{Choices of the tuning parameter $\lambda$ of the sampling method for $\tau=0.2$. The resampling size $M$ is set as $1000$. We shall increase the value of $\lambda$ such that there are $1000\cdot \texttt{prop}$ sampled intervals are non-empty. For example, for $\texttt{prop}=10\%$, we increase $\lambda$ until 100 sampled intervals are non-empty.}
\label{tab: dependence on lambda}
\end{table}

\begin{table}[H]
\centering
\resizebox{0.9\linewidth}{!}{
\begin{tabular}[t]{|c|c|c c|c c|c c|c c|c c|}
\hline
\multicolumn{2}{|c|}{ } & \multicolumn{2}{c|}{$\texttt{prop}=1\%$} & \multicolumn{2}{c|}{$\texttt{prop}=5\%$} & \multicolumn{2}{c|}{$\texttt{prop}=10\%$} & \multicolumn{2}{c|}{$\texttt{prop}=20\%$} & \multicolumn{2}{c|}{$\texttt{prop}=30\%$} \\
\hline
Set & n & Cov & Len & Cov & Len & Cov & Len & Cov & Len & Cov & Len\\
\hline
 & 500 & 0.978 & 0.214 & 1.000 & 0.290 & 1.000 & 0.330 & 1.000 & 0.389 & 1.000 & 0.423\\
 & 1000 & 0.980 & 0.168 & 0.998 & 0.198 & 1.000 & 0.222 & 1.000 & 0.254 & 1.000 & 0.277\\
\multirow{-3}{*}{\centering\arraybackslash S1} & 2000 & 0.992 & 0.133 & 1.000 & 0.143 & 1.000 & 0.159 & 1.000 & 0.179 & 1.000 & 0.194\\
\cline{1-12}
 & 500 & 0.884 & 0.268 & 0.954 & 0.354 & 0.958 & 0.410 & 0.964 & 0.467 & 0.964 & 0.495\\
 & 1000 & 0.968 & 0.190 & 0.980 & 0.214 & 0.980 & 0.234 & 0.982 & 0.269 & 0.982 & 0.288\\
\multirow{-3}{*}{\centering\arraybackslash S2} & 2000 & 0.986 & 0.141 & 0.996 & 0.148 & 0.996 & 0.163 & 1.000 & 0.185 & 1.000 & 0.200\\
\cline{1-12}
 & 500 & 0.864 & 0.345 & 0.916 & 0.478 & 0.906 & 0.518 & 0.902 & 0.560 & 0.904 & 0.583\\
 & 1000 & 0.878 & 0.229 & 0.938 & 0.287 & 0.936 & 0.316 & 0.936 & 0.347 & 0.936 & 0.364\\
\multirow{-3}{*}{\centering\arraybackslash S3} & 2000 & 0.986 & 0.148 & 0.996 & 0.158 & 0.994 & 0.172 & 0.994 & 0.197 & 0.994 & 0.212\\
\cline{1-12}
 & 500 & 0.890 & 0.791 & 0.882 & 0.787 & 0.888 & 0.799 & 0.896 & 0.827 & 0.898 & 0.849\\
 & 1000 & 0.944 & 0.563 & 0.962 & 0.572 & 0.970 & 0.589 & 0.982 & 0.612 & 0.988 & 0.630\\
\multirow{-3}{*}{\centering\arraybackslash S4} & 2000 & 0.942 & 0.199 & 0.948 & 0.206 & 0.974 & 0.223 & 0.978 & 0.244 & 0.982 & 0.257\\
\cline{1-12}
 & 500 & 0.620 & 0.697 & 0.660 & 0.711 & 0.614 & 0.701 & 0.590 & 0.702 & 0.584 & 0.705\\
 & 1000 & 0.940 & 0.784 & 0.948 & 0.786 & 0.962 & 0.795 & 0.962 & 0.808 & 0.966 & 0.819\\
\multirow{-3}{*}{\centering\arraybackslash S5} & 2000 & 0.942 & 0.283 & 0.958 & 0.291 & 0.968 & 0.303 & 0.980 & 0.321 & 0.984 & 0.333\\
\hline
\end{tabular}}
\caption{Comparison of the sampling CIs with different $\lambda$ for $\tau=0.4$. The resampling size $M$ is set as $1000$. The columns indexed with ``Cov" and ``Len" denote the empirical coverage and the average lengths, respectively.  We shall increase the value of $\lambda$ such that there are $1000\cdot \texttt{prop}$ sampled intervals are non-empty. For example, for $\texttt{prop}=10\%$, we increase $\lambda$ until 100 sampled intervals are non-empty.}
\label{tab: prop tuning}
\end{table}

\begin{table}[H]
\centering
\resizebox{0.6\linewidth}{!}{
\begin{tabular}[t]{|c|c c|c c|c c|}
\hline
\multicolumn{3}{|c|}{ } & \multicolumn{2}{c|}{$\CIsample(\mathcal{M}_0)$ in \eqref{eq: sampling CI theory}} & \multicolumn{2}{c|}{$\CIsample$ in \eqref{eq: sampling CI}} \\
\hline
set & $\tau$ & n & Cov & Len & Cov & Len\\
\hline
 &  & 500 & 1.000 & 0.339 & 1.000 & 0.344\\
 &  & 1000 & 0.998 & 0.240 & 0.998 & 0.243\\
 & \multirow{-3}{*}{\centering\arraybackslash 0.2} & 2000 & 0.998 & 0.167 & 1.000 & 0.168\\
\cline{2-7}
 &  & 500 & 1.000 & 0.332 & 1.000 & 0.339\\
 &  & 1000 & 0.998 & 0.225 & 0.998 & 0.228\\
\multirow{-6}{*}{\centering\arraybackslash S1} & \multirow{-3}{*}{\centering\arraybackslash 0.4} & 2000 & 1.000 & 0.160 & 1.000 & 0.161\\
\cline{1-7}
 &  & 500 & 0.998 & 0.374 & 0.998 & 0.377\\
 &  & 1000 & 0.996 & 0.263 & 0.996 & 0.266\\
 & \multirow{-3}{*}{\centering\arraybackslash 0.2} & 2000 & 0.984 & 0.188 & 0.988 & 0.190\\
\cline{2-7}
 &  & 500 & 0.960 & 0.413 & 0.960 & 0.417\\
 &  & 1000 & 0.980 & 0.234 & 0.980 & 0.238\\
\multirow{-6}{*}{\centering\arraybackslash S2} & \multirow{-3}{*}{\centering\arraybackslash 0.4} & 2000 & 1.000 & 0.163 & 1.000 & 0.164\\
\cline{1-7}
 &  & 500 & 0.986 & 0.453 & 0.986 & 0.460\\
 &  & 1000 & 0.990 & 0.286 & 0.988 & 0.291\\
 & \multirow{-3}{*}{\centering\arraybackslash 0.2} & 2000 & 0.972 & 0.196 & 0.976 & 0.197\\
\cline{2-7}
 &  & 500 & 0.908 & 0.527 & 0.906 & 0.533\\
 &  & 1000 & 0.936 & 0.316 & 0.936 & 0.320\\
\multirow{-6}{*}{\centering\arraybackslash S3} & \multirow{-3}{*}{\centering\arraybackslash 0.4} & 2000 & 0.994 & 0.173 & 0.994 & 0.176\\
\cline{1-7}
 &  & 500 & 0.940 & 0.487 & 0.946 & 0.497\\
 &  & 1000 & 0.990 & 0.381 & 0.990 & 0.386\\
 & \multirow{-3}{*}{\centering\arraybackslash 0.2} & 2000 & 0.968 & 0.227 & 0.962 & 0.233\\
\cline{2-7}
 &  & 500 & 0.892 & 0.799 & 0.888 & 0.807\\
 &  & 1000 & 0.968 & 0.587 & 0.976 & 0.593\\
\multirow{-6}{*}{\centering\arraybackslash S4} & \multirow{-3}{*}{\centering\arraybackslash 0.4} & 2000 & 0.974 & 0.224 & 0.966 & 0.226\\
\cline{1-7}
 &  & 500 & 0.880 & 0.408 & 0.882 & 0.418\\
 &  & 1000 & 0.756 & 0.307 & 0.766 & 0.316\\
 & \multirow{-3}{*}{\centering\arraybackslash 0.2} & 2000 & 0.860 & 0.254 & 0.854 & 0.261\\
\cline{2-7}
 &  & 500 & 0.608 & 0.697 & 0.620 & 0.709\\
 &  & 1000 & 0.958 & 0.794 & 0.962 & 0.798\\
\multirow{-6}{*}{\centering\arraybackslash S5} & \multirow{-3}{*}{\centering\arraybackslash 0.4} & 2000 & 0.974 & 0.303 & 0.978 & 0.307\\
\hline
\end{tabular}}
\caption{Comparison of $\CIsample$ defined in \eqref{eq: sampling CI} and $\CIsample(\mathcal{M}_0)$ defined in \eqref{eq: sampling CI theory}  for settings {\bf S1} to {\bf S5} with $\tau=0.2$ and $\tau=0.4$. The columns indexed with ``Cov" and ``Len" denote the empirical coverage and the average lengths, respectively.}
\label{tab: filtering}
\end{table}


\section{Proofs}
\label{sec: main proof}
\subsection{Proof Preparation}
Throughout the proof, we focus on the low-dimensional setting with heteroscedastic errors. Define $\Omega=\Sigma^{-1}$ and $\widehat{\Omega}=\widehat{\Sigma}^{-1}.$  The OLS estimators $\widehat{\gamma}$ and $\widehat{\Gamma}$ defined in \eqref{eq: OLS} satisfy the following expression, 
\begin{equation}
\widehat{\gamma}_j-\gammab_j=\widehat{\Omega}^{\intercal}_{j\cdot} \frac{1}{{n}} W^{\intercal}\delta \quad \text{and} \quad  \widehat{\Gamma}_j-\Gammab_j=\widehat{\Omega}^{\intercal}_{j\cdot} \frac{1}{{n}} W^{\intercal} \epsilon \quad \text{for}\quad 1\leq j\leq p.
\label{eq: expression for OLS}
\end{equation}
The following lemma states the asymptotic properties of the reduced-form estimators. This lemma was proved in the proof of Theorem 4.2 in \citet{wooldridge2010econometric}; see also Section 4.2.3 in  \citet{wooldridge2010econometric}.
\begin{Lemma}
Consider the model \eqref{eq: reduced-form}. Suppose that Conditions {\rm (C1)} and {\rm (C2)} hold, then, with $n\rightarrow \infty,$ we have 
\begin{equation*} 
\sqrt{n}\begin{pmatrix}\widehat{\Gamma}-\Gammap\\ \widehat{\gamma}-\gammap\end{pmatrix}
\cid N\left({\bf 0},\Cov\right) \quad \text{with}\quad \Cov=\begin{pmatrix}\V^{\Gamma}& {\C}\\ {\C}^{\intercal}& \V^{\gamma}\end{pmatrix},
\end{equation*}
where $
{\V}^{\Gamma}=\left[\Sigma^{-1}\left(\E \epsilon_i^2 W_{i\cdot} W_{i\cdot}^{\intercal}\right)\Sigma^{-1}\right]_{1:\pz,1:\pz}$, ${\V}^{\gamma}=\left[\Sigma^{-1}\left(\E \delta_i^2 W_{i\cdot} W_{i\cdot}^{\intercal}\right)\Sigma^{-1}\right]_{1:\pz,1:\pz}$, and 
${\C}=\left[\Sigma^{-1}\left(\E \epsilon_i\delta_i W_{i\cdot} W_{i\cdot}^{\intercal}\right)\Sigma^{-1}\right]_{1:\pz,1:\pz}.
$ 
\label{lem: asymp normal}
\end{Lemma}

 Define the agumented covariance matrix $${\rm Cov}^{A}=\begin{pmatrix}\Sigma^{-1}\left(\E \epsilon_i^2 W_{i\cdot} W_{i\cdot}^{\intercal}\right)\Sigma^{-1}&\Sigma^{-1}\left(\E \epsilon_i\delta_i W_{i\cdot} W_{i\cdot}^{\intercal}\right)\Sigma^{-1}\\
\Sigma^{-1}\left(\E \epsilon_i\delta_i W_{i\cdot} W_{i\cdot}^{\intercal}\right)\Sigma^{-1}&\Sigma^{-1}\left(\E \delta_i^2 W_{i\cdot} W_{i\cdot}^{\intercal}\right)\Sigma^{-1}\end{pmatrix}\in \R^{2p\times 2p}.$$
For any $u,v\in \R^{p},$ we have 
$$(u^{\intercal},v^{\intercal}){\rm Cov}^{A}(u^{\intercal},v^{\intercal})^{\intercal}= \E (U_i,V_i)(\epsilon_i,\delta_i)^{\intercal}(\epsilon_i,\delta_i)(U_i,V_i)^{\intercal},$$
where $U_i=W_{i\cdot}^{\intercal}\Sigma^{-1}u$ and $V_i=W_{i\cdot}^{\intercal}\Sigma^{-1}v.$
Since $\lambda_{\min}\left(\E\left[(\epsilon_i,\delta_i)^{\intercal}(\epsilon_i,\delta_i)\mid W_{i\cdot}\right]\right)\geq c_1>0$, $\E U^2_i=u^{\intercal}\Sigma^{-1}u$, and $\E V^2_i=v^{\intercal}\Sigma^{-1}v$,  we have 
$$(u^{\intercal},v^{\intercal}){\rm Cov}^{A}(u^{\intercal},v^{\intercal})^{\intercal}\geq c_1 \left(u^{\intercal}\Sigma^{-1}u+v^{\intercal}\Sigma^{-1}v\right)\geq c_1\lambda_{\min}(\Sigma^{-1}) \left(\|u\|_2^2+\|v\|_2^2\right).$$
This further implies 
\begin{equation}
\lambda_{\min}(\Cov)\geq \lambda_{\min}(\Cov^{A})\geq c_1\cdot \lambda_{\min}(\Sigma^{-1})\geq c_1/C_0.
\label{eq: lower bound on Cov}
\end{equation}
With a similar argument, we establish 
\begin{equation}
\lambda_{\max}(\Cov)\leq \lambda_{\max}(\Cov^{A})\leq C_1\cdot \lambda_{\max}(\Sigma^{-1})\leq C_1/c_0.
\label{eq: upper bound on Cov}
\end{equation}
Define the matrix 
\begin{equation}
{\bf R}(\beta)={\V}^{\Gamma}+\beta^2{\V}^{\gamma}-2\beta {\C} \quad \text{and} \quad  \widehat{\bf R}(\beta)=\widehat{\V}^{\Gamma}+\beta^2\widehat{\V}^{\gamma}-2\beta \widehat{\C}.
\label{eq: R def}
\end{equation}
We apply \eqref{eq: lower bound on Cov} and establish
\begin{equation}
\lambda_{\min}\left({\bf R}(\beta)\right)\geq \lambda_{\min}({\rm Cov})(1+\beta^2)\geq c_1\lambda_{\min}(\Sigma^{-1}).
\label{eq: R bound 2}
\end{equation}
Define $\mathcal{U}(a)=\{\beta\in \R:|\beta-\betap|\leq n^{-a}\}$ for some $a>0.5$ and the events 
\begin{equation}
\begin{aligned}
\mathcal{E}_0(\alpha)&=\left\{\max_{\beta \in \mathcal{U}(a)}\max_{j\in \widehat{\mathcal{S}}} \frac{|\widehat{\Gamma}_j-\Gammap_j-\beta(\widehat{\gamma}_j-\gammap_j)|}{\sqrt{(\widehat{\V}^{\Gamma}_{jj}+\beta^2\widehat{\V}^{\gamma}_{jj}-2\beta \widehat{\C}_{jj})/n}}\leq \Phi^{-1}\left(1-\frac{\alpha}{2|\Shat|}\right) \right\},\\
\widetilde{\mathcal{E}}_0(\alpha)&=\left\{\max_{\beta \in \mathcal{U}(a)}\max_{j\in \widehat{\mathcal{S}}} \frac{|\widehat{\Gamma}_j-\beta\widehat{\gamma}_j-\pip_j|}{\sqrt{(\widehat{\V}^{\Gamma}_{jj}+\beta^2\widehat{\V}^{\gamma}_{jj}-2\beta \widehat{\C}_{jj})/n}}\leq \Phi^{-1}\left(1-\frac{\alpha}{2|\Shat|}\right) \right\}.
\end{aligned}
\label{eq: good event}
\end{equation}
We present the following lemma to control the probability of the above events, whose proof is postponed to Section \ref{sec: good event threshold proof}.
\begin{Lemma}
Suppose that the conditions {\rm (C1)} and {\rm (C2)} hold, then the events $\mathcal{E}_0(\alpha)$ and $\widetilde{\mathcal{E}}_0(\alpha)$ defined in \eqref{eq: good event}  satisfy $$\liminf_{n\rightarrow \infty}\PP(\mathcal{E}_0(\alpha))\geq 1-\alpha \quad \text{and} \quad \liminf_{n\rightarrow \infty}\PP(\widetilde{\mathcal{E}}_0(\alpha))\geq 1-\alpha.$$ 
\label{lem: good event threshold}
\end{Lemma}

Define the following subset of relevant IVs,
\begin{equation}
{\mathcal{S}}^{0}=\left\{1\leq j\leq \pz: |\gammap_j|\geq (\sqrt{\log n}-C (\log n)^{1/4})\cdot \sqrt{\widehat{\V}^{\gamma}_{jj}/n}\right\}.
\label{eq: relevant est proof}
\end{equation}

Define the following events 
\begin{equation}
\begin{aligned}
\mathcal{G}_1&=\left\{\max\left\{\left\|\frac{1}{{n}} W^{\intercal} \epsilon\right\|_{\infty}, \left\|\frac{1}{{n}} W^{\intercal} \delta\right\|_{\infty}\right\}\leq C \frac{(\log n)^{1/4}}{\sqrt{n}}\right\}\\
\mathcal{G}_2&=\left\{\max_{1\leq j\leq p}\max\left\{{|\widehat{\gamma}_j-\gammab_j|}/{\sqrt{{\V}^{\gamma}_{jj}/n}},{|\widehat{\Gamma}_j-\Gammab_j|}/{\sqrt{{\V}^{\Gamma}_{jj}/n}}\right\}\leq C (\log n)^{1/4}\right\}\\
\mathcal{G}_3&=\left\{\|\widehat{\Omega}-\Sigma^{-1}\|_2\leq C\sqrt{{\log n}/{n}}\right\}\\
\mathcal{G}_4&=\left\{\max\left\{\|\widehat{\V}^{\Gamma}-{\V}^{\Gamma}\|_2,\|\widehat{\V}^{\gamma}-{\V}^{\gamma}\|_2,\|\widehat{\C}-\C\|_2\right\}\leq C {{(\log n)^{3/2}}/{\sqrt{n}}}\right\}\\
\mathcal{G}_5&=\left\{\mathcal{S}_{\rm str}\subset \widehat{\mathcal{S}}\subset \mathcal{S}^{0}\subset\mathcal{S}\right\}\\
\mathcal{G}_6&=\left\{\max_{j,k\in \widehat{\mathcal{S}}}\left|\frac{\widehat{\gamma}_k/\widehat{\gamma}_j}{\gammap_k/\gammap_j}-1\right|\leq C \frac{1}{(\log n)^{1/4}}\right\}\\
\mathcal{G}_7&=\left\{\max_{j\in \widehat{\mathcal{S}}}\left|\frac{\widehat{\Gamma}_j}{\widehat{\gamma}_j}-\frac{\Gammap_j}{\gammap_j} \right|\leq C \left(1+\left|\frac{\Gammap_j}{\gammap_j}\right|\right)\frac{1}{(\log n)^{1/4}} \right\}.
\end{aligned}
\label{eq: events}
\end{equation}
Define $\mathcal{G}=\cap_{j=1}^{7}\mathcal{G}_j.$ The following lemmas control the probability of $\mathcal{G}$, whose proof is presented in Section 
\ref{sec: good event 1 proof}.
\begin{Lemma}
Suppose that the conditions {\rm (C1)} and {\rm (C2)} hold, then for a sufficiently large $n$,
\begin{equation*}
\PP(\mathcal{G})\geq 1-\exp(-c\sqrt{\log n}),
\end{equation*}
for some positive constant $c>0.$
\label{lem: good event 1}
\end{Lemma}
Note that 
\begin{equation}
\max_{\beta\in \mathcal{U}(a)}\left\|{\bf R}(\betap)-{\bf R}(\beta)\right\|_2\leq C n^{-a}.
\label{eq: R bound 1}
\end{equation}
On the event $\mathcal{G}_4,$ we have
\begin{equation}
\max_{1\leq j\leq \pz}\left|\widehat{\bf R}_{jj}(\beta)-{\bf R}_{jj}(\beta)\right|\leq C \frac{(1+\beta^2)(\log n)^{3/2}}{\sqrt{n}},
\label{eq: R bound 3}
\end{equation}
for some positive constant $C>0.$ Together with \eqref{eq: R bound 2}, we have 
\begin{equation}
\min_{1\leq j\leq \pz} {\bf R}_{jj}(\beta)\geq  \lambda_{\min}({\rm Cov})/2(1+\beta^2).
\label{eq: R bound est}
\end{equation}

On the event $\mathcal{G},$
\begin{equation*}
\begin{aligned}
\max_{j\in \Shat}\widehat{\bf R}_{jj}(\beta)=\max_{j\in \Shat}\left[\widehat{\V}^{\Gamma}_{jj}+\beta^2\widehat{\V}^{\gamma}_{jj}-2\beta \widehat{\C}_{jj}\right]\leq C (1+\beta^2).
\end{aligned}
\end{equation*}
and hence 
\begin{equation}
\max_{j\in \Shat}\widehat{\rho}_j(\beta)\lesssim \sqrt{\frac{1+\beta^2}{n}}.
\label{eq: threshold bound}
\end{equation}

\subsection{Proof of Theorem \ref{thm: inference-searching}}
\subsubsection{The coverage property of $\CIsearchdef$ defined in \eqref{eq: searching CI def}}
\label{sec: searching CI proof}
By the decomposition \eqref{eq: error decomposition}, if $\beta$ is taken as $\betap$, then 
\begin{equation*}
(\widehat{\Gamma}_j-\betap\widehat{\gamma}_j)-\pip_j=\widehat{\Gamma}_j-\Gammap_j-\betap(\widehat{\gamma}_j-\gammap_j).
\label{eq: true value result}
\end{equation*}
Hence, on the event $\mathcal{E}_0(\alpha)$ defined in \eqref{eq: good event}, for all $j\in \mathcal{V}\cap \widehat{\mathcal{S}},$
$$
\left|\widehat{\Gamma}_{j}-\betap\widehat{\gamma}_j\right|\leq \widehat{\rho}_j(\betap) $$
where $\widehat{\rho}_j(\betap)$ is defined in \eqref{eq: key threshold}. 
This leads to
\begin{equation}
\left|\left\{j\in \widehat{\mathcal{S}}: \left|\widehat{\Gamma}_{j}-\betap\widehat{\gamma}_j\right|\leq \widehat{\rho}_j(\betap)\right\}\right|\geq \left|\mathcal{V}\cap\widehat{\mathcal{S}}\right|.
\label{eq: inter a}
\end{equation}
On the event $\mathcal{G}_5$ defined in \eqref{eq: events}, we have
\begin{equation}
|\mathcal{V}\cap\widehat{\mathcal{S}}|\geq \left|\mathcal{V}\cap{\mathcal{S}_{\rm str}}\right|>\frac{|\mathcal{S}|}{2}\geq \frac{|\widehat{\mathcal{S}}|}{2},
\label{eq: inter b}
\end{equation}
where the second inequality follows from the finite-sample majority rule and the last inequality follows from the definition of $\mathcal{G}_5$ in \eqref{eq: events}.

By combining \eqref{eq: inter a} and \eqref{eq: inter b}, we show that, on the event $\mathcal{E}_0(\alpha)\cap\mathcal{G}_5,$ 
\begin{equation}
\|\widehat{\pi}_{\widehat{\mathcal{S}}}(\betap)\|_0\leq |\widehat{\mathcal{S}}|-|\mathcal{V}\cap\widehat{\mathcal{S}}|<\frac{|\widehat{\mathcal{S}}|}{2}.
\label{eq: sparse vector}
\end{equation}
Hence, on the event $\mathcal{E}_0(\alpha)\cap \mathcal{G}_5,$ 
$\betap\in \CIsearchdef,$ that is, $$\mathbf{P}\left(\betap\in \CIsearchdef\right)\geq \mathbf{P}\left(\mathcal{E}_0(\alpha)\cap \mathcal{G}_5\right).$$
We establish the coverage property by applying Lemmas \ref{lem: good event threshold} and \ref{lem: good event 1}.

\subsubsection{The coverage property of  $\CIsearch$ defined in \eqref{eq: searching CI}}
If $j\in \mathcal{V}\cap \Shat,$ we have 
$$\PP\left(\widehat{\Gamma}_j/\widehat{\gamma}_j -\sqrt{\log n\cdot \widehat{\rm Var}\left(\widehat{\Gamma}_j/\widehat{\gamma}_j\right)}\leq \betap\leq \widehat{\Gamma}_j/\widehat{\gamma}_j +\sqrt{\log n\cdot \widehat{\rm Var}\left(\widehat{\Gamma}_j/\widehat{\gamma}_j\right)}\right)\rightarrow 1,$$
where $\widehat{\rm Var}\left(\widehat{\Gamma}_j/\widehat{\gamma}_j\right)= \frac{1}{n}\left({\widehat{\V}^\Gamma_{jj}}/{\widehat{\gamma}_j^2} + {\widehat{\V}^\gamma_{jj} \widehat{\Gamma}_j^2 }/{\widehat{\gamma}_j^4} - 2 {\widehat{\C}_{jj} \widehat{\Gamma}_j}/{\widehat{\gamma}_j^3}\right).$ Then this implies $$\PP(\betap\in [L,U])\rightarrow 1.$$
We consider two cases 
\begin{enumerate}
\item $\betap\in \mathcal{B};$
\item $\betap\not\in \mathcal{B}.$ By the construction of $\mathcal{B},$ there exists $\beta^{L},\beta^{U}\in \mathcal{B}$ such that $\beta^{L}\leq \betap\leq \beta^{U}$ and $\beta^{U}-\beta^{L}\leq n^{-a}$ for $a>0.5.$
\end{enumerate}
For the case (a), we can apply the same proof as that in Section \ref{sec: searching CI proof}. In the following, we shall modify the proof in Section \ref{sec: searching CI proof} and establish the coverage property of $\CIsearch$ for the case (b). It follows from \eqref{eq: error decomposition} that 
\begin{equation*}
\begin{aligned}
(\widehat{\Gamma}_j-\beta^{L}\widehat{\gamma}_j)-\pip_j
=\widehat{\Gamma}_j-\Gammap_j-\beta^{L}(\widehat{\gamma}_j-\gammap_j)+(\betap-\beta^{L}) \gammap_j.
\end{aligned}
\end{equation*}
On the event $\widetilde{\mathcal{E}}_0(\alpha)$ defined in \eqref{eq: good event}, we have
\begin{equation*}
\left|\widehat{\Gamma}_j-\beta^{L}\widehat{\gamma}_j\right|\leq \widehat{\rho}_j(\beta^{L}) \quad \text{for all}\quad j\in \mathcal{V}\cap \widehat{\mathcal{S}},
\label{eq: part upper}
\end{equation*}
which implies 
\begin{equation}
\left|\left\{j\in \widehat{\mathcal{S}}: \left|\widehat{\Gamma}_{j}-\beta^{L}\widehat{\gamma}_j\right|\leq \widehat{\rho}_j(\beta^{L})\right\}\right|\geq \left|\mathcal{V}\cap\widehat{\mathcal{S}}\right|.
\label{eq: inter a prime}
\end{equation}

By the same argument as in \eqref{eq: inter b} and \eqref{eq: sparse vector}, we establish that, on the event $\widetilde{\mathcal{E}}_0(\alpha)\cap\mathcal{G},$ 
\begin{equation*}
\|\widehat{\pi}_{\widehat{\mathcal{S}}}(\beta^{L})\|_0\leq |\widehat{\mathcal{S}}|-|\mathcal{V}\cap\widehat{\mathcal{S}}|<\frac{|\widehat{\mathcal{S}}|}{2}.
\end{equation*}
That is, $\beta^{L}\in \left(\beta_{\min},\beta_{\max}\right).$ With a similar argument, on the event $\widetilde{\mathcal{E}}_0(\alpha)\cap\mathcal{G},$ we have $\beta^{U}\in \left(\beta_{\min},\beta_{\max}\right).$ Then we establish $$\mathbf{P}\left(\betap\in\left(\beta^{L},\beta^{U}\right)\subset \left(\beta_{\min},\beta_{\max}\right)\right)\geq \mathbf{P}\left(\widetilde{\mathcal{E}}_0(\alpha)\cap\mathcal{G} \right).$$
We establish the coverage property by applying Lemmas \ref{lem: good event threshold} and \ref{lem: good event 1}.

\subsubsection{Length of $\CIsearchdef$ and $\CIsearch$} For the $j$-th IV with $\pip_j=0$, we simplify the decomposition in \eqref{eq: error decomposition} as 
\begin{equation*}
\widehat{\Gamma}_j-\beta\widehat{\gamma}_j=\widehat{\Gamma}_j-\Gammap_j-\beta(\widehat{\gamma}_j-\gammap_j)+(\betap-\beta) \gammap_j.
\end{equation*}
In the remaining of the proof, we assume that the event 
 $\mathcal{E}_0(\alpha)\cap \mathcal{G}$ happens. 
For $\beta$ satisfying $|\gammap_j|\cdot\left|\beta-\beta^{*}\right|\geq 2 \widehat{\rho}_j(\beta)$, we have $\widehat{\pi}_j(\beta)\neq 0$. Consequently, if $\beta$ satisfies 
\begin{equation}
\left|\beta-\beta^{*}\right|\geq \max_{j\in \Shat\cap \mathcal{V}}\frac{2 \widehat{\rho}_j(\beta)}{|\gammap_j|},
\label{eq: weak and invalid}
\end{equation}
then 
$$\|\widehat{\pi}_{\widehat{\mathcal{S}}}(\beta)\|_0\geq |\Shat\cap \mathcal{V}|>\frac{|\widehat{\mathcal{S}}|}{2},$$
where the second inequality follows from \eqref{eq: inter b}. That is, for $\beta$ satisfying \eqref{eq: weak and invalid}, $\beta\not\in \CIsearchdef$ and $\beta\not\in \CIsearch.$ If $\beta\in \CIsearchdef$ or $\beta\in \CIsearch$, then we apply \eqref{eq: weak and invalid} and obtain
\begin{equation}
\left|\beta-\beta^{*}\right|\leq \max_{j\in \Shat\cap \mathcal{V}}\frac{2 \widehat{\rho}_j(\beta)}{|\gammap_j|}\leq C \frac{\sqrt{1+\beta^2}}{\sqrt{n}\min_{j\in \Shat\cap \mathcal{V}}|\gammap_j|},
\label{eq: length upper immediate}
\end{equation}
where the last inequality follows from \eqref{eq: threshold bound}. 
By the definition of $\mathcal{S}^0$ and the event $\mathcal{G}_5,$ we further simplify the above inequality as 
\begin{equation*}
\left|\beta-\beta^{*}\right|\leq C \frac{\sqrt{1+\beta^2}}{\sqrt{\log n}}.
\label{eq: initial conclusion}
\end{equation*}
If $|\beta|\geq 2 |\betap|,$ then the above inequality implies $|\beta|\leq 2C/\sqrt{\log n-4C^2}.$
This implies that, if $\beta\in \CIsearchdef$ or $\beta\in \CIsearch$, then there exists some positive constant $C>0$ such that 
\begin{equation}
|\beta|\leq \max\{2 |\betap|,\frac{2C}{\log n}\}\leq C.
\label{eq: boundness conclu}
\end{equation}
We apply \eqref{eq: length upper immediate} again and establish that, if $\beta\in \CIsearchdef$ or $\beta\in \CIsearch$, then
\begin{equation*}
|\beta-\betap|\leq \max_{|\beta|\leq C}\max_{j\in \Shat\cap \mathcal{V}}\frac{4 \widehat{\rho}_j(\beta)}{|\gammap_j|}.
\end{equation*}
That is,   
$$\max\left\{{\bf L}(\CIsearchdef),{\bf L}(\CIsearch)\right\}\leq \max_{|\beta|\leq C} \max_{j\in \Shat\cap \mathcal{V}}\frac{4 \widehat{\rho}_j(\beta)}{|\gammap_j|},$$
where $C>0$ is some positive constant. 
We apply \eqref{eq: threshold bound} and establish 
\begin{equation*}
\max_{|\beta|\leq C}\max_{j\in \Shat\cap \mathcal{V}}\widehat{\rho}_j(\beta)\lesssim 1/\sqrt{n}.
\end{equation*}
 Then we establish the upper bound for the length of $\CIsearchdef$ and $\CIsearch$.

\subsection{Proof of Proposition \ref{prop: sampling}}
We shall establish the following result, which will imply Proposition \ref{prop: sampling}.
\begin{equation}
\liminf_{n\rightarrow\infty}\PP\left(\min_{m\in \mathcal{M}_0}\left[\max_{\beta\in \mathcal{U}(a)}\max_{j\in \Shat} \frac{|\widehat{\Gamma}^{[m]}_j-\Gammap_j-\beta(\widehat{\gamma}^{[m]}_j-\gammap_j)|}{\sqrt{(\widehat{\V}^{\Gamma}_{jj}+\beta^2\widehat{\V}^{\gamma}_{jj}-2\beta \widehat{\C}_{jj})/n}}\right]\leq C \err_n(M,\alpha_0)\right)\geq 1- \alpha_0,
\label{eq: sample quantile strong}
\end{equation}
where $\mathcal{M}_0$ is defined as
{\small
\begin{equation}
\mathcal{M}_0=\left\{1\leq m\leq M: \max_{j\in \widehat{\mathcal{S}}}\max\left\{{\left|\widehat{\gamma}^{\m}_j-\widehat{\gamma}_j\right|}/{\sqrt{\widehat{\V}^{\gamma}_{jj}/n}},{\left|\widehat{\Gamma}^{\m}_j-\widehat{\Gamma}_j\right|}/{\sqrt{\widehat{\V}^{\Gamma}_{jj}/n}}\right\}\leq 1.1 \Phi^{-1}\left(1-\frac{\alpha_0}{4|\Shat|}\right)\right\}.
\label{eq: screening set}
\end{equation}}
\noindent We take $\err_n(M,\alpha_0) =\frac{1}{2}\left[\frac{2 \log n}{c^*(\alpha_0) M}\right]^{\frac{1}{2|\Shat|}},$ and assume  $\err_n(M,\alpha_0)$ to be smaller than a sufficiently small positive constant $c$, ensuring that the following \eqref{eq: M condition 2} holds and \begin{equation}
\err_n(M,\alpha_0)<0.1 \min_{j\in \Shat}\min\{\widehat{\V}^{\Gamma}_{jj},\widehat{\V}^{\gamma}_{jj}\} \Phi^{-1}\left(1-\frac{\alpha_0}{4|\Shat|}\right). 
\label{eq: M condition 1}
\end{equation}
Note that a sufficiently large resampling size $M$ will guarantee $\err_n(M,\alpha_0)$ satisfying both \eqref{eq: M condition 1} and \eqref{eq: M condition 2}.

Denote the observed data by $\Data$, that is, $\Data=\{Y_{i\cdot},D_{i\cdot},Z_{i\cdot},X_{i\cdot}\}_{1\leq i\leq n}.$
Define $$\widehat{U}=\sqrt{n}\left[\begin{pmatrix}\widehat{\Gamma}_{\Shat}\\ \widehat{\gamma}_{\Shat}\end{pmatrix}-\begin{pmatrix}\Gammap_{\Shat}\\ \gammap_{\Shat}\end{pmatrix}\right] \quad \text{and}\quad {U}^{\m}=\sqrt{n}\left[\begin{pmatrix}\widehat{\Gamma}_{\Shat}\\ \widehat{\gamma}_{\Shat}\end{pmatrix}-\begin{pmatrix}\widehat{\Gamma}^{\m}_{\Shat}\\ \widehat{\gamma}^{\m}_{\Shat}\end{pmatrix}\right] \quad \text{for}\quad 1\leq m \leq M.$$ By slightly abusing the notation, we use the following notations throughout the rest of the proof of Proposition \ref{prop: sampling}, $$\Cov=\begin{pmatrix}\V^{\Gamma}_{\Shat,\Shat}& {\C}_{\Shat,\Shat}\\ {\C}_{\Shat,\Shat}^{\intercal}& \V^{\gamma}_{\Shat,\Shat}\end{pmatrix} \quad \text{and}\quad \widehat{\Cov}=\begin{pmatrix}\widehat{\V}_{\Shat,\Shat}^{\Gamma}& \widehat{\C}_{\Shat,\Shat}\\ \widehat{\C}_{\Shat,\Shat}^{\intercal}& \widehat{\V}^{\gamma}_{\Shat,\Shat}\end{pmatrix}.$$
Recall that $\widehat{U}$ and $\Shat$ are functions of the observed data $\Data,$ and 
\begin{equation*} 
{U}^{\m}{\mid} \Data \stackrel{i.i.d.}{\sim} N\left({\bf 0},\widehat{\Cov}\right) \quad \text{for}\quad 1\leq m\leq M.
\label{eq: re-writing} 
\end{equation*}
Let
$f(\cdot\mid {\Data})$ denote the conditional density function of $U^{\m}$ given the data $\Data$, that is, \begin{equation*}
f({U}^{m}=U\mid {\Data})=\frac{1}{\sqrt{(2\pi)^{2|\Shat|} {\rm det}(\widehat{\Cov}) }} \exp\left(-\frac{1}{2}{U}^{\intercal}\widehat{\Cov}^{-1}{U}\right).
\end{equation*}
We define the following event for the data $\Data$,
\begin{equation}
\mathcal{E}_1=\left\{\|\widehat{\Cov}-\Cov\|_2<c_2\right\}
\end{equation}
where $\|\widehat{\Cov}-\Cov\|_2$ denotes the spectral norm of the matrix  $\widehat{\Cov}-\Cov$ and $0<c_2<\lambda_{\min}(\Cov)/2$ is a small positive constant.

We define the following function to facilitate the proof,
\begin{equation}
g({U})=\frac{1}{\sqrt{(2\pi)^{2|\Shat|} {\rm det}({\Cov}+c_2 {\bf I}) }} \exp\left(-\frac{1}{2}{U}^{\intercal}({\Cov}-c_2 {\bf I})^{-1}{U}\right).
\label{eq: expression density}
\end{equation}
On the event $\Data\in \mathcal{E}_1$, we have
$
\Cov+c_2 {\bf I}\succ \widehat{\Cov}
\succ \Cov-c_2 {\bf I}\succ \frac{1}{2}\lambda_{\min}(\Cov)\cdot {\bf I},
$ where two matrices $A$ and $B$ satisfy $A\succ B$ if $A-B$ is a positive definite matrix. Hence
\begin{equation}
f(U^{\m}=U\mid {\Data}) \cdot {\bf 1}_{ \Data\in \mathcal{E}_1} \geq g(U) \cdot {\bf 1}_{ \Data\in \mathcal{E}_1}.
\label{eq: lower density}
\end{equation}

We define the following event for the data $\Data$,
\begin{equation*}
\mathcal{E}_2=\left\{\max_{j\in \widehat{\mathcal{S}}}\max\left\{{\left|\widehat{\gamma}_j-\gammap_j\right|}/{\sqrt{\widehat{\V}^{\gamma}_{jj}/n}},{\left|\widehat{\Gamma}_j-\Gammap_j\right|}/{\sqrt{\widehat{\V}^{\Gamma}_{jj}/n}}\right\}\leq \Phi^{-1}\left(1-\frac{\alpha_0}{4|\Shat|}\right)\right\}.
\end{equation*}

By the definition of $\mathcal{G}$ and Lemma \ref{lem: asymp normal}, we apply the union bound and establish
\begin{equation}
\liminf_{n\rightarrow \infty} \PP(\mathcal{E}_1\cap\mathcal{E}_2)\geq 1-\alpha_0.
\label{eq: high prob general}
\end{equation}
By the definition of the event $\mathcal{E}_1\cap \mathcal{E}_2$, we now establish a lower bound for  $g(\widehat{U}).$ On the event $\mathcal{E}_1\cap\mathcal{E}_2,$ we have 
\begin{equation*}
\begin{aligned}
\widehat{U}^{\intercal}({\Cov}-c_2 {\bf I})^{-1}\widehat{U}&\leq \frac{\sum_{j\in \Shat}\widehat{\V}^{\gamma}_{jj}+\sum_{j\in \Shat}\widehat{\V}^{\Gamma}_{jj}}{\lambda_{\min}(\Cov)/2}\cdot \left[\Phi^{-1}\left(1-\frac{\alpha_0}{4|\Shat|}\right)\right]^2\\
&\leq \frac{|\Shat|\cdot 3\lambda_{\max}(\Cov)}{\lambda_{\min}(\Cov)}\cdot \left[\Phi^{-1}\left(1-\frac{\alpha_0}{4|\Shat|}\right)\right]^2.
\end{aligned}
\end{equation*}
Hence, we apply the definition of the $g$ function in \eqref{eq: expression density} and establish
\begin{equation}
g(\widehat{U})\cdot {\bf 1}_{\mathcal{E}_1\cap\mathcal{E}_2}\geq c^{*}(\alpha_0) \quad \text{with}\; c^{*}(\alpha_0)\; \text{defined in}\; \eqref{eq: key constant 2}.
\label{eq: event}
\end{equation}
If $m\not\in \mathcal{M}_0,$ there exists $j\in \Shat$ such that 
\begin{equation*}
\frac{\left|\widehat{\gamma}^{\m}_j-\widehat{\gamma}_j\right|}{\sqrt{\widehat{\V}^{\gamma}_{jj}/n}}\geq 1.1 \Phi^{-1}\left(1-\frac{\alpha_0}{4|\Shat|}\right) \quad \text{or}\quad \frac{\left|\widehat{\Gamma}^{\m}_j-\widehat{\Gamma}_j\right|}{\sqrt{\widehat{\V}^{\Gamma}_{jj}/n}}\geq 1.1 \Phi^{-1}\left(1-\frac{\alpha_0}{4|\Shat|}\right).
\end{equation*}
On the event $\mathcal{E}_2,$ we further establish that, there exists $j\in \Shat$ such that 
\begin{equation*}
\frac{\left|\widehat{\gamma}^{\m}_j-\gammap_j\right|}{\sqrt{\widehat{\V}^{\gamma}_{jj}/n}}\geq 0.1 \Phi^{-1}\left(1-\frac{\alpha_0}{4|\Shat|}\right) \quad \text{or}\quad \frac{\left|\widehat{\Gamma}^{\m}_j-\Gammap_j\right|}{\sqrt{\widehat{\V}^{\Gamma}_{jj}/n}}\geq 0.1 \Phi^{-1}\left(1-\frac{\alpha_0}{4|\Shat|}\right).
\end{equation*}
This further implies that, on the event $\mathcal{E}_1\cap\mathcal{E}_2,$ 
\begin{equation*}
\min_{m\not\in \mathcal{M}_0}\|U^{\m}-\widehat{U}\|_{\infty}\geq 0.1 \min_{j\in \Shat}\min\{\widehat{\V}^{\Gamma}_{jj},\widehat{\V}^{\gamma}_{jj}\} \Phi^{-1}\left(1-\frac{\alpha_0}{4|\Shat|}\right)> \err_n(M,\alpha_0),
\end{equation*}
where the last inequality follows from \eqref{eq: M condition 1}.
That is, on the event $\mathcal{E}_1\cap \mathcal{E}_2,$
\begin{equation}
\left\{\min_{m\not\in \mathcal{M}_0}\|U^{\m}-\widehat{U}\|_{\infty}\right\}=\left\{\min_{1\leq m\leq M}\|U^{\m}-\widehat{U}\|_{\infty}\right\}.
\label{eq: equi smaller set}
\end{equation}

We use $\PP(\cdot \mid \Data)$ to denote the conditional probability with respect to the observed data $\Data$.
Note that
\begin{equation*}
\begin{aligned}
&\PP\left(\min_{1\leq m\leq M}\|U^{\m}-\widehat{U}\|_{\infty} \leq \err_n(M,\alpha_0)\mid \Data\right)\\
&=1-\PP\left(\min_{1\leq m\leq M}\|U^{\m}-\widehat{U}\|_{\infty} \geq \err_n(M,\alpha_0)\mid \Data\right)\\
&=1-\prod_{m=1}^{M}\left[1-\PP\left(\|U^{\m}-\widehat{U}\|_{\infty} \leq \err_n(M,\alpha_0)\mid \Data\right)\right]\\
&\geq 1-\exp\left[-\sum_{m=1}^{M}  \PP\left(\|U^{\m}-\widehat{U}\|_{\infty} \leq \err_n(M,\alpha_0)\mid \Data\right)\right],
\end{aligned}
\label{eq: exp lower}
\end{equation*}
where the second equality follows from the conditional independence of $\{U^{\m}\}_{1\leq m\leq M}$ given the data $\mathcal{O}$ and the last inequality follows from $1-x\leq e^{-x}.$ 
By applying the above inequality and \eqref{eq: equi smaller set}, we establish 
\begin{equation}
\begin{aligned}
&\PP\left(\min_{m\in \mathcal{M}_0}\|U^{\m}-\widehat{U}\|_{\infty} \leq \err_n(M,\alpha_0)\mid \Data\right)\cdot {\bf 1}_{ \Data\in \mathcal{E}_1\cap \mathcal{E}_2}\\
&=\PP\left(\min_{1\leq m\leq M}\|U^{\m}-\widehat{U}\|_{\infty} \leq \err_n(M,\alpha_0)\mid \Data\right)\cdot {\bf 1}_{ \Data\in \mathcal{E}_1\cap \mathcal{E}_2}\\
&\geq \left(1-\exp\left[-\sum_{m=1}^{M}\PP\left(\|U^{\m}-\widehat{U}\|_{\infty} \leq \err_n(M,\alpha_0)\mid \Data\right)\right]\right)\cdot {\bf 1}_{ \Data\in \mathcal{E}_1\cap \mathcal{E}_2}\\
&=1-\exp\left[-\sum_{m=1}^{M}\PP\left(\|U^{\m}-\widehat{U}\|_{\infty} \leq \err_n(M,\alpha_0)\mid \Data\right)\cdot {\bf 1}_{ \Data\in \mathcal{E}_1\cap \mathcal{E}_2}\right].
\end{aligned}
\label{eq: connection}
\end{equation}

For the remaining of the proof, we establish a lower bound for 
\begin{equation}
\PP\left(\|U^{\m}-\widehat{U}\|_{\infty} \leq \err_n(M,\alpha_0)\mid \Data\right)\cdot {\bf 1}_{ \Data\in \mathcal{E}_1\cap \mathcal{E}_2},
\label{eq: target}
\end{equation}
and then apply \eqref{eq: connection} to establish a lower bound for $$\PP\left(\min_{m\in \mathcal{M}_0}\|U^{\m}-\widehat{U}\|_{\infty} \leq \err_n(M,\alpha_0)\mid \Data\right).$$

We apply \eqref{eq: lower density} and further lower bound the targeted probability in \eqref{eq: target} as
\begin{equation}
\begin{aligned}
&\PP\left(\|U^{\m}-\widehat{U}\|_{\infty} \leq \err_n(M,\alpha_0)\mid \Data\right)\cdot {\bf 1}_{ \Data\in \mathcal{E}_1\cap \mathcal{E}_2}
\\
=& \int f(U^{\m}=U\mid \Data) \cdot {\bf 1}_{\left\{\|U-\widehat{U}\|_{\infty} \leq \err_n(M,\alpha_0)\right\}}d U \cdot {\bf 1}_{ \Data\in \mathcal{E}_1\cap \mathcal{E}_2}\\
\geq &\int g(U) \cdot {\bf 1}_{\left\{\|U-\widehat{U}\|_{\infty} \leq \err_n(M,\alpha_0)\right\}}d U\cdot {\bf 1}_{ \Data\in \mathcal{E}_1\cap \mathcal{E}_2}\\
=& \int g(\widehat{U}) \cdot {\bf 1}_{\left\{\|U-\widehat{U}\|_{\infty} \leq \err_n(M,\alpha_0)\right\}}d U \cdot {\bf 1}_{ \Data\in \mathcal{E}_1\cap \mathcal{E}_2}\\
&+\int [g(U)-g(\widehat{U})] \cdot {\bf 1}_{\left\{\|U-\widehat{U}\|_{\infty} \leq \err_n(M,\alpha_0)\right\}}d U\cdot {\bf 1}_{ \Data\in \mathcal{E}_1\cap \mathcal{E}_2}. 
\end{aligned}
\label{eq: decomposition density}
\end{equation}
By \eqref{eq: event}, we establish 
\begin{equation}
\begin{aligned}
&\int g(\widehat{U}) \cdot {\bf 1}_{\left\{\|U-\widehat{U}\|_{\infty} \leq \err_n(M,\alpha_0)\right\}}d U \cdot {\bf 1}_{ \Data\in \mathcal{E}_1\cap \mathcal{E}_2}\\
&\geq c^{*}(\alpha_0) \cdot \int {\bf 1}_{\left\{\|U-\widehat{U}\|_{\infty} \leq \err_n(M,\alpha_0)\right\}}d U \cdot {\bf 1}_{ \Data\in \mathcal{E}_1\cap \mathcal{E}_2}\\
&\geq c^{*}(\alpha_0)\cdot [2\err_n(M,\alpha_0)]^{2|\Shat|} \cdot {\bf 1}_{ \Data\in \mathcal{E}_1\cap \mathcal{E}_2}.
\end{aligned}
\label{eq: main density}
\end{equation}
There exists $t\in (0,1)$ such that
$$g(U)-g(\widehat{U})=[\triangledown g(\widehat{U}+t(U-\widehat{U}))]^{\intercal} (U-\widehat{U}),$$
with $$\triangledown g(u)=\frac{1}{\sqrt{(2\pi)^{2|\Shat|} {\rm det}({\Cov}+c_2 {\bf I}) }} \exp\left(-\frac{1}{2}{u}^{\intercal}({\Cov}-c_2 {\bf I})^{-1}{u}\right)^{-1}({\Cov}-c_2 {\bf I})^{-1}{u}.$$ 
Since $\lambda_{\min}({\Cov}-c_2 {\bf I})\geq \lambda_{\min}(\Cov)/2,$ 
then $\triangledown g$ is bounded and there exists a positive constant $C>0$ such that $$\left|g(U)-g(\widehat{U})\right|\leq C \sqrt{2 |\Shat|}\|U-\widehat{U}\|_{\infty}.$$ Then we establish  
\begin{equation}
\begin{aligned}
&\left|\int [g(U)-g(\widehat{U})] \cdot {\bf 1}_{\left\{\|U-\widehat{U}\|_{\infty} \leq \err_n(M,\alpha_0)\right\}}d U\cdot {\bf 1}_{ \Data\in \mathcal{E}_1\cap \mathcal{E}_2}
\right|\\
&\leq C \sqrt{2 |\Shat|}\cdot \err_n(M,\alpha_0)\cdot \int {\bf 1}_{\left\{\|U-\widehat{U}\|_{\infty} \leq \err_n(M,\alpha_0)\right\}}d U \cdot {\bf 1}_{ \Data\in \mathcal{E}_1\cap \mathcal{E}_2}\\
&= C \sqrt{2 |\Shat|}\cdot \err_n(M,\alpha_0)\cdot [2\err_n(M,\alpha_0)]^{2|\Shat|} \cdot {\bf 1}_{ \Data\in \mathcal{E}_1\cap \mathcal{E}_2}.
\end{aligned}
\label{eq: approx density}
\end{equation}
Note that we assume $\err_n(M,\alpha_0)$ to be sufficiently small such that 
\begin{equation}
C\sqrt{2|\Shat|} \cdot \err_n(M,\alpha_0) \leq \frac{1}{2}c^{*}(\alpha_0),
\label{eq: M condition 2}
\end{equation}
where $c^{*}(\alpha_0)$ is a positive constant. We combine the above inequality, \eqref{eq: decomposition density}, \eqref{eq: main density} and \eqref{eq: approx density} and obtain 
\begin{equation*}
\begin{aligned}
\PP\left(\|U-\widehat{U}\|_{\infty} \leq \err_n(M,\alpha_0)\mid \Data\right)\cdot {\bf 1}_{ \Data\in \mathcal{E}_1\cap \mathcal{E}_2} \geq \frac{1}{2}c^{*}(\alpha_0)\cdot[2\err_n(M,\alpha_0)]^{2|\Shat|} \cdot {\bf 1}_{ \Data\in \mathcal{E}_1\cap \mathcal{E}_2}.
\end{aligned}
\end{equation*}
Together with \eqref{eq: connection}, we establish
\begin{equation}
\begin{aligned}
&\PP\left(\min_{m\in \mathcal{M}_0}\|U^{\m}-\widehat{U}\|_{\infty} \leq \err_n(M,\alpha_0)\mid \Data\right)\cdot {\bf 1}_{ \Data\in \mathcal{E}_1\cap \mathcal{E}_2}\\
&\geq 1-\exp\left[-M\cdot\frac{1}{2}c^{*}(\alpha_0)\cdot [2\err_n(M,\alpha_0)]^{2|\Shat|} \cdot {\bf 1}_{\Data\in \mathcal{E}_1\cap \mathcal{E}_2}\right]\\
&=\left(1-\exp\left[-M\cdot\frac{1}{2}c^{*}(\alpha_0)\cdot [2\err_n(M,\alpha_0)]^{2|\Shat|} \right]\right)
\cdot {\bf 1}_{ \Data\in \mathcal{E}_1\cap \mathcal{E}_2}.
\end{aligned}
\label{eq: connection 2}
\end{equation}
With $\E_{\Data}$ denoting the expectation taken with respect to the observed data $\Data,$ we further integrate with respect to $\Data$ and establish
\begin{equation*}
\begin{aligned}
&\PP\left(\min_{m\in \mathcal{M}_0}\|U^{\m}-\widehat{U}\|_{\infty} \leq \err_n(M,\alpha_0)\right)\\
&=\E_{\Data}\left[\PP\left(\min_{m\in \mathcal{M}_0}\|U^{\m}-\widehat{U}\|_{\infty} \leq \err_n(M,\alpha_0)\mid \Data\right)\right]\\
&\geq \E_{\Data}\left[\PP\left(\min_{m\in \mathcal{M}_0}\|U^{\m}-\widehat{U}\|_{\infty} \leq \err_n(M,\alpha_0)\mid \Data\right)\cdot {\bf 1}_{ \Data\in \mathcal{E}_1\cap \mathcal{E}_2}\right]\\
&\geq \E_{\Data}\left[\left(1-\exp\left[-M\cdot \frac{1}{2}c^*(\alpha_0)\cdot [2\err_n(M,\alpha_0)]^{2|\Shat|} \right]\right)\cdot {\bf 1}_{ \Data\in \mathcal{E}_1\cap \mathcal{E}_2}\right].
\end{aligned}
\end{equation*}
By plugging in the expression $$\err_n(M,\alpha_0) =\frac{1}{2}\left[\frac{2 \log n}{c^*(\alpha_0) M}\right]^{\frac{1}{2|\Shat|}},$$ we establish 
$$\PP\left(\min_{m\in \mathcal{M}_0}\|U^{\m}-\widehat{U}\|_{\infty} \leq \err_n(M,\alpha_0)\right)\geq (1-n^{-1})\cdot \PP\left(\mathcal{E}_1\cap \mathcal{E}_2\right).$$
We further apply \eqref{eq: high prob general} and establish
\begin{equation*}
\begin{aligned}
\liminf_{n\rightarrow\infty}\PP\left(\min_{m\in \mathcal{M}_0}\|U^{\m}-\widehat{U}\|_{\infty} \leq \err_n(M,\alpha_0)\right)\geq \PP\left(\mathcal{E}_1\cap \mathcal{E}_2\right)\geq 1-\alpha_0.
\end{aligned}
\end{equation*}
We use $m^*$ to denote the index such that 
$$\|U^{[m^*]}-\widehat{U}\|_{\infty}=\min_{m\in \mathcal{M}_0}\|U^{\m}-\widehat{U}\|_{\infty}.$$
Then we have 
\begin{equation}
\max_{j\in \Shat}\left|\widehat{\gamma}^{[m^{*}]}_j-\gammap_j\right|\leq \frac{\err_n(M,\alpha_0)}{\sqrt{n}}, \quad
\max_{j\in \Shat}\left|\widehat{\Gamma}^{[m^{*}]}_j-\Gammap_j-\beta(\widehat{\gamma}^{[m^{*}]}_j-\gammap_j)\right|\leq (1+|\beta|) \frac{\err_n(M,\alpha_0)}{\sqrt{n}}.
\label{eq: fastest bound}
\end{equation}
By combining the above inequality and \eqref{eq: R bound est}, we establish that there exists some positive constant $C>0$ such that
\begin{equation}
\max_{j\in \Shat} \frac{|\widehat{\Gamma}^{[m^*]}_j-\Gammap_j-\beta(\widehat{\gamma}^{[m^*]}_j-\gammap_j)|}{\sqrt{(\widehat{\V}^{\Gamma}_{jj}+\beta^2\widehat{\V}^{\gamma}_{jj}-2\beta \widehat{\C}_{jj})/n}}\leq 0.9 C {\err_n(M,\alpha_0)}.
\label{eq: important inter bound}
\end{equation}
Note that 
\begin{equation*}
\begin{aligned}
&\left|\max_{j\in \Shat} \frac{|\widehat{\Gamma}^{[m^*]}_j-\Gammap_j-\beta(\widehat{\gamma}^{[m^*]}_j-\gammap_j)|}{\sqrt{(\widehat{\V}^{\Gamma}_{jj}+\beta^2\widehat{\V}^{\gamma}_{jj}-2\beta \widehat{\C}_{jj})/n}}-\max_{j\in \Shat}\frac{|\widehat{\Gamma}^{[m^*]}_j-\Gammap_j-\betap(\widehat{\gamma}^{[m^*]}_j-\gammap_j)|}{\sqrt{(\widehat{\V}^{\Gamma}_{jj}+[\betap]^2\widehat{\V}^{\gamma}_{jj}-2\betap \widehat{\C}_{jj})/n}}\right|\\
&\leq\max_{j\in \Shat}\left| \frac{|\widehat{\Gamma}^{[m^*]}_j-\Gammap_j-\beta(\widehat{\gamma}^{[m^*]}_j-\gammap_j)|}{\sqrt{(\widehat{\V}^{\Gamma}_{jj}+\beta^2\widehat{\V}^{\gamma}_{jj}-2\beta \widehat{\C}_{jj})/n}}-\frac{|\widehat{\Gamma}^{[m^*]}_j-\Gammap_j-\betap(\widehat{\gamma}^{[m^*]}_j-\gammap_j)|}{\sqrt{(\widehat{\V}^{\Gamma}_{jj}+[\betap]^2\widehat{\V}^{\gamma}_{jj}-2\betap \widehat{\C}_{jj})/n}}\right|\\
&\leq\max_{j\in \Shat}\left| \frac{(\beta-\betap)(\widehat{\gamma}^{[m^*]}_j-\gammap_j)}{\sqrt{\widehat{\RR}_{jj}(\beta)/{n}}}\right|+\max_{j\in \Shat}\left|\frac{|\widehat{\Gamma}^{[m^*]}_j-\Gammap_j-\betap(\widehat{\gamma}^{[m^*]}_j-\gammap_j)|}{{\sqrt{\widehat{\RR}_{jj}(\beta)/{n}}}}\left(\sqrt{\frac{\widehat{\RR}_{jj}(\beta)}{\widehat{\RR}_{jj}(\betap)}}-1\right)\right|\\
&\leq C {\err_n(M,\alpha_0)}\left(\left|\beta-\betap\right|+\left|\frac{\widehat{\RR}_{jj}(\beta)}{\widehat{\RR}_{jj}(\betap)}-1\right|\right)\leq 0.1\cdot C {\err_n(M,\alpha_0)},
\end{aligned}
\end{equation*}
where the third inequality follows from \eqref{eq: fastest bound} and the last inequality follows from the fact $\left|\beta-\betap\right|\leq n^{-a}$ and \eqref{eq: R bound 1}, \eqref{eq: R bound 2}, and \eqref{eq: R bound 3}. 
Together with \eqref{eq: important inter bound}, we have 
\begin{equation*}
\max_{\beta\in \mathcal{U}(a)}\max_{j\in \Shat} \frac{|\widehat{\Gamma}^{[m^*]}_j-\Gammap_j-\beta(\widehat{\gamma}^{[m^*]}_j-\gammap_j)|}{\sqrt{(\widehat{\V}^{\Gamma}_{jj}+\beta^2\widehat{\V}^{\gamma}_{jj}-2\beta \widehat{\C}_{jj})/n}}\leq C {\err_n(M,\alpha_0)}.
\end{equation*}
Hence, we have established \eqref{eq: sample quantile strong}.

\subsection{Proof of Theorem \ref{thm: inference-sampling}}

We apply the definitions of 
 $\CIsample$ in \eqref{eq: sampling CI} and $\CIsample(\mathcal{M}_0)$  in \eqref{eq: sampling CI theory} and establish 
$$
\liminf_{n\rightarrow \infty}\PP\left(\betap\in \CIsample\right)\geq  \liminf_{n\rightarrow \infty}\PP\left(\betap\in \CIsample(\mathcal{M}_0)\right).$$
Hence, it is sufficient to control the probability $\PP\left(\betap\in \CIsample(\mathcal{M}_0)\right).$ 



\subsubsection{The coverage property of  $\CIsample(\mathcal{M}_0)$ in \eqref{eq: sampling CI theory}} Recall the definition of  $\CIsample(\mathcal{M}_0)$ in \eqref{eq: sampling CI theory},
\begin{equation*}
\CIsample(\mathcal{M}_0)=\left(\min_{m\in \mathcal{M}'}\beta^{\m}_{\min}(\lambda),\max_{m\in \mathcal{M}'}\beta^{\m}_{\max}(\lambda)\right),
\end{equation*}
where $\mathcal{M}'=\{ m\in \mathcal{M}_0: [\beta^{\m}_{\min}(\lambda),\beta^{\m}_{\max}(\lambda)]\neq \emptyset \},
$ with $\mathcal{M}_0$ defined in \eqref{eq: screening set}.

We define the event
{\small
\begin{equation*}
\begin{aligned}
\mathcal{E}_3=\left\{\text{there exists} \; m^*\in \mathcal{M}_0\;\text{such that}\; \max_{\beta\in \mathcal{U}(a)}\max_{j\in \Shat} \frac{|\widehat{\Gamma}^{[m^*]}_j-\Gammap_j-\beta(\widehat{\gamma}^{[m^*]}_j-\gammap_j)|}{\sqrt{(\widehat{\V}^{\Gamma}_{jj}+\beta^2\widehat{\V}^{\gamma}_{jj}-2\beta \widehat{\C}_{jj})/n}}\leq \frac{\lambda}{2}\cdot \Phi^{-1}\left(1-\frac{\alpha}{2|\Shat|}\right)\right\}.
\end{aligned}
\end{equation*}}
By assuming
$$\frac{1}{2}\lambda\cdot \Phi^{-1}\left(1-\frac{\alpha}{2|\Shat|}\right)\geq C {\err_n(M,\alpha_0)},$$
we apply Proposition \ref{prop: sampling} and establish that 
\begin{equation}
\liminf_{n\rightarrow \infty}\PP\left(\mathcal{E}_3\right)\geq 1-\alpha_0.
\label{eq: E3 coverage}
\end{equation}
Using a similar decomposition as \eqref{eq: error decomposition}, we have
\begin{equation}
\begin{aligned}
\left(\widehat{\Gamma}^{\m}_j-\beta\widehat{\gamma}^{\m}_j\right)-\pip_j
=\widehat{\Gamma}^{\m}_j-\Gammap_j-\beta(\widehat{\gamma}^{\m}_j-\gammap_j)+(\betap-\beta) \gammap_j \quad \text{for}\quad 1\leq m\leq M.
\end{aligned}
\label{eq: error decomposition sample}
\end{equation}
We consider two cases 
\begin{enumerate}
\item[(a)] $\betap\in \mathcal{B};$
\item[(b)] $\betap\not\in \mathcal{B}.$ By the construction of $\mathcal{B},$ there exists $\beta^{L},\beta^{U}\in \mathcal{B}$ such that $\beta^{L}\leq \betap\leq \beta^{U}$ and $\beta^{U}-\beta^{L}\leq n^{-a}$ for $a>0.5.$
\end{enumerate}
\noindent \textbf{Case (a).} If $\beta$ is taken as $\betap$, then 
\begin{equation}
(\widehat{\Gamma}^{\m}_j-\betap\widehat{\gamma}^{\m}_j)-\pip_j=\widehat{\Gamma}^{\m}_j-\Gammap_j-\betap(\widehat{\gamma}^{\m}_j-\gammap_j).
\label{eq: true value result sample}
\end{equation}

On the event $\mathcal{E}_3$, we show that there exists $1\leq m^{*}\leq M$ such that 
\begin{equation}
\left|\widehat{\Gamma}^{[m^{*}]}_{j}-\betap\widehat{\gamma}^{[m^{*}]}_j\right|\leq  \lambda \widehat{\rho}_j(\betap) \quad \text{for any}\; j\in \mathcal{V}\cap \widehat{\mathcal{S}}.
\label{eq: key upper sample}
\end{equation}
The above inequality implies 
\begin{equation}
\left|\left\{j\in \widehat{\mathcal{S}}: \left|\widehat{\Gamma}^{[m^{*}]}_{j}-\betap\widehat{\gamma}^{[m^{*}]}_j\right|\leq \lambda \widehat{\rho}_j(\beta)\right\}\right|\geq \left|\mathcal{V}\cap\widehat{\mathcal{S}}\right|.
\label{eq: inter 1 sample}
\end{equation}
Together with the definition in \eqref{eq: thresholding sample} and \eqref{eq: inter b}, we show that, on the event $\mathcal{G}_5\cap\mathcal{E}_3,$ $$\|\widehat{\pi}^{[m^{*}]}_{\widehat{\mathcal{S}}}(\betap,\lambda)\|_0\leq |\widehat{\mathcal{S}}|-|\mathcal{V}\cap\widehat{\mathcal{S}}|<\frac{|\widehat{\mathcal{S}}|}{2}.$$
Then on the event $ \mathcal{G}_5\cap \mathcal{E}_3,$ 
\begin{equation}
\betap\in \left(\beta^{[m^*]}_{\min},\beta^{[m^{*}]}_{\max}\right)\subset \CIsample(\mathcal{M}_0).
\label{eq: key covering step}
\end{equation} Hence the definition of  $\CIsample(\mathcal{M}_0)$ in \eqref{eq: sampling CI theory} implies $$\mathbf{P}\left(\betap\in \CIsample(\mathcal{M}_0)\right)\geq \mathbf{P}\left(\mathcal{G}_5\cap \mathcal{E}_{3}\right).$$
Together with Lemma \ref{lem: good event 1} and \eqref{eq: E3 coverage}, we establish $\PP\left(\betap\in \CIsample(\mathcal{M}_0)\right)\geq 1-\alpha_0.$

\noindent \textbf{Case (b).} It follows from \eqref{eq: error decomposition sample} that 
\begin{equation}
\begin{aligned}
\widehat{\Gamma}^{\m}_j-\beta^{L}\widehat{\gamma}^{\m}_j-\pip_j 
=\widehat{\Gamma}^{\m}_j-\Gammap_j-\beta^{L}(\widehat{\gamma}^{\m}_j-\gammap_j)+(\betap-\beta^{L}) \gammap_j\quad \text{for}\quad 1\leq m\leq M.
\end{aligned}
\label{eq: error decomposition betaL}
\end{equation}
On the event $\mathcal{E}_3$, there exists $1\leq m^{*}\leq M$ such that, for any  $j\in \mathcal{V}\cap \widehat{\mathcal{S}},$
\begin{equation}
\left|\widehat{\Gamma}^{[m^{*}]}_j-\Gammap_j-\beta^{L}(\widehat{\gamma}^{[m^{*}]}_j-\gammap_j)\right|\leq \frac{1}{2}\lambda \widehat{\rho}_j(\beta^{L}).
\label{eq: key upper sample betaL}
\end{equation}
Note that $\lambda \gg n^{1/2-a}$ implies $$|(\betap-\beta^{L}) \gammap_j|\leq n^{-a}|\gammap_j|\leq \frac{1}{2}\lambda \widehat{\rho}_j(\beta^{L}).$$
Combined with \eqref{eq: error decomposition betaL} and \eqref{eq: key upper sample betaL}, we establish 
\begin{equation*}
\left|\widehat{\Gamma}^{[m^{*}]}_{j}-\beta^{L}\widehat{\gamma}^{[m^{*}]}_j\right|
\leq \lambda \widehat{\rho}_j(\beta^{L}) \quad \text{for any}\; j\in \mathcal{V}\cap \widehat{\mathcal{S}}.
\end{equation*}
This is similar to the result in \eqref{eq: key upper sample} with replacing $\betap$ by $\beta^{L}.$ Then we apply a similar argument as that of \eqref{eq: key covering step} and establish that, on the event $\mathcal{G}\cap\mathcal{E}_3,$
$$
\beta^{L}\in \left(\beta^{[m^*]}_{\min},\beta^{[m^{*}]}_{\max}\right)\subset \CIsample(\mathcal{M}_0).
$$ Similarly, we can show that, on the event $\mathcal{G}\cap\mathcal{E}_3,$
 $\beta^{U}\in \CIsample(\mathcal{M}_0).$ It follows from the definition of $\CIsample(\mathcal{M}_0)$ in \eqref{eq: sampling CI theory} that on the event $\mathcal{G}\cap\mathcal{E}_3,$
$\betap\in \CIsample(\mathcal{M}_0).$ Hence $$\mathbf{P}\left(\betap\in \CIsample(\mathcal{M}_0)\right)\geq \mathbf{P}\left(\mathcal{G}\cap \mathcal{E}_3\right).$$
Together with Lemma \ref{lem: good event 1} and \eqref{eq: E3 coverage}, we establish $\PP\left(\betap\in \CIsample(\mathcal{M}_0)\right)\geq 1-\alpha_0.$


\subsubsection{Length of $\CIsample$ in \eqref{eq: sampling CI} and $\CIsample(\mathcal{M}_0)$ in \eqref{eq: sampling CI theory}} 

Recall the definition of the index sets $$\mathcal{M}'=\{ m\in \mathcal{M}_0: [\beta^{\m}_{\min}(\lambda),\beta^{\m}_{\max}(\lambda)]\neq \emptyset \}, \quad \mathcal{M}=\{ 1\leq m\leq M: [\beta^{\m}_{\min}(\lambda),\beta^{\m}_{\max}(\lambda)]\neq \emptyset\},
$$ with $\mathcal{M}_0$ defined in \eqref{eq: screening set}.
We define the following events 
\begin{equation*}
\begin{aligned}
&\mathcal{E}_4=
\left\{\max_{m\in \mathcal{M}'}\max_{j\in \widehat{\mathcal{S}}} \left(\left|\widehat{\Gamma}^{\m}_j-\Gammap_j\right|+\left|\widehat{\gamma}^{\m}_j-\gammap_j\right|\right)\leq C \frac{1}{\sqrt{n}}\right\},\\
&\mathcal{E}_5=
\left\{\max_{m\in \mathcal{M}}\max_{j\in \widehat{\mathcal{S}}} \left(\left|\widehat{\Gamma}^{\m}_j-\Gammap_j\right|+\left|\widehat{\gamma}^{\m}_j-\gammap_j\right|\right)\leq C \frac{\sqrt{\log |\mathcal{M}|}}{\sqrt{n}}\right\},
\end{aligned}
\end{equation*}
where 
for some positive constant $C>0$ independent of $n$. 
For any $1\leq m\leq M,$ we have 
\begin{equation}
\begin{aligned}
\max_{j\in \widehat{\mathcal{S}}} \left|\widehat{\Gamma}^{\m}_j-\Gammap_j\right|
\leq \max_{j\in \widehat{\mathcal{S}}} \left|\widehat{\Gamma}^{\m}_j-\widehat{\Gamma}_j\right|
+\max_{j\in \widehat{\mathcal{S}}} \left|\widehat{\Gamma}_j-\Gammap_j\right|.
\end{aligned}
\label{eq: key triangle}
\end{equation}
On the event $\mathcal{E}_2$ and the definition of $\mathcal{M}_0,$ we apply \eqref{eq: key triangle} and establish  
\begin{equation*}
\begin{aligned}
\max_{m\in \mathcal{M}'}\max_{j\in \widehat{\mathcal{S}}} \left|\widehat{\Gamma}^{\m}_j-\Gammap_j\right|
\lesssim \max_{j\in \Shat}\sqrt{\widehat{\V}^{\Gamma}_{jj}/n}\cdot \Phi^{-1}\left(1-\frac{\alpha_0}{4|\Shat|}\right).
\end{aligned}
\end{equation*}
By a similar argument, we can establish that, on the event $\mathcal{E}_2\cap \mathcal{G},$ $$\max_{m\in \mathcal{M}'}\max_{j\in \widehat{\mathcal{S}}} \left|\widehat{\gamma}^{\m}_j-\gammap_j\right|\lesssim \max_{j\in \Shat}\sqrt{\widehat{\V}^{\gamma}_{jj}/n}\cdot \Phi^{-1}\left(1-\frac{\alpha_0}{4|\Shat|}\right).$$ 
The above two inequalities imply 
\begin{equation}
\liminf_{n\rightarrow \infty}\PP(\mathcal{E}_4)\geq \liminf_{n\rightarrow \infty}\PP(\mathcal{E}_2\cap \mathcal{G})\geq 1-\alpha_0.
\label{eq: event 4 prob}
\end{equation}
We now consider the event $\mathcal{E}_5$ and the only difference is that 
\begin{equation*}
\PP\left(\max_{m\in \mathcal{M}}\max_{j\in \widehat{\mathcal{S}}} \left|\widehat{\Gamma}^{\m}_j-\widehat{\Gamma}_j\right|\geq C \max_{j\in \Shat}\sqrt{\widehat{\V}^{\Gamma}_{jj}/n} \sqrt{\log |\mathcal{M}|}\right)\leq |\mathcal{M}|^{-c}
\end{equation*}
for some positive constants $c>0$ and $C>0.$ Hence we establish 
\begin{equation}
\liminf_{n\rightarrow \infty}\PP(\mathcal{E}_5)\geq 1-\alpha_0.
\label{eq: event 5 prob}
\end{equation}

We now control the length of the sampling interval. For $1\leq m\leq M$ and $j\in \mathcal{V}\cap \widehat{\mathcal{S}},$ we have
\begin{equation*}
\widehat{\Gamma}^{\m}_j-\beta\widehat{\gamma}^{\m}_j=\widehat{\Gamma}^{\m}_j-\Gammap_j-\beta(\widehat{\gamma}^{\m}_j-\gammap_j)+(\betap-\beta) \gammap_j.
\end{equation*}

For $\beta$ satisfying $$|\gammap_j|\cdot\left|\beta-\beta^{*}\right|\geq \left|\widehat{\Gamma}^{\m}_j-\Gammap_j-\beta(\widehat{\gamma}^{\m}_j-\gammap_j)\right|+\lambda \widehat{\rho}_j(\beta),$$ we have $\widehat{\pi}_j^{\m}(\beta,\lambda)\neq 0$ for $j\in \mathcal{V}\cap \widehat{\mathcal{S}}.$ 
Consequently, if $\beta$ satisfies
\begin{equation}
\left|\beta-\beta^{*}\right|\geq \max_{j\in \Shat\cap \mathcal{V}}\frac{\left|\widehat{\Gamma}^{\m}_j-\Gammap_j-\beta(\widehat{\gamma}^{\m}_j-\gammap_j)\right|+\lambda \widehat{\rho}_j(\beta)}{|\gammap_j|},
\label{eq: contradict assumption}
\end{equation}
then we have $$\|\widehat{\pi}^{\m}_{\Shat}(\beta,\lambda)\|_0\geq |\Shat\cap \mathcal{V}|>{|\widehat{\mathcal{S}}|}/{2},$$
where the second inequality follows from \eqref{eq: inter b}. The above inequality implies that if $\beta$ satisfies \eqref{eq: contradict assumption}, then $\beta\not\in \left[\beta^{\m}_{\min},\beta^{\m}_{\max}\right].$ That is, 
$$
\left|\beta^{\m}_{\min}-\beta^{*}\right|\leq \max_{j\in \Shat\cap \mathcal{V}}\frac{\left|\widehat{\Gamma}^{\m}_j-\Gammap_j-\beta^{\m}_{\min}(\widehat{\gamma}^{\m}_j-\gammap_j)\right|+\lambda \widehat{\rho}_j(\beta^{\m}_{\min})}{|\gammap_j|},$$
and 
$$
\left|\beta^{\m}_{\max}-\beta^{*}\right|\leq \max_{j\in \Shat\cap \mathcal{V}}\frac{\left|\widehat{\Gamma}^{\m}_j-\Gammap_j-\beta^{\m}_{\max}(\widehat{\gamma}^{\m}_j-\gammap_j)\right|+\lambda \widehat{\rho}_j(\beta^{\m}_{\max})}{|\gammap_j|}.$$
On the event $\mathcal{G}\cap\mathcal{E}_4,$ we apply \eqref{eq: threshold bound} and establish  
\begin{equation}
\left|\beta^{\m}_{\min}-\beta^{*}\right|\leq C \max_{j\in \Shat\cap \mathcal{V}}\frac{(1+\lambda) \sqrt{1+(\beta^{\m}_{\min})^2}}{\sqrt{n}|\gammap_j|},\quad \left|\beta^{\m}_{\max}-\beta^{*}\right|\leq C \max_{j\in \Shat\cap \mathcal{V}}\frac{(1+\lambda) \sqrt{1+(\beta^{\m}_{\max})^2}}{\sqrt{n}|\gammap_j|}.
\label{eq: sampling upper}
\end{equation}
The above inequalities are similar to \eqref{eq: length upper immediate} for the searching CI.
By a similar argument as that of \eqref{eq: boundness conclu}, we establish that $|\beta^{\m}_{\max}|\leq C$ and $|\beta^{\m}_{\min}|\leq C.$ On the event $\mathcal{E}_4\cap \mathcal{G},$
we apply \eqref{eq: sampling upper} and establish 
{
\begin{equation*}
\begin{aligned}
\max_{m\in \mathcal{M}'}\max\left\{\left|\beta^{\m}_{\max}-\beta^{*}\right|,\left|\beta^{\m}_{\min}-\beta^{*}\right|\right\}\leq 
\max_{j\in \Shat\cap \mathcal{V}}\frac{C}{\sqrt{n}|\gammap_j|}.
\end{aligned}
\end{equation*}}
Similarly, on the event $\mathcal{E}_5\cap \mathcal{G}$, we have {
\begin{equation*}
\begin{aligned}
\max_{m\in \mathcal{M}}\max\left\{\left|\beta^{\m}_{\max}-\beta^{*}\right|,\left|\beta^{\m}_{\min}-\beta^{*}\right|\right\}\leq 
\max_{j\in \Shat\cap \mathcal{V}}\frac{C\sqrt{\log |\mathcal{M}|}}{\sqrt{n}|\gammap_j|}.
\end{aligned}
\end{equation*}}

\subsection{Proof of Proposition \ref{prop: voting}
} 
\subsubsection{Finite-sample analysis of the voting method}
For $j,k\in \widehat{\mathcal{S}},$ we have the following error decomposition of $\betatemp=\widehat{\Gamma}_j/\widehat{\gamma}_j$ and $\widehat{\pi}^{[j]}_k$ defined in \eqref{eq:pi_est},
\begin{equation}
\betatemp-\frac{\Gammap_j}{\gammap_j}=\frac{1}{\gammap_j}\widehat{\Omega}_{j\cdot} \frac{1}{n}W^{\intercal}\left(\epsilon-\frac{\Gammap_j}{\gammap_j} \cdot \delta \right)+\frac{1}{\gammap_j}\left(\frac{\widehat{\Gamma}_j}{\widehat{\gamma}_j}-\frac{\Gammap_j}{\gammap_j}\right)(\gammap_j-\widehat{\gamma}_j),
\label{eq: robust beta 1}
\end{equation}
and
\begin{equation}
\begin{aligned}
&\widehat{\pi}^{[j]}_k-\left(\Gammap_k-\frac{\Gammap_j}{\gammap_j}\gammap_k\right)=\left(\widehat{\Gamma}_k-\betatemp\widehat{\gamma}_k\right)-\left(\Gammap_k-\frac{\Gammap_j}{\gammap_j}\gammap_k\right)\\
&=\left(\widehat{\Gamma}_k-\Gammab_{k}\right)-\frac{\Gammap_j}{\gammap_j}\left(\widehat{\gamma}_k-\gammab_k\right)-\gammab_k\left(\betatemp-\frac{\Gammap_j}{\gammap_j}\right)-\left(\betatemp-\frac{\Gammap_j}{\gammap_j}\right)\left(\widehat{\gamma}_k-\gammab_k\right).
\end{aligned}
\label{eq: decomposition of eta robust}
\end{equation}
Note that 
\begin{equation}
\Gammap_k-\frac{\Gammap_j}{\gammap_j}\gammap_k=\pib_k-\frac{\pib_j}{\gammab_j} \gammab_k.
\label{eq: equi relation}
\end{equation}

By plugging \eqref{eq: equi relation} and \eqref{eq: robust beta 1} into \eqref{eq: decomposition of eta robust}, we have the following decomposition of $\widehat{\pi}^{[j]}_k-\pib_k$
\begin{equation}
\widehat{\pi}^{[j]}_k-\left(\pib_k-\frac{\pib_j}{\gammab_j} \gammab_k\right)=\mathcal{M}^{[j]}_{k}+\mathcal{A}^{[j]}_{k},
\label{eq: robust decomposition}
\end{equation}
where
$$\mathcal{M}^{[j]}_{k}=\left(\widehat{\Omega}_{k\cdot} -\frac{\gammab_k}{\gammab_j} \widehat{\Omega}_{j\cdot} \right)^{\intercal}\frac{1}{n}W^{\intercal} \left(\epsilon-\frac{\Gammap_j}{\gammap_j} \delta\right),$$
and
\begin{equation}
\mathcal{A}^{[j]}_{k}=- \frac{\gammab_k}{\gammap_j}\left(\betatemp-\frac{\Gammap_j}{\gammap_j}\right)(\gammap_j-\widehat{\gamma}_j)-\left(\betatemp-\frac{\Gammap_j}{\gammap_j}\right)\left(\widehat{\gamma}_k-\gammab_k\right).
\label{eq: explicit expression of selection procedure error}
\end{equation}

The remaining proof requires the following lemma, whose proof can be found in Section \ref{sec: good event 3 proof}.
\begin{Lemma} Suppose that Conditions {\rm (C1)} and {\rm (C2)} hold, then for any $j,k\in \mathcal{S}^0,$ we have 
\begin{equation}
\frac{1}{{\rm SE}(\widehat{\pi}_{k}^{[j]})}\left(\widehat{\Omega}_{k\cdot} -\frac{\gammab_k}{\gammab_j} \widehat{\Omega}_{j\cdot} \right)^{\intercal}\frac{1}{n}W^{\intercal} \left(\epsilon-\frac{\Gammap_j}{\gammap_j} \delta\right)\cid N(0, 1)
\label{eq: diff normal}
\end{equation}
where $\mathcal{S}^0$ is defined in \eqref{eq: relevant est proof} and
$${\rm SE}(\widehat{\pi}_{k}^{[j]})=|\gammap_k|\cdot \Diff_{j,k}^{0}=\sqrt{\frac{1}{n}\left(\RR^{[j]}_{k,k}+\frac{[\gammap_k]^2}{[\gammap_j]^2}\RR^{[j]}_{j,j}-\frac{2\gammap_k}{\gammap_j}\RR^{[j]}_{j,k}\right)}$$
with $\Diff_{j,k}^{0}$  defined in \eqref{eq: exact difference}. In addition, we have 
\begin{equation}
{\rm SE}(\widehat{\pi}_{k}^{[j]})\geq \sqrt{\frac{c_1\cdot \lambda_{\min}(\Sigma^{-1})}{{n}}}\sqrt{1+\left(\frac{\Gammap_j}{\gammap_j}\right)^2}\cdot \sqrt{1+\left(\frac{\gammap_k}{\gammap_j}\right)^2}.
\label{eq: lower SE}
\end{equation}
On the event $\mathcal{G},$  $\widehat{\rm SE}(\widehat{\pi}_{k}^{[j]})$ defined in \eqref{eq: SE pi} satsifies
\begin{equation}
\left|\left(\frac{\widehat{\rm SE}(\widehat{\pi}_{k}^{[j]})}{{\rm SE}(\widehat{\pi}_{k}^{[j]})}\right)^2-1\right|\lesssim \frac{1}{(\log n)^{1/4}}.
\label{eq: SE bound}
\end{equation}
\label{lem: diff normal}
\end{Lemma}

Define the event 
\begin{equation}
 \mathcal{F}=\cap_{j,k \in \widehat{\mathcal{S}}}\mathcal{F}^{[j]}_{k} \quad \text{with}\quad \mathcal{F}^{[j]}_{k}=\left\{
 \left|\mathcal{M}^{[j]}_{k}\right| \leq 0.9 \sqrt{{\log n}}\cdot \widehat{\rm SE}(\widehat{\pi}_{k}^{[j]})
\right\}.
 \label{eq: F event}
\end{equation}
By \eqref{eq: diff normal} and \eqref{eq: SE bound}, we apply the union bound and establish $$
\liminf_{n\rightarrow \infty}\PP\left(\mathcal{F}\right)=1.$$
On the event $\mathcal{G},$
we apply the expression \eqref{eq: explicit expression of selection procedure error} and establish 
$$\left|\mathcal{A}^{[j]}_{k}\right|\lesssim \left(1+\left|\frac{\gammab_k}{\gammap_j}\right|\right)\left(1+\left|\frac{\Gammap_j}{\gammap_j}\right|\right)\frac{1}{(\log n)^{1/4}} \cdot \frac{(\log n)^{1/4}}{\sqrt{n}}.$$
On the event $\mathcal{G}$, we further apply \eqref{eq: lower SE} and \eqref{eq: SE bound} and establish that
\begin{equation}
 \max_{j,k \in \widehat{\mathcal{S}}}\left|\mathcal{A}^{[j]}_{k}\right| \leq 0.05 \sqrt{{\log n}}\widehat{\rm SE}(\widehat{\pi}_{k}^{[j]}),
\label{eq: bias term robust}
\end{equation}
for a sufficiently large $n.$

We shall consider two cases in the following.\\
\noindent \textbf{Case (a).}
When $$\left|\frac{\pib_k}{\gammab_k}-\frac{\pib_j}{\gammab_j}\right|=0,$$ we simplify \eqref{eq: robust decomposition} as
\begin{equation}
\piesta_k=\mathcal{M}^{[j]}_{k}+\mathcal{A}^{[j]}_{k}.
\label{eq: first case1}
\end{equation}
By \eqref{eq: bias term robust}, on the event $\mathcal{G}\cap \mathcal{F},$
\begin{equation}
\max_{j,k \in \widehat{\mathcal{S}} } \left|\mathcal{M}^{[j]}_{k}+\mathcal{A}^{[j]}_{k}\right|\leq 0.95\sqrt{\log n}\cdot \widehat{\rm SE}(\widehat{\pi}_{k}^{[j]}).
\label{eq: first case goal1}
\end{equation}
Similarly, we establish on the event $\mathcal{G}\cap \mathcal{F},$
\begin{equation}
\max_{j,k \in \widehat{\mathcal{S}} } \left|\mathcal{M}^{[k]}_{j}+\mathcal{A}^{[k]}_{j}\right|\leq 0.95\sqrt{\log n}\cdot \widehat{\rm SE}(\widehat{\pi}_{j}^{[k]}).
\label{eq: first case goal1 switch}
\end{equation}
 By the definition in \eqref{eq: col def}, we have $$\liminf_{n\rightarrow \infty}\PP\left(\widehat{\Pi}_{k,j}=\widehat{\Pi}_{j,k}=1\right)\geq \liminf_{n\rightarrow \infty}\PP\left(\mathcal{F}\cap \mathcal{G}\right)=1.$$

\noindent \textbf{Case (b).}
By \eqref{eq: robust decomposition}, the event $\{\widehat{\Pi}_{k,j}=\widehat{\Pi}_{j,k}=0\}$ is equivalent to that at least one of the following two events happens
\begin{equation}
\left|\pib_k-\frac{\pib_j}{\gammab_j} \gammab_k+\mathcal{M}^{[j]}_{k}+\mathcal{A}^{[j]}_{k}\right| > \sqrt{\log n}\cdot \widehat{\rm SE}(\widehat{\pi}_{k}^{[j]});
\label{eq: case a}
\end{equation}
\begin{equation}
\left|\pib_j-\frac{\pib_k}{\gammab_k} \gammab_j+\mathcal{M}^{[k]}_{j}+\mathcal{A}^{[k]}_{j}\right| > \sqrt{\log n}\cdot \widehat{\rm SE}(\widehat{\pi}_{j}^{[k]}).
\label{eq: case b}
\end{equation}
By the upper bound in \eqref{eq: first case goal1}, on the event $\mathcal{G}\cap \mathcal{F},$ the event in \eqref{eq: case a} happens if 
\begin{equation*}
\left|\frac{\pib_k}{ \gammab_k}-\frac{\pib_j}{\gammab_j}\right|\geq 1.95\sqrt{\log n}\cdot \frac{\widehat{\rm SE}(\widehat{\pi}_{k}^{[j]})}{{ \gammab_k}};
\label{eq: second case goal1}
\end{equation*}
the event
\eqref{eq: case b} happens if 
\begin{equation*}
\left|\frac{\pib_j}{\gammab_j}-\frac{\pib_k}{\gammab_k} \right|\geq 1.95\sqrt{\log n}\cdot \frac{\widehat{\rm SE}(\widehat{\pi}_{j}^{[k]})}{{\gammab_j}}.
\label{eq: second case goal2}
\end{equation*}
That is, the event $\{\widehat{\Pi}_{k,j}=\widehat{\Pi}_{j,k}=0\}$ happens if $$\left|\frac{\pib_k}{\gammab_k}-\frac{\pib_j}{\gammab_j} \right|\geq 1.95 \sqrt{\log n}\cdot \min\left\{\frac{\widehat{\rm SE}(\widehat{\pi}_{j}^{[k]})}{{\gammab_j}},\frac{\widehat{\rm SE}(\widehat{\pi}_{k}^{[j]})}{{\gammab_k}}\right\}.$$
Together with \eqref{eq: SE bound}, we obtain that, on the event $\mathcal{G}\cap \mathcal{F}$, the event $\{\widehat{\Pi}_{k,j}=\widehat{\Pi}_{j,k}=0\}$ happens if $$\left|\frac{\pib_k}{\gammab_k}-\frac{\pib_j}{\gammab_j} \right|\geq 2 \sqrt{\log n}\cdot \min\left\{\frac{{\rm SE}(\widehat{\pi}_{j}^{[k]})}{{\gammab_j}},\frac{{\rm SE}(\widehat{\pi}_{k}^{[j]})}{{\gammab_k}}\right\}.$$
Hence, we establish 
\begin{equation*}
\begin{aligned}
\liminf_{n\rightarrow \infty}\PP\left(\widehat{\Pi}_{k,j}=\widehat{\Pi}_{j,k}=0\right)\geq \liminf_{n\rightarrow \infty}\PP\left(\mathcal{G}\cap \mathcal{F}\right)=1.
\end{aligned}
\end{equation*}

\subsubsection{Analysis of $\VTSHT$ in \eqref{eq: initial valid}}
We apply the definition of $\VTSHT$ in \eqref{eq: initial valid} and write $\VTSHT$ as $\widehat{\mathcal{V}}$ in the following proof. By the construction, we have $\widehat{\mathcal{W}}\subset \widehat{\mathcal{V}}.$  Throughout the proof, we condition on the event $\mathcal{G}\cap \mathcal{F}$, which has asymptotic probability $1$. Then we can directly apply case (a) and case (b) of Proposition \ref{prop: voting}. In particular, we use the following results: for any $l\in \widehat{\mathcal{V}},$ there exists $j\in \widehat{\mathcal{W}} $ such that \begin{equation}
\left|\frac{\pip_l}{\gammap_l}-\frac{\pip_{j}}{\gammap_{j}}\right|\leq 2\sep.
\label{eq: combined}
\end{equation}
Similarly, for any $j\in \widehat{\mathcal{W}},$ there exists $l\in \widehat{\mathcal{V}}$ such that \eqref{eq: combined} holds.
In the following, we apply \eqref{eq: combined} and consider two cases.
\paragraph{\bf Case 1.} We first consider the case where $\widehat{\mathcal{W}}\subset \mathcal{V}$. It follows from \eqref{eq: combined} that 
\begin{equation}
\widehat{\mathcal{V}}\subset \mathcal{I}(0, 2\sep).
\label{eq: case 1-b}
\end{equation} 
For $k\in \mathcal{V}\cap \widehat{\mathcal{S}}$ and $j\in \widehat{\mathcal{W}}\subset \mathcal{V},$ we have $\frac{\pib_k}{\gammab_k}=\frac{\pib_{j}}{\gammab_{j}}=0.$  Case (a) of Proposition \ref{prop: voting} implies 
$$
\widehat{\Pi}_{k,j}=\widehat{\Pi}_{j,k}=1.
$$ Then we apply the definition of $\widehat{\mathcal{V}}=\VTSHT$ in \eqref{eq: initial valid} and establish that $k\in \widehat{\mathcal{V}},$ which implies
\begin{equation}
\mathcal{V}\cap\widehat{\mathcal{S}}\subset \widehat{\mathcal{V}}.
\label{eq: case 1-a}
\end{equation}
Since $\mathcal{V}\cap\mathcal{S}_{\rm Str}\subset \mathcal{V}\cap\widehat{\mathcal{S}},$ we combine \eqref{eq: case 1-a} and \eqref{eq: case 1-b} and establish 
\begin{equation}
\mathcal{V}\cap\mathcal{S}_{\rm Str}\subset\widehat{\mathcal{V}}\subset \mathcal{I}(0, 2\sep).
\label{eq: case 1}
\end{equation}

\paragraph{\bf Case 2.} We then consider the case where $\widehat{\mathcal{W}}\not\subset \mathcal{V}$. Then there exists $j$ such that $j\in \widehat{\mathcal{W}}$ but $j\not\in \mathcal{V}.$ Let ${\rm supp}(\widehat{\Pi}_{j\cdot})$ denote the support of $\widehat{\Pi}_{j\cdot}$.

We define $l\in \mathcal{V}\cap\Shat$. Case (a) of Proposition \ref{prop: voting} implies 
 \begin{equation}
\mathcal{V}\cap \Shat \subset {\rm supp}(\widehat{\Pi}_{l\cdot}).
\label{eq: inter 2}
\end{equation}
Case (b) of Proposition \ref{prop: voting} implies
\begin{equation}
{\rm supp}(\widehat{\Pi}_{j\cdot})\subset\mathcal{I}\left(\frac{\pip_j}{\gammap_j},2\sep\right)\cap \Shat.
\label{eq: inter 0}
\end{equation} 
In the following, we show by contradiction that ${\rm supp}(\widehat{\Pi}_{j\cdot})\cap \mathcal{V}\neq \emptyset.$ Assume that \begin{equation}
{\rm supp}(\widehat{\Pi}_{j\cdot})\cap \mathcal{V}=\emptyset.
\label{eq: contra assump}
\end{equation}

The definition of $\widehat{\mathcal{W}}$ implies $\left|{\rm supp}(\widehat{\Pi}_{l\cdot})\right|\leq \left|{\rm supp}(\widehat{\Pi}_{j\cdot})\right|.$ We apply \eqref{eq: inter 2} and establish $$|\mathcal{V}\cap \Shat |\leq \left|{\rm supp}(\widehat{\Pi}_{j\cdot})\right|=\left|{\rm supp}(\widehat{\Pi}_{j\cdot})\backslash \mathcal{V}\right|\leq \left|\mathcal{I}\left(\frac{\pip_j}{\gammap_j},2\sep\right)\backslash \mathcal{V}
\right|,$$
where the first equality follows from the assumption \eqref{eq: contra assump}  and the last inequality follows from \eqref{eq: inter 0}. The above inequality implies that $$|\mathcal{V}\cap \mathcal{S}_{\rm str}| \leq |\mathcal{V}\cap \Shat|\leq  \left|\mathcal{I}\left(\frac{\pip_j}{\gammap_j},2\sep\right)\backslash \mathcal{V}
\right|,$$ which contradicts the finite-sample plurality rule and hence the conjecture \eqref{eq: contra assump} does not hold. That is, for any $j\in \widehat{\mathcal{W}},$
\begin{equation}
\text{there exists} \; k\in {\rm supp}(\widehat{\Pi}_{j\cdot})\cap \mathcal{V}.
\label{eq: key result}
\end{equation} 
Together with case (b) of Proposition \ref{prop: voting}, we establish 
$$
\left|{\pip_j}/{\gammap_j}\right|\leq \sep \quad \text{for}\quad j\in \widehat{\mathcal{W}}.
$$
Combined with \eqref{eq: combined}, we establish that 
\begin{equation}
\widehat{\mathcal{V}}\subset \mathcal{I}(0, 3\sep).
\label{eq: case 2-b}
\end{equation}

For any $k\in \mathcal{V}\cap\Shat,$ we have  $\mathcal{V}\cap\Shat \subset {\rm supp}(\widehat{\Pi}_{k\cdot}).$ By combining this with \eqref{eq: key result} and the definition of $\widehat{\mathcal{V}}=\VTSHT$ in \eqref{eq: initial valid}, we establish
\begin{equation}
\mathcal{V}\cap\widehat{\mathcal{S}}\subset \widehat{\mathcal{V}}.
\label{eq: case 2-a}
\end{equation}

Since $\mathcal{V}\cap\mathcal{S}_{\rm Str}\subset \mathcal{V}\cap\widehat{\mathcal{S}},$ we combine \eqref{eq: case 2-a} and \eqref{eq: case 2-b} and establish 
\begin{equation}
\mathcal{V}\cap\mathcal{S}_{\rm Str}\subset\widehat{\mathcal{V}}\subset \mathcal{I}(0, 3\sep).
\label{eq: case 2}
\end{equation}
We combine \eqref{eq: case 1} and \eqref{eq: case 2} and establish that $\widehat{\mathcal{V}}=\VTSHT$ defined in \eqref{eq: initial valid}  satisfies \eqref{eq: initial condition}.
\subsection{Proofs of Theorem \ref{thm: inference-searching plurality}}

The proof of the searching interval follows from that of Theorem \ref{thm: inference-searching},  and Proposition \ref{prop: voting}. Note that, on the event $\mathcal{G},$ we have $\mathcal{V}\cap\mathcal{S}_{\rm Str}\subset \widehat{\mathcal{V}}\subset \mathcal{I}(0,3 \sep).$
By the finite sample plurality rule (Condition \ref{cond: plurality-finite}), $\mathcal{V}\cap \mathcal{S}_{\rm Str}$ is the majority of the set $\mathcal{I}(0,3 \sep)$ and also the majority of $\Vhat.$ 
We then apply the same argument for Theorem \ref{thm: inference-searching} by replacing $\widehat{\mathcal{S}}$ with $\widehat{\mathcal{V}}$ and $\mathcal{S}$ with $\mathcal{I}(0,3 \sep)$. 

The proof of the sampling interval follows from that of Theorem \ref{thm: inference-sampling} and Proposition \ref{prop: voting}. Note that, on the event $\mathcal{G},$ we have $\mathcal{V}\cap\mathcal{S}_{\rm Str}\subset \widehat{\mathcal{V}}\subset \mathcal{I}(0,3 \sep).$
By the finite sample plurality rule (Condition \ref{cond: plurality-finite}), $\mathcal{V}\cap \mathcal{S}_{\rm Str}$ is the majority of the set $\mathcal{I}(0,3 \sep)$ and also the majority of $\Vhat.$ We then apply the same argument for Theorem \ref{thm: inference-sampling} by replacing $\widehat{\mathcal{S}}$ with $\widehat{\mathcal{V}}$ and $\mathcal{S}$ with $\mathcal{I}(0,3 \sep)$.

\section{Proofs of Lemmas}
\label{sec: lemma proof}

\subsection{Proof of Lemma \ref{lem: good event 1}}
\label{sec: good event 1 proof}
\underline{Control of $\mathcal{G}_1.$} 
We present the proof of controlling $\left\|\frac{1}{{n}} W^{\intercal} \epsilon\right\|_{\infty}$ in the following. The proof of $\left\|\frac{1}{{n}} W^{\intercal} \delta\right\|_{\infty}$ is similar to that of $\left\|\frac{1}{{n}} W^{\intercal} \epsilon\right\|_{\infty}.$
Note that $\E W_{ij}\delta_i=0$ for any $1\leq i\leq n$ and $1\leq j\leq p$ and $W_{ij}\delta_i$ is Sub-exponential random variable. Since $(W^{\intercal}\delta)_{j}=\sum_{i=1}^{n}W_{ij}\delta_i$, we apply Proposition 5.16 of \citet{vershynin2010introduction} with the corresponding $t=C \sqrt{n\sqrt{\log n}}$ and establish 
\begin{equation*}
\PP\left(\left|\sum_{i=1}^{n}W_{ij}\delta_i\right|\leq C\sqrt{n\sqrt{\log n}}\right)\geq 1-\exp(-c'\sqrt{\log n}),
\end{equation*}
where $C$ and $c'$ are positive constants independent of $n.$
For the fixed $p$ setting, we apply the union bound and establish that 
\begin{equation}
\begin{aligned}
\PP\left(\|W^{\intercal}\delta\|_{\infty}=\max_{1\leq j\leq \pz}\left|\sum_{i=1}^{n}W_{ij}\delta_i\right|\leq C\sqrt{n\sqrt{\log n}}\right)\geq &1-\pz\exp(-c'\sqrt{\log n})\\
\geq &1-\exp(-c\sqrt{\log n}),
\end{aligned}
\label{eq: result 4}
\end{equation}
for some positive constant $c>0.$\\
\noindent \underline{Control of $\mathcal{G}_3.$} 
Since $\{W_{i\cdot}\}_{1\leq i\leq n}$ are i.i.d Sub-gaussian random vectors and the dimension $p$ is fixed, then we apply equation (5.25) in \citet{vershynin2010introduction} and establish the following concentration results for $\widehat{\Sigma}-\Sigma:$ with probability larger than $1-n^{-c},$
\begin{equation}
\|\widehat{\Sigma}-\Sigma\|_2\leq C\sqrt{{\log n}/{n}},
\label{eq: result 1}
\end{equation}
where $c$ and $C$ are positive constants independent of $n$. As a consequence, we have 
\begin{equation}
|\lambda_{\min}(\widehat{\Sigma})-\lambda_{\min}({\Sigma})|\leq \|\widehat{\Sigma}-\Sigma\|_2\leq C\sqrt{{\log n}/{n}}.
\label{eq: result 2}
\end{equation}
Since $\widehat{\Sigma}^{-1}-\Sigma^{-1}=\widehat{\Sigma}^{-1}(\Sigma-\widehat{\Sigma})\Sigma^{-1},$
we have 
$$\|\widehat{\Sigma}^{-1}-\Sigma^{-1}\|_2\leq \frac{1}{\lambda_{\min}(\widehat{\Sigma})\cdot \lambda_{\min}({\Sigma})}\|\Sigma-\widehat{\Sigma}\|_2\leq \frac{C\sqrt{{\log n}/{n}}}{\left(\lambda_{\min}({\Sigma})-C\sqrt{{\log n}/{n}}\right)\cdot \lambda_{\min}({\Sigma})}.$$
Then we have 
\begin{equation}
\PP(\mathcal{G}_3)\geq 1-n^{-c}.
\label{eq: control 1}
\end{equation}
\noindent\underline{Control of $\mathcal{G}_2$.} We shall focus on the analysis of $\widehat{\gamma}_j-\gammab_j$ and the analysis of $\widehat{\Gamma}_j-\Gammab_j$ is similar. We apply the expression \eqref{eq: expression for OLS} and establish
\begin{equation}
\frac{\widehat{\gamma}_j-\gammab_j}{{\sqrt{{\V}^{\gamma}_{jj}/n}}}=\frac{\widehat{\Omega}_{j\cdot}^{\intercal} \frac{1}{{n}} W^{\intercal}\delta}{{\sqrt{{\V}^{\gamma}_{jj}/n}}}=\frac{{\Omega}_{j\cdot}^{\intercal} \frac{1}{{n}} W^{\intercal}\delta}{{\sqrt{{\V}^{\gamma}_{jj}/n}}}+\frac{(\widehat{\Omega}_{j\cdot} -\Omega_{j\cdot})^{\intercal}\frac{1}{{n}} W^{\intercal}\delta}{{\sqrt{{\V}^{\gamma}_{jj}/n}}}.
\label{eq: result 3}
\end{equation}
By Lemma \ref{lem: asymp normal}, we have 
$${{\Omega}_{j\cdot}^{\intercal} \frac{1}{{n}} W^{\intercal}\delta}/{{\sqrt{{\V}^{\gamma}_{jj}/n}}}\cid N(0,1).$$ 
On the event $\mathcal{G}_1\cap \mathcal{G}_3$, we apply the decomposition \eqref{eq: result 2} together with \eqref{eq: lower bound on Cov}, \eqref{eq: result 4}  and \eqref{eq: result 3} and establish 
$$\PP\left(\max_{1\leq j\leq p_z}{|\widehat{\gamma}_j-\gammab_j|}/{\sqrt{{\V}^{\gamma}_{jj}/n}}\leq C (\log n)^{1/4}\right)\geq 1-\exp(-c\sqrt{\log n}).$$
We can apply a similar argument to control $\widehat{\Gamma}_j-\Gammab_j$ and then establish 
\begin{equation}
\PP(\mathcal{G}_2)\geq 1-\exp(-c\sqrt{\log n}).
\label{eq: control 2}
\end{equation}
By a similar argument, for the fixed $p$ setting, we can establish
\begin{equation}
\PP\left(\|\Psi^*-\widehat{\Psi}\|_{1}+\|\psi^*-\widehat{\psi}\|_{1}\leq C\frac{(\log n)^{1/4}}{\sqrt{n}}\right)\geq 1-\exp(-c\sqrt{\log n}).
\label{eq: result 7}
\end{equation}

\noindent \underline{Control of $\mathcal{G}_4$.}
We shall control the difference $\|\widehat{\C}-\C\|_2$ and the upper bounds for $\|\widehat{\V}^{\Gamma}-{\V}^{\Gamma}\|_2$ and $\|\widehat{\V}^{\gamma}-{\V}^{\gamma}\|_2$ can be established in a similar way. Define $A=\E \epsilon_i\delta_i W_{i\cdot} W_{i\cdot}^{\intercal}$ and $\widehat{A}=\frac{1}{n}\sum_{i=1}^{n} \widehat{\epsilon}_i\widehat{\delta}_i W_{i\cdot} W_{i\cdot}^{\intercal}.$ We first control $\|\widehat{A}-A\|_2.$ Note that $\widehat{\epsilon}_i=\epsilon_i+Z_{i\cdot}^{\intercal}(\Gammap-\widehat{\Gamma})+X_{i\cdot}^{\intercal}(\Psi^*-\widehat{\Psi})$ and $\widehat{\delta}_i=\delta_i+Z_{i\cdot}^{\intercal}(\gammap-\widehat{\gamma})+X_{i\cdot}^{\intercal}(\psi^*-\widehat{\psi}).$ 

For $u\in \R^{p}$, we define $U_i=u^{\intercal}W_{i\cdot}$ and  decompose the error $u^{\intercal}(\widehat{A}-A)u$ as 
\begin{equation*}
\begin{aligned}
&\frac{1}{n}\sum_{i=1}^{n}\epsilon_i\delta_i U^2_i -\E \epsilon_i\delta_i U^2_i
+\frac{1}{n}\sum_{i=1}^{n}\epsilon_i\left(Z_{i\cdot}^{\intercal}(\gammap-\widehat{\gamma})+X_{i\cdot}^{\intercal}(\psi^*-\widehat{\psi})\right)U^2_i\\
&+\frac{1}{n}\sum_{i=1}^{n}\delta_i\left(Z_{i\cdot}^{\intercal}(\Gammap-\widehat{\Gamma})+X_{i\cdot}^{\intercal}(\Psi^*-\widehat{\Psi})\right)U_i^2\\
&+\frac{1}{n}\sum_{i=1}^{n}\left(Z_{i\cdot}^{\intercal}(\gammap-\widehat{\gamma})+X_{i\cdot}^{\intercal}(\psi^*-\widehat{\psi})\right)\left(Z_{i\cdot}^{\intercal}(\Gammap-\widehat{\Gamma})+X_{i\cdot}^{\intercal}(\Psi^*-\widehat{\Psi})\right)U_i^2.
\end{aligned}
\end{equation*}
Since $\epsilon_,\delta_i$ and $Z_{i\cdot}$ and $X_{i\cdot}$ are Sub-gaussian random variables, then with probability larger than $1-n^{-c},$ $\max_{1\leq i\leq n}\max\left\{|\epsilon_i|,|\delta_i|,\|Z_{i\cdot}\|_{\infty},\|X_{i\cdot}\|_{\infty}\right\}\leq C\sqrt{\log n}$ where $C>0$ and $c>0$ are positive constants independent of $n.$ Together with  \eqref{eq: result 7}, we establish that, with probability larger than $1-\exp(-c\sqrt{\log n}),$
we have 
\begin{equation}
\left|u^{\intercal}(\widehat{A}-A)u-\left(\frac{1}{n}\sum_{i=1}^{n}\epsilon_i\delta_i U^2_i -\E \epsilon_i\delta_i U^2_i\right)\right|\leq C \frac{(\log n)^{3/2}}{\sqrt{n}} \cdot \frac{1}{n} \sum_{i=1}^{n} U^2_i\leq C \frac{(\log n)^{3/2}}{\sqrt{n}},
\label{eq: decomp bound}
\end{equation}
where the last inequality follows from \eqref{eq: result 1}.

We define $\mathcal{H}_i=\{\max\{|\epsilon_i|,|\delta_i|\}\leq 2\sqrt{\log n}\}$ and we have $\PP\left(\cap_{i=1}^{n} \mathcal{H}_i\right)\geq 1-C n^{-1}$ for some positive constant $C>0.$ We decompose $\frac{1}{n}\sum_{i=1}^{n}\epsilon_i\delta_i U^2_i -\E \epsilon_i\delta_i U^2_i$ as 
\begin{equation*}
\begin{aligned}
\frac{1}{n}\sum_{i=1}^{n}\epsilon_i\delta_i U^2_i\cdot \mathcal{H}_i -\E \epsilon_i\delta_i U^2_i\cdot \mathcal{H}_i +\frac{1}{n}\sum_{i=1}^{n}\epsilon_i\delta_i U^2_i\cdot \mathcal{H}_i^{c} -\E \epsilon_i\delta_i U^2_i\cdot \mathcal{H}_i^{c}.
\end{aligned}
\end{equation*}
Since $\epsilon_i$ and $\delta_i$ are sub-gaussian, $\PP\left(\frac{1}{n}\sum_{i=1}^{n}\epsilon_i\delta_i U^2_i\cdot \mathcal{H}_i^{c}\neq 0\right)\leq C n^{-1},$ for some positive constant $C>0.$ Note that $\E \epsilon_i\delta_i U^2_i\cdot \mathcal{H}_i^{c}\leq \sqrt{\E \epsilon^2_i\delta^2_i U^4_i} \sqrt{\PP\left(\mathcal{H}_i^{c}\right)}\leq C n^{-1}$ for some positive constant $C>0.$
Since $\frac{1}{4\log n}\left|\epsilon_i\delta_i\right|\cdot \mathcal{H}_i\leq 1$, we generalize (5.25) in \citet{vershynin2010introduction} and establish
\begin{equation*}
\PP\left(\max_{u\in\R^{p},\|u\|_2=1}\frac{1}{4\log n}\left(\frac{1}{n}\sum_{i=1}^{n}\epsilon_i\delta_i U^2_i\cdot \mathcal{H}_i -\E \epsilon_i\delta_i U^2_i\cdot \mathcal{H}_i\right)\leq C \sqrt{\frac{p+\log n}{n}}\right)\geq 1-n^{-c},
\end{equation*}
for some positive constant $c>0.$
Combined with \eqref{eq: decomp bound}, we have 
\begin{equation}
\PP\left(\|\widehat{\C}-\C\|_2\leq C \frac{(\log n)^{3/2}}{\sqrt{n}}\right)\geq 1-\exp(-c\sqrt{\log n}),
\label{eq: key bound C}
\end{equation}
for some positive constants $C>0$ and $c>0.$
Note that 
\begin{equation*}
\begin{aligned}
&\|\widehat{\C}-\C\|_2\leq \left\| \widehat{\Sigma}^{-1}\widehat{A}\widehat{\Sigma}^{-1}-\Sigma^{-1}A\Sigma^{-1}\right\|_2\\
&\leq \|\widehat{\Sigma}^{-1}-\Sigma^{-1}\|_2 \|\widehat{A}\|_2\|\widehat{\Sigma}^{-1}\|_2+\|{\Sigma}^{-1}\|_2\|\widehat{A}-A\|_2\|\widehat{\Sigma}^{-1}\|_2+\|{\Sigma}^{-1}\|_2\|{A}\|_2\|\widehat{\Sigma}^{-1}-\Sigma^{-1}\|_2.
\end{aligned}
\end{equation*}
On the event $\mathcal{G}_3$, we apply \eqref{eq: key bound C} and \eqref{eq: upper bound on Cov} and establish 
\begin{equation}
\PP\left(\mathcal{G}_3\right)\geq 1-\exp(-c\sqrt{\log n}).
\label{eq: control 3}
\end{equation}

\noindent \underline{Control of $\mathcal{G}_5.$} 
Recall that $\widehat{\mathcal{S}}$ is defined in \eqref{eq: relevant est} and $\mathcal{S}_{\rm str}$ is defined in \eqref{eq: str relevant}. Then for $j\in \mathcal{S}_{\rm str},$ on the event $\mathcal{G}_2\cap \mathcal{G}_4,$ if $\sqrt{\log n}>C(\log n)^{1/4},$ we have 
$$|\widehat{\gamma}_j|\geq \sqrt{\log n} \cdot \sqrt{\widehat{\V}^{\gamma}_{jj}/n},$$
that is $\mathcal{S}_{\rm str}\subset \Shat.$
Furthermore, for $j\in \Shat,$ on the event $\mathcal{G}_2\cap \mathcal{G}_4,$ we have
$$|\gammap_j|\geq (\sqrt{\log n}-C (\log n)^{1/4})\cdot \sqrt{\widehat{\V}^{\gamma}_{jj}/n}$$
that is $\Shat\subset \mathcal{S}^{0}.$
Hence, we have 
%
\begin{equation}
\PP(\mathcal{G}_4\cap\mathcal{G}_5)\geq \PP\left(\mathcal{G}_2\cap\mathcal{G}_4\cap\mathcal{G}_3\right).
\label{eq: control 4}
\end{equation}

\noindent \underline{Control of events $\mathcal{G}_6$ and $\mathcal{G}_7.$}
On the event $\mathcal{G}_2\cap\mathcal{G}_4\cap\mathcal{G}_5,$ for $j\in \mathcal{S}^0$ 
\begin{equation*}
\left|\frac{\widehat{\gamma}_j}{\gammap_j}-1\right|\leq \frac{C(\log n)^{1/4}\sqrt{{\V}^{\gamma}_{jj}/n}}{(\sqrt{\log n}-C(\log n)^{1/4})\sqrt{\widehat{\V}^{\gamma}_{jj}/n}}\lesssim \frac{1}{(\log n)^{1/4}},
\end{equation*}
and hence 
\begin{equation}
\max_{j\in \mathcal{S}^0}\left|\frac{\gammap_j}{\widehat{\gamma}_j}-1\right|\lesssim \frac{1}{(\log n)^{1/4}}.
\label{eq: fact 1}
\end{equation}
By the decomposition 
\begin{equation*}
\begin{aligned}
(\widehat{\gamma}_k/\widehat{\gamma}_j-\gammap_k/\gammap_j)\gammap_j=&(\widehat{\gamma}_k-\gammap_k)\frac{\gammap_j}{\widehat{\gamma}_j}+\gammap_k\left(\frac{\gammap_j}{\widehat{\gamma}_j}-1\right)\\
=&\left(\frac{\widehat{\gamma}_k}{\gammap_k}-1\right)\cdot \gammap_k\cdot\frac{\gammap_j}{\widehat{\gamma}_j}+\gammap_k\left(\frac{\gammap_j}{\widehat{\gamma}_j}-1\right),
\end{aligned}
\end{equation*}
we apply \eqref{eq: fact 1} and establish 
\begin{equation*}
\left|(\widehat{\gamma}_k/\widehat{\gamma}_j-\gammap_k/\gammap_j)\gammap_j\right|\lesssim \frac{|\gammap_k|}{(\log n)^{1/4}}\quad \text{and}\quad \left|\frac{\widehat{\gamma}_k}{\widehat{\gamma}_j}-\frac{\gammap_k}{\gammap_j} \right|\lesssim \left|\frac{\gammap_k}{\gammap_j}\right|\cdot \frac{1}{(\log n)^{1/4}} .
\end{equation*}
That is, the event $\mathcal{G}_6$ holds and \begin{equation}
\PP(\mathcal{G}_6)\geq \PP(\mathcal{G}_2\cap\mathcal{G}_4\cap\mathcal{G}_5).
\end{equation}
By the decomposition 
$$(\widehat{\Gamma}_j/\widehat{\gamma}_j-\Gammap_j/\gammap_j)\gammap_j=(\widehat{\Gamma}_j-\Gammap_j)\frac{\gammap_j}{\widehat{\gamma}_j}+\Gammap_j\left(\frac{\gammap_j}{\widehat{\gamma}_j}-1\right),$$
we apply \eqref{eq: fact 1} and establish on the event $\mathcal{G}_2$ that
\begin{equation*}
\left|(\widehat{\Gamma}_j/\widehat{\gamma}_j-\Gammap_j/\gammap_j)\gammap_j \right|\leq  C {\sqrt{{\V}^{\Gamma}_{jj}/n}}(\log n)^{1/4}+\left|\Gammap_j\right|\frac{1}{(\log n)^{1/4}}.
\end{equation*}
Hence, for $j\in \mathcal{S}^{0},$ we have 
$
\left|\widehat{\Gamma}_j/\widehat{\gamma}_j-\Gammap_j/\gammap_j \right|\leq C \left(1+\left|{\Gammap_j}/{\gammap_j}\right|\right)\frac{1}{(\log n)^{1/4}}.
$
That is, the event $\mathcal{G}_7$ holds and 
\begin{equation}
\PP(\mathcal{G}_7)\geq \PP(\mathcal{G}_2\cap\mathcal{G}_4\cap\mathcal{G}_5).
\label{eq: control 5}
\end{equation}

We establish the lemma by combining \eqref{eq: control 1}, \eqref{eq: control 2}, \eqref{eq: control 3}, \eqref{eq: control 4} and \eqref{eq: control 5}.
\subsection{Proof of Lemma \ref{lem: good event threshold}}
\label{sec: good event threshold proof}

Recall the definitions of $\widehat{\bf R}(\beta)$ and ${\bf R}(\beta)$ in \eqref{eq: R def}. Note that 
$
\frac{\widehat{\Gamma}_j-\Gammap_j-\betap(\widehat{\gamma}_j-\gammap_j)}{\sqrt{{\bf R}_{jj}(\betap)/n}}\cid N(0,1).
$
Together with \eqref{eq: R bound 2} and \eqref{eq: R bound 3}, we establish 
\begin{equation}
\frac{\widehat{\Gamma}_j-\Gammap_j-\betap(\widehat{\gamma}_j-\gammap_j)}{\sqrt{\widehat{\bf R}_{jj}(\betap)/n}}\cid N(0,1).
\label{eq: limiting}
\end{equation}
For any $\beta\in \mathcal{U}(a),$ we have 
\begin{equation}
\begin{aligned}
&\frac{\widehat{\Gamma}_j-\Gammap_j-\beta(\widehat{\gamma}_j-\gammap_j)}{\sqrt{\widehat{\bf R}_{jj}(\beta)/n}}-\frac{\widehat{\Gamma}_j-\Gammap_j-\betap(\widehat{\gamma}_j-\gammap_j)}{\sqrt{{\bf R}_{jj}(\betap)/n}}\\
&= \frac{\widehat{\Gamma}_j-\Gammap_j-\betap(\widehat{\gamma}_j-\gammap_j)}{\sqrt{\widehat{\bf R}_{jj}(\betap)/n}}\cdot\left(1-\frac{\sqrt{\widehat{\bf R}_{jj}(\beta)/n}}{\sqrt{{\bf R}_{jj}(\betap)/n}}\right)-\frac{(\beta-\betap)(\widehat{\gamma}_j-\gammap_j)}{\sqrt{\widehat{\bf R}_{jj}(\beta)/n}}.
\end{aligned}
\label{eq: separation key}
\end{equation}
By the triangle inequality, we further establish 
\begin{equation*}
\begin{aligned}
&\left|\max_{\beta\in \mathcal{U}_a}\frac{\left|\widehat{\Gamma}_j-\Gammap_j-\beta(\widehat{\gamma}_j-\gammap_j)\right|}{\sqrt{\widehat{\bf R}_{jj}(\beta)/n}}-\frac{\left|\widehat{\Gamma}_j-\Gammap_j-\betap(\widehat{\gamma}_j-\gammap_j)\right|}{\sqrt{{\bf R}_{jj}(\betap)/n}}\right|\\
&\leq  \max_{\beta\in \mathcal{U}_a}\left| \frac{\widehat{\Gamma}_j-\Gammap_j-\betap(\widehat{\gamma}_j-\gammap_j)}{\sqrt{\widehat{\bf R}_{jj}(\betap)/n}}\cdot\left(1-\frac{\sqrt{\widehat{\bf R}_{jj}(\beta)/n}}{\sqrt{{\bf R}_{jj}(\betap)/n}}\right)-\frac{(\beta-\betap)(\widehat{\gamma}_j-\gammap_j)}{\sqrt{\widehat{\bf R}_{jj}(\beta)/n}}\right|.
\end{aligned}
\end{equation*}
By applying \eqref{eq: R bound 1}, \eqref{eq: R bound 2}, \eqref{eq: R bound 3}, and \eqref{eq: limiting}, we obtain 
\begin{equation}
\left|\max_{\beta\in \mathcal{U}_a}\frac{\left|\widehat{\Gamma}_j-\Gammap_j-\beta(\widehat{\gamma}_j-\gammap_j)\right|}{\sqrt{\widehat{\bf R}_{jj}(\beta)/n}}-\frac{\left|\widehat{\Gamma}_j-\Gammap_j-\betap(\widehat{\gamma}_j-\gammap_j)\right|}{\sqrt{{\bf R}_{jj}(\betap)/n}}\right|\cip 0
\label{eq: approximation bound}
\end{equation}
 Together with \eqref{eq: limiting}, we establish 
\begin{equation*}
\PP\left[\max_{\beta\in \mathcal{U}(a)} \frac{|\widehat{\Gamma}_j-\Gammap_j-\beta(\widehat{\gamma}_j-\gammap_j)|}{\sqrt{(\widehat{\V}^{\Gamma}_{jj}+\beta^2\widehat{\V}^{\gamma}_{jj}-2\beta \widehat{\C}_{jj})/n}}\geq \Phi^{-1}\left(1-\frac{\alpha}{2|\Shat|}\right)\right]\leq \alpha/|\Shat|.
\end{equation*}
We further apply the union bound and establish
\begin{equation}
\begin{aligned}
&\PP\left[\max_{\beta\in \mathcal{U}(a)}\max_{j\in \Shat} \frac{|\widehat{\Gamma}_j-\Gammap_j-\beta(\widehat{\gamma}_j-\gammap_j)|}{\sqrt{(\widehat{\V}^{\Gamma}_{jj}+\beta^2\widehat{\V}^{\gamma}_{jj}-2\beta \widehat{\C}_{jj})/n}}\geq \Phi^{-1}\left(1-\frac{\alpha}{2|\Shat|}\right)\right]\\
&\leq \sum_{j\in \Shat} \PP\left[\max_{\beta\in \mathcal{U}(a)} \frac{|\widehat{\Gamma}_j-\Gammap_j-\beta(\widehat{\gamma}_j-\gammap_j)|}{\sqrt{(\widehat{\V}^{\Gamma}_{jj}+\beta^2\widehat{\V}^{\gamma}_{jj}-2\beta \widehat{\C}_{jj})/n}}\geq \Phi^{-1}\left(1-\frac{\alpha}{2|\Shat|}\right)\right]\leq \alpha.
\end{aligned}
\label{eq: union bound}
\end{equation}
This implies $\liminf_{n\rightarrow \infty}\PP(\mathcal{E}_0(\alpha))\geq 1-\alpha.$ 

Note that 
$
\widehat{\Gamma}_j-\beta\widehat{\gamma}_j-\pip_j=\widehat{\Gamma}_j-\Gammap_j-\beta(\widehat{\gamma}_j-\gammap_j)-(\betap-\beta) \gammap_j.$
Similarly to \eqref{eq: separation key}, we have 
\begin{equation*}
\begin{aligned}
&\frac{\widehat{\Gamma}_j-\beta\widehat{\gamma}_j-\pip_j}{\sqrt{\widehat{\bf R}_{jj}(\beta)/n}}-\frac{\widehat{\Gamma}_j-\Gammap_j-\betap(\widehat{\gamma}_j-\gammap_j)}{\sqrt{{\bf R}_{jj}(\betap)/n}}\\
&=  \frac{\widehat{\Gamma}_j-\Gammap_j-\betap(\widehat{\gamma}_j-\gammap_j)}{\sqrt{\widehat{\bf R}_{jj}(\betap)/n}}\cdot\left(1-\frac{\sqrt{\widehat{\bf R}_{jj}(\beta)/n}}{\sqrt{{\bf R}_{jj}(\betap)/n}}\right)-\frac{(\beta-\betap)(\widehat{\gamma}_j-\gammap_j)}{\sqrt{\widehat{\bf R}_{jj}(\beta)/n}}+\frac{(\beta-\betap)\gammap_j}{\sqrt{\widehat{\bf R}_{jj}(\beta)/n}}.
\end{aligned}
\end{equation*}
Applying the similar argument as \eqref{eq: approximation bound} and \eqref{eq: union bound}, we establish 
$\liminf_{n\rightarrow \infty}\PP(\widetilde{\mathcal{E}}_0(\alpha))\geq 1-\alpha$ since $n^{1/2-a}\ll \Phi^{-1}\left(1-\frac{\alpha}{2|\Shat|}\right).$

\subsection{Proof of Lemma \ref{lem: diff normal}}
\label{sec: good event 3 proof}
We shall establish the lower bound for ${\rm SE}(\widehat{\pi}_{k}^{[j]})$ in \eqref{eq: lower SE}. 
\begin{equation}
\RR^{[j]}_{k,k}+\frac{[\gammap_k]^2}{[\gammap_j]^2}\RR^{[j]}_{j,j}-\frac{2\gammap_k}{\gammap_j}\RR^{[j]}_{j,k}\geq \lambda_{\min}\left(\RR^{[j]}_{\mathcal{J},\mathcal{J}}\right)\left(1+\frac{[\gammap_k]^2}{[\gammap_j]^2}\right),
\label{eq: 1st lower}
\end{equation}
where $\mathcal{J}=\{j,k\}.$
For any $\|u\|_2=1$ and $u\in \R^2$, we have 
$$u^{\intercal}\RR^{[j]}_{\mathcal{J},\mathcal{J}}u=u^{\intercal}\V^{\Gamma}_{\mathcal{J},\mathcal{J}}u+\left(\frac{\Gammap_j}{\gammap_j}\right)^2u^{\intercal}\V^{\gamma}_{\mathcal{J},\mathcal{J}}u-2\frac{\Gammap_j}{\gammap_j} u^{\intercal}\C_{\mathcal{J},\mathcal{J}}u\geq \left[1+\left(\frac{\Gammap_j}{\gammap_j}\right)^2\right]\lambda_{\min}({\rm Cov}).$$
Combined with \eqref{eq: 1st lower} and
\eqref{eq: lower bound on Cov}, we establish \eqref{eq: lower SE}.

We now establish the asymptotic normality of $\mathcal{M}^{[j]}_{k}.$ Note that 
\begin{equation}
\begin{aligned}
&\left(\widehat{\Omega}_{k\cdot} -\frac{\gammab_k}{\gammab_j} \widehat{\Omega}_{j\cdot} \right)^{\intercal}\frac{1}{n}W^{\intercal} \left(\epsilon-\frac{\Gammap_j}{\gammap_j} \delta\right)=\left({\Omega}_{k,\cdot} -\frac{\gammab_k}{\gammab_j} {\Omega}_{j,\cdot} \right)^{\intercal}\frac{1}{n}W^{\intercal} \left(\epsilon-\frac{\Gammap_j}{\gammap_j} \delta\right)\\
&+\left(\widehat{\Omega}_{k\cdot}-{\Omega}_{k,\cdot} \right)^{\intercal}\frac{1}{n}W^{\intercal} \left(\epsilon-\frac{\Gammap_j}{\gammab_j}\delta\right)- \left(\widehat{\Omega}_{j\cdot}-{\Omega}_{j,\cdot} \right)^{\intercal}\frac{1}{n}W^{\intercal} \left(\epsilon-\frac{\Gammap_j}{\gammap_j} \delta\right)\cdot \frac{\gammap_k}{\gammap_j}.
\end{aligned}
\label{eq: diff decomp}
\end{equation}
On the event $\mathcal{G},$ we have 
\begin{equation*}
\begin{aligned}
&\left|\left(\widehat{\Omega}_{k\cdot}-{\Omega}_{k,\cdot} \right)^{\intercal}\frac{1}{n}W^{\intercal} \left(\epsilon-\frac{\Gammap_j}{\gammab_j}\delta\right)+ \frac{\gammab_k}{\gammab_j}\left(\widehat{\Omega}_{j\cdot}-{\Omega}_{j,\cdot} \right)^{\intercal}\frac{1}{n}W^{\intercal} \left(\epsilon-\frac{\Gammap_j}{\gammap_j} \delta\right)\right|\\
&\lesssim {\frac{\log n}{n}}\left(1+\left|\frac{\gammab_k}{\gammab_j}\right|\right)\left(1+\left|\frac{\Gammap_j}{\gammab_j}\right|\right).
\end{aligned}
\end{equation*}
Combined with \eqref{eq: lower SE}, we have 
\begin{equation*}
\begin{aligned}
\frac{1}{{\rm SE}(\widehat{\pi}_{k}^{[j]})}\left|\left(\widehat{\Omega}_{k\cdot}-{\Omega}_{k,\cdot} \right)^{\intercal}\frac{1}{n}W^{\intercal} \left(\epsilon-\frac{\Gammap_j}{\gammab_j}\delta\right)+ \frac{\gammab_k}{\gammab_j}\left(\widehat{\Omega}_{j\cdot}-{\Omega}_{j,\cdot} \right)^{\intercal}\frac{1}{n}W^{\intercal} \left(\epsilon-\frac{\Gammap_j}{\gammap_j} \delta\right)\right|\lesssim {\frac{\log n}{\sqrt{n}}}.
\end{aligned}
\end{equation*}
Together with $\frac{1}{{\rm SE}(\widehat{\pi}_{k}^{[j]})}\left({\Omega}_{k,\cdot} -\frac{\gammab_k}{\gammab_j} {\Omega}_{j,\cdot} \right)^{\intercal}\frac{1}{n}W^{\intercal} \left(\epsilon-\frac{\Gammap_j}{\gammap_j} \delta\right)\cid N(0,1),$
we apply the decomposition \eqref{eq: diff decomp} and establish \eqref{eq: diff normal}.

We now establish the consistency of $\widehat{\rm SE}(\widehat{\pi}_{k}^{[j]}).$
We control the difference between $\widehat{\RR}^{[j]}$ and $\RR^{[j]}=\V^{\Gamma}+(\beta^{[j]})^2\V^{\gamma}-2\beta^{[j]}\C.$
On the event $\mathcal{G}_4,$ we have 
\begin{equation}
\max_{j,k\in \Shat}\left|\widehat{\RR}^{[j]}_{j,k}-{\RR}^{[j]}_{j,k}\right|\lesssim \frac{(\log n)^{3/2}}{\sqrt{n}}\left((\beta^{[j]})^2+1\right)+\left|\widehat{\beta}^{[j]}-\beta^{[j]}\right|\cdot \left(1+\left|\beta^{[j]}\right|\right)+ (\widehat{\beta}^{[j]}-\beta^{[j]})^2.
\label{eq: R error}
\end{equation}
Then we have the following upper bound for $n\cdot \left|\widehat{\rm SE}^2(\widehat{\pi}_{k}^{[j]})-{\rm SE}^2(\widehat{\pi}_{k}^{[j]})\right|$
\begin{equation}
C\max_{j,k\in \Shat}\left|\widehat{\RR}^{[j]}_{j,k}-{\RR}^{[j]}_{j,k}\right|\cdot\left(1+\left(\frac{\widehat{\gamma}_k}{\widehat{\gamma}_j}\right)^2\right)+C\max_{j,k\in \Shat}\left|{\RR}^{[j]}_{j,k}\right|\cdot \left(\left|\left(\frac{\widehat{\gamma}_k}{\widehat{\gamma}_j}\right)^2-\left(\frac{\gammap_k}{\gammap_j}\right)^2\right|+\left|\frac{\widehat{\gamma}_k}{\widehat{\gamma}_j}-\frac{\gammap_k}{\gammap_j}\right|\right).
\label{eq: SE error decomp}
\end{equation}
On the event $\mathcal{G}_6\cap\mathcal{G}_7$, we apply \eqref{eq: R error} 
and establish 
\begin{equation}
\max_{j,k\in \Shat}\left|\widehat{\RR}^{[j]}_{j,k}-{\RR}^{[j]}_{j,k}\right|\cdot\left(1+\left(\frac{\widehat{\gamma}_k}{\widehat{\gamma}_j}\right)^2\right)\lesssim \left(1+\left(\frac{\Gammap_j}{\gammap_j}\right)^2\right)\cdot \left(1+\left(\frac{\gammap_k}{\gammap_j}\right)^2\right)\frac{1}{(\log n)^{1/4}}.
\label{eq: se bound part 1}
\end{equation}

Note that 
\begin{equation*}
\begin{aligned}
\RR^{[j]}_{j,k}=u^{\intercal}\V^{\Gamma}_{j,k}+\left(\frac{\Gammap_j}{\gammap_j}\right)^2\V^{\gamma}_{j,k}-2\frac{\Gammap_j}{\gammap_j} \C_{j,k}&\leq \left[1+\left(\frac{\Gammap_j}{\gammap_j}\right)^2\right]\lambda_{\max}({\rm Cov})\\
&\leq \left[1+\left(\frac{\Gammap_j}{\gammap_j}\right)^2\right]C_1\cdot \lambda_{\max}(\Sigma^{-1}),
\end{aligned}
\end{equation*}
where the least inequality follows from \eqref{eq: upper bound on Cov}. On the event $\mathcal{G}_6,$ we have 
\begin{equation*}
\max_{j,k\in \Shat}\left|{\RR}^{[j]}_{j,k}\right|\cdot \left(\left|\left(\frac{\widehat{\gamma}_k}{\widehat{\gamma}_j}\right)^2-\left(\frac{\gammap_k}{\gammap_j}\right)^2\right|+\left|\frac{\widehat{\gamma}_k}{\widehat{\gamma}_j}-\frac{\gammap_k}{\gammap_j}\right|\right)\lesssim \left[1+\left(\frac{\Gammap_j}{\gammap_j}\right)^2\right]\left(1+\left(\frac{\gammap_k}{\gammap_j}\right)^2\right)\frac{1}{(\log n)^{1/4}}.
\end{equation*}
Together with the error decomposition \eqref{eq: SE error decomp} and the inequalities \eqref{eq: se bound part 1} and \eqref{eq: lower SE}, we establish \eqref{eq: SE bound}.
\section{Additional Numerical Results}
\label{sec: sim supplement}

\subsection{Additional Simulation Results for Settings {\bf S1} to {\bf S5}}
\label{sec: extra S settings}
We present the extra simulation results for settings {\bf S1} to {\bf S5}. In Table \ref{tab: CI homo tau0.4}, we report the coverage and length properties of different CIs for Settings {\bf S1} to {\bf S5} with $\tau=0.4.$ 

We also consider the heteroscedastic errors demonstrate the robustness of our method. For $1\leq i\leq n,$ generate $\delta_i \sim N(0,1)$ and  
\begin{equation}
e_i=0.3 \delta_i+\sqrt{{[1-0.3^2]}/[0.86^4+1.38072^2]}(1.38072\cdot \tau_{1,i}+0.86^2\cdot \tau_{2,i}),
\label{eq: hetero}
\end{equation}
where conditioning on $Z_i$, $\tau_{1,i}\sim N(0, [0.5\cdot Z_{i,1}^2+0.25]^2),$ $\tau_{2,i} \sim N(0,1),$ and $\tau_{1,i}$ and $\tau_{2,i}$ are independent of $\delta_i.$ In Table \ref{tab: CI hetero 0.2}, we report the coverage and length properties of different CIs for Settings {\bf S1} to {\bf S5} with heteroscedastic errors and $\tau=0.2.$ In Table \ref{tab: CI hetero 0.4}, we report the coverage and length properties of different CIs for Settings {\bf S1} to {\bf S5} with heteroscedastic errors and $\tau=0.4.$

The main observation is similar to those in Section \ref{sec: simulation} in the main paper, which is summarized in the following.  

\begin{enumerate}
\item The CIs by \texttt{TSHT} \citep{guo2018confidence} and \texttt{CIIV} \citep{windmeijer2019confidence} achieve the 95\% coverage level for a large sample size and a relatively large violation level, such as $n=5000$ and $\tau= 0.4.$ For many settings with $\tau= 0.2$, the CIs by \texttt{TSHT} and \texttt{CIIV} do not even have coverage when $n=5000.$ The CI by \texttt{CIIV} is more robust in the sense that its validity may require a smaller sample size than \texttt{TSHT}. 
\item The CIs by the \texttt{Union} method \citep{kang2020two} with $\bar{s}=\pz-1$ (assuming there are two valid IVs) achieve the desired coverage levels for all settings. The CIs by the \texttt{Union} method with $\bar{s}=\lceil \pz/2\rceil$ (assuming the majority rule) do not achieve the desired coverage level, except for the setting {\bf S1} where the majority rule holds.
\item Among all CIs achieving the desired coverage level, the sampling CIs are typically the shortest CIs achieving the desired coverage levels. Both searching and sampling CIs are in general shorter than the CIs by the \texttt{Union} method. The sampling CI can have a comparable length to the oracle bias-aware confidence interval.
\end{enumerate}
\begin{table}[H]
\centering
\resizebox{0.85\linewidth}{!}{
\begin{tabular}[t]{|c|c|c c|c c|c c|c c|c|c c|}
\multicolumn{13}{c}{Empirical Coverage of Confidence Intervals for $\tau=0.4$} \\
\hline
\multicolumn{2}{|c|}{ } & \multicolumn{2}{c|}{Oracle} & \multicolumn{2}{c|}{ } & \multicolumn{2}{c|}{Searching} & \multicolumn{2}{c|}{Sampling} & \multicolumn{1}{c|}{ } & \multicolumn{2}{c|}{Union} \\
\hline
Set & n & \texttt{TSLS} & \texttt{BA} & \texttt{TSHT} & \texttt{CIIV} & $\VTSHT$ & $\VCIIV$ & $\VTSHT$ & $\VCIIV$ & Check & $p_z-1$ & $\lceil p_z/2 \rceil$\\
\hline
 & 500 & 0.95 & 0.95 & 0.79 & 0.82 & 1.00 & 1.00 & 1.00 & 1.00 & 1.00 & 1.00 & 0.97\\
 & 1000 & 0.94 & 0.95 & 0.89 & 0.89 & 1.00 & 1.00 & 1.00 & 1.00 & 1.00 & 1.00 & 0.95\\
 & 2000 & 0.96 & 0.96 & 0.96 & 0.95 & 1.00 & 1.00 & 1.00 & 1.00 & 1.00 & 1.00 & 0.95\\
\multirow{-4}{*}{\centering\arraybackslash {\bf S1}} & 5000 & 0.94 & 0.95 & 0.94 & 0.94 & 1.00 & 1.00 & 1.00 & 1.00 & 1.00 & 1.00 & 0.96\\
\cline{1-13}
 & 500 & 0.95 & 0.95 & 0.72 & 0.72 & 0.97 & 0.98 & 0.96 & 0.96 & 0.98 & 1.00 & 0.00\\
 & 1000 & 0.94 & 0.93 & 0.79 & 0.87 & 0.98 & 1.00 & 0.98 & 0.99 & 0.99 & 1.00 & 0.00\\
 & 2000 & 0.94 & 0.96 & 0.92 & 0.93 & 1.00 & 1.00 & 1.00 & 1.00 & 1.00 & 1.00 & 0.00\\
\multirow{-4}{*}{\centering\arraybackslash {\bf S2}} & 5000 & 0.96 & 0.95 & 0.96 & 0.95 & 1.00 & 1.00 & 1.00 & 1.00 & 1.00 & 1.00 & 0.00\\
\cline{1-13}
 & 500 & 0.95 & 0.93 & 0.67 & 0.71 & 0.90 & 0.96 & 0.91 & 0.94 & 0.97 & 1.00 & 0.26\\
 & 1000 & 0.94 & 0.93 & 0.71 & 0.83 & 0.94 & 0.99 & 0.94 & 0.98 & 0.94 & 1.00 & 0.01\\
 & 2000 & 0.94 & 0.93 & 0.88 & 0.91 & 0.99 & 1.00 & 0.99 & 1.00 & 1.00 & 1.00 & 0.00\\
\multirow{-4}{*}{\centering\arraybackslash {\bf S3}} & 5000 & 0.96 & 0.98 & 0.96 & 0.95 & 1.00 & 1.00 & 1.00 & 1.00 & 1.00 & 1.00 & 0.00\\
\cline{1-13}
 & 500 & 0.95 & 0.96 & 0.53 & 0.53 & 0.90 & 0.83 & 0.89 & 0.81 & 0.75 & 0.95 & 0.00\\
 & 1000 & 0.94 & 0.91 & 0.64 & 0.84 & 0.98 & 0.93 & 0.97 & 0.92 & 0.75 & 0.94 & 0.00\\
 & 2000 & 0.93 & 0.95 & 0.88 & 0.92 & 0.98 & 0.98 & 0.97 & 0.96 & 0.94 & 0.94 & 0.00\\
\multirow{-4}{*}{\centering\arraybackslash {\bf S4}} & 5000 & 0.95 & 0.99 & 0.94 & 0.94 & 0.98 & 0.98 & 0.98 & 0.98 & 0.99 & 0.94 & 0.00\\
\cline{1-13}
 & 500 & 0.95 & 0.96 & 0.19 & 0.45 & 0.59 & 0.74 & 0.61 & 0.73 & 0.61 & 0.95 & 0.00\\
 & 1000 & 0.94 & 0.89 & 0.30 & 0.77 & 0.96 & 0.85 & 0.96 & 0.84 & 0.44 & 0.94 & 0.00\\
 & 2000 & 0.93 & 0.94 & 0.74 & 0.91 & 0.99 & 0.98 & 0.97 & 0.97 & 0.78 & 0.94 & 0.00\\
\multirow{-4}{*}{\centering\arraybackslash {\bf S5}} & 5000 & 0.95 & 0.99 & 0.94 & 0.94 & 0.98 & 0.98 & 0.98 & 0.98 & 0.99 & 0.94 & 0.00\\
\hline
\multicolumn{13}{c}{}\\
\multicolumn{13}{c}{Average Lengths of Confidence Intervals for $\tau=0.4$} \\
\hline
\multicolumn{2}{|c|}{ } & \multicolumn{2}{c|}{Oracle} & \multicolumn{2}{c|}{ } & \multicolumn{2}{c|}{Searching} & \multicolumn{2}{c|}{Sampling} & \multicolumn{1}{c|}{ } & \multicolumn{2}{c|}{\texttt{Union}} \\
\hline
Set & n & \texttt{TSLS} & \texttt{BA} & \texttt{TSHT} & \texttt{CIIV} & $\VTSHT$ & $\VCIIV$ & $\VTSHT$ & $\VCIIV$ & Check & $p_z-1$ & $\lceil p_z/2 \rceil$\\

\cline{1-13}
 & 500 & 0.10 & 0.16 & 0.09 & 0.09 & 0.57 & 0.62 & 0.33 & 0.35 & - & 1.28 & 0.17\\
 & 1000 & 0.07 & 0.10 & 0.07 & 0.07 & 0.40 & 0.41 & 0.22 & 0.23 & - & 0.70 & 0.08\\
 & 2000 & 0.05 & 0.05 & 0.05 & 0.05 & 0.29 & 0.29 & 0.16 & 0.16 & - & 0.41 & 0.05\\
\multirow{-4}{*}{\centering\arraybackslash {\bf S1}} & 5000 & 0.03 & 0.03 & 0.03 & 0.03 & 0.18 & 0.18 & 0.10 & 0.10 & - & 0.25 & 0.03\\
\cline{1-13}
 & 500 & 0.13 & 0.48 & 0.27 & 0.12 & 0.56 & 0.59 & 0.41 & 0.37 & - & 2.58 & 0.00\\
 & 1000 & 0.09 & 0.33 & 0.16 & 0.09 & 0.37 & 0.39 & 0.23 & 0.24 & - & 1.49 & 0.00\\
 & 2000 & 0.06 & 0.09 & 0.07 & 0.07 & 0.27 & 0.27 & 0.16 & 0.16 & - & 0.73 & 0.00\\
\multirow{-4}{*}{\centering\arraybackslash {\bf S2}} & 5000 & 0.04 & 0.04 & 0.04 & 0.04 & 0.17 & 0.17 & 0.10 & 0.10 & - & 0.26 & 0.00\\
\cline{1-13}
 & 500 & 0.13 & 0.53 & 0.21 & 0.13 & 0.61 & 0.64 & 0.53 & 0.41 & - & 1.90 & 0.04\\
 & 1000 & 0.09 & 0.44 & 0.21 & 0.09 & 0.42 & 0.41 & 0.31 & 0.26 & - & 1.40 & 0.00\\
 & 2000 & 0.06 & 0.24 & 0.14 & 0.07 & 0.27 & 0.27 & 0.17 & 0.16 & - & 0.83 & 0.00\\
\multirow{-4}{*}{\centering\arraybackslash {\bf S3}} & 5000 & 0.04 & 0.05 & 0.04 & 0.04 & 0.17 & 0.17 & 0.10 & 0.10 & - & 0.33 & 0.00\\
\cline{1-13}
 & 500 & 0.23 & 1.35 & 0.31 & 0.24 & 0.86 & 0.54 & 0.79 & 0.40 & - & 0.80 & 0.00\\
 & 1000 & 0.16 & 1.14 & 0.18 & 0.16 & 0.64 & 0.32 & 0.58 & 0.23 & - & 0.34 & 0.00\\
 & 2000 & 0.11 & 0.46 & 0.11 & 0.11 & 0.27 & 0.21 & 0.22 & 0.16 & - & 0.14 & 0.00\\
\multirow{-4}{*}{\centering\arraybackslash {\bf S4}} & 5000 & 0.07 & 0.12 & 0.07 & 0.07 & 0.14 & 0.14 & 0.11 & 0.11 & - & 0.08 & 0.00\\
\cline{1-13}
 & 500 & 0.23 & 1.05 & 0.24 & 0.23 & 0.70 & 0.52 & 0.69 & 0.39 & - & 0.98 & 0.00\\
 & 1000 & 0.16 & 1.01 & 0.14 & 0.16 & 0.83 & 0.31 & 0.79 & 0.23 & - & 0.40 & 0.00\\
 & 2000 & 0.11 & 0.52 & 0.11 & 0.11 & 0.34 & 0.21 & 0.30 & 0.16 & - & 0.15 & 0.00\\
\multirow{-4}{*}{\centering\arraybackslash {\bf S5}} & 5000 & 0.07 & 0.14 & 0.07 & 0.07 & 0.14 & 0.14 & 0.11 & 0.10 & - & 0.08 & 0.00\\
\hline
\end{tabular}}
\caption{Settings {\bf S1} to {\bf S5} with $\tau=0.4$ and homoscadastic errors. The columns indexed with \texttt{TSLS}, \texttt{BA} \texttt{TSHT} and \texttt{CIIV} represent the oracle TSLS CI with the knowledge of $\mathcal{V}$, the oracle bias-aware confidence interval, the CI by \texttt{TSHT}, and the CI by \texttt{CIIV}, respectively. Under the columns indexed with ``Searching'' (or ``Sampling''), the columns indexed with $\VTSHT$ and $\VCIIV$ represent our proposed searching (or sampling) CI in Algorithm \ref{algo: USS-plurality} with $\VTSHT$ and $\VCIIV$, respectively. The column indexed with ``Check" reports the proportion of simulations passing the Plurality rule check in Algorithm \ref{algo: USS-plurality}. The columns indexed with \texttt{Union} represent the union of TSLS estimators, which pass the Sargan test. The columns indexed with $\pz-1$ and $\lceil \pz/2\rceil$ correspond to the \texttt{Union} methods assuming two valid IVs and the majority rule, respectively.}
\label{tab: CI homo tau0.4}
\end{table}


\begin{table}[H]
\centering
\resizebox{0.8\linewidth}{!}{
\begin{tabular}[t]{|c|c|c c|c c|c c|c c|c|}
\multicolumn{11}{c}{Empirical Coverage of Confidence Intervals for $\tau=0.2$}\\
\hline
\multicolumn{2}{|c|}{ } & \multicolumn{2}{c|}{Oracle} & \multicolumn{2}{c|}{ } & \multicolumn{2}{c|}{Searching} & \multicolumn{2}{c|}{Sampling} & \multicolumn{1}{c|}{ } \\
\hline
Set & n & \texttt{TSLS} & \texttt{BA} & \texttt{TSHT} & \texttt{CIIV} & $\VTSHT$ & $\VCIIV$ & $\VTSHT$ & $\VCIIV$ & Check\\
\hline
 & 500 & 0.92 & 0.96 & 0.49 & 0.51 & 1.00 & 1.00 & 1.00 & 1.00 & 1.00\\
 & 1000 & 0.94 & 0.98 & 0.38 & 0.61 & 1.00 & 1.00 & 1.00 & 0.99 & 1.00\\
 & 2000 & 0.95 & 0.95 & 0.52 & 0.75 & 1.00 & 1.00 & 1.00 & 1.00 & 1.00\\
\multirow{-4}{*}{\centering\arraybackslash {\bf S1}} & 5000 & 0.94 & 0.94 & 0.83 & 0.90 & 1.00 & 1.00 & 1.00 & 1.00 & 1.00\\
\cline{1-11}
 & 500 & 0.95 & 0.95 & 0.58 & 0.41 & 1.00 & 1.00 & 0.99 & 1.00 & 1.00\\
 & 1000 & 0.92 & 0.95 & 0.47 & 0.45 & 1.00 & 1.00 & 0.98 & 0.98 & 1.00\\
 & 2000 & 0.94 & 0.94 & 0.52 & 0.61 & 0.99 & 0.97 & 0.99 & 0.94 & 1.00\\
\multirow{-4}{*}{\centering\arraybackslash {\bf S2}} & 5000 & 0.94 & 0.94 & 0.76 & 0.86 & 0.98 & 0.99 & 0.99 & 0.99 & 1.00\\
\cline{1-11}
 & 500 & 0.95 & 0.96 & 0.65 & 0.52 & 0.99 & 0.99 & 0.98 & 0.98 & 1.00\\
 & 1000 & 0.92 & 0.96 & 0.57 & 0.46 & 0.98 & 1.00 & 0.97 & 0.97 & 1.00\\
 & 2000 & 0.94 & 0.95 & 0.60 & 0.61 & 0.98 & 0.97 & 0.97 & 0.94 & 1.00\\
\multirow{-4}{*}{\centering\arraybackslash {\bf S3}} & 5000 & 0.94 & 0.95 & 0.76 & 0.86 & 0.98 & 0.99 & 0.99 & 0.99 & 1.00\\
\cline{1-11}
 & 500 & 0.94 & 0.96 & 0.68 & 0.63 & 0.97 & 0.97 & 0.97 & 0.95 & 0.99\\
 & 1000 & 0.96 & 0.99 & 0.54 & 0.45 & 0.98 & 0.94 & 0.98 & 0.94 & 0.99\\
 & 2000 & 0.94 & 0.98 & 0.48 & 0.42 & 0.99 & 0.93 & 0.97 & 0.92 & 1.00\\
\multirow{-4}{*}{\centering\arraybackslash {\bf S4}} & 5000 & 0.94 & 0.97 & 0.72 & 0.75 & 0.97 & 0.93 & 0.96 & 0.92 & 1.00\\
\cline{1-11}
 & 500 & 0.94 & 0.97 & 0.47 & 0.37 & 0.83 & 0.86 & 0.84 & 0.81 & 0.98\\
 & 1000 & 0.96 & 0.97 & 0.33 & 0.33 & 0.55 & 0.75 & 0.67 & 0.73 & 0.98\\
 & 2000 & 0.94 & 0.97 & 0.19 & 0.36 & 0.51 & 0.85 & 0.61 & 0.84 & 0.89\\
\multirow{-4}{*}{\centering\arraybackslash {\bf S5}} & 5000 & 0.94 & 0.95 & 0.57 & 0.75 & 0.97 & 0.92 & 0.96 & 0.92 & 0.84\\
\hline
\multicolumn{11}{c}{}\\
\multicolumn{11}{c}{Average Lengths of Confidence Intervals for $\tau=0.2$}\\
\hline
\multicolumn{2}{|c|}{ } & \multicolumn{2}{c|}{Oracle} & \multicolumn{2}{c|}{ } & \multicolumn{2}{c|}{Searching} & \multicolumn{2}{c|}{Sampling} & \multicolumn{1}{c|}{ } \\
\hline
Set & n & \texttt{TSLS} & \texttt{BA} & \texttt{TSHT} & \texttt{CIIV} & $\VTSHT$ & $\VCIIV$ & $\VTSHT$ & $\VCIIV$ & Check\\
\hline
\cline{1-11}
 & 500 & 0.10 & 0.18 & 0.09 & 0.09 & 0.63 & 0.65 & 0.37 & 0.37 & -\\
 & 1000 & 0.08 & 0.16 & 0.06 & 0.07 & 0.44 & 0.47 & 0.27 & 0.28 & -\\
 & 2000 & 0.05 & 0.13 & 0.05 & 0.05 & 0.31 & 0.33 & 0.20 & 0.20 & -\\
\multirow{-4}{*}{\centering\arraybackslash {\bf S1}} & 5000 & 0.03 & 0.05 & 0.03 & 0.03 & 0.19 & 0.20 & 0.11 & 0.11 & -\\
\cline{1-11}
 & 500 & 0.15 & 0.28 & 0.17 & 0.11 & 0.64 & 0.65 & 0.41 & 0.41 & -\\
 & 1000 & 0.11 & 0.27 & 0.15 & 0.09 & 0.43 & 0.47 & 0.30 & 0.31 & -\\
 & 2000 & 0.08 & 0.27 & 0.16 & 0.07 & 0.30 & 0.33 & 0.24 & 0.23 & -\\
\multirow{-4}{*}{\centering\arraybackslash {\bf S2}} & 5000 & 0.05 & 0.21 & 0.13 & 0.05 & 0.18 & 0.20 & 0.13 & 0.13 & -\\
\cline{1-11}
 & 500 & 0.15 & 0.27 & 0.13 & 0.11 & 0.63 & 0.68 & 0.45 & 0.42 & -\\
 & 1000 & 0.11 & 0.25 & 0.13 & 0.09 & 0.43 & 0.46 & 0.32 & 0.31 & -\\
 & 2000 & 0.08 & 0.26 & 0.15 & 0.07 & 0.31 & 0.33 & 0.24 & 0.23 & -\\
\multirow{-4}{*}{\centering\arraybackslash {\bf S3}} & 5000 & 0.05 & 0.21 & 0.13 & 0.05 & 0.18 & 0.20 & 0.13 & 0.13 & -\\
\cline{1-11}
 & 500 & 0.32 & 0.60 & 0.31 & 0.21 & 0.69 & 0.70 & 0.61 & 0.59 & -\\
 & 1000 & 0.23 & 0.40 & 0.20 & 0.16 & 0.51 & 0.49 & 0.45 & 0.42 & -\\
 & 2000 & 0.16 & 0.30 & 0.17 & 0.13 & 0.36 & 0.32 & 0.32 & 0.28 & -\\
\multirow{-4}{*}{\centering\arraybackslash {\bf S4}} & 5000 & 0.10 & 0.22 & 0.13 & 0.10 & 0.21 & 0.18 & 0.19 & 0.15 & -\\
\cline{1-11}
 & 500 & 0.32 & 0.58 & 0.32 & 0.20 & 0.54 & 0.61 & 0.50 & 0.51 & -\\
 & 1000 & 0.23 & 0.57 & 0.27 & 0.16 & 0.34 & 0.46 & 0.36 & 0.40 & -\\
 & 2000 & 0.16 & 0.49 & 0.18 & 0.13 & 0.27 & 0.32 & 0.28 & 0.28 & -\\
\multirow{-4}{*}{\centering\arraybackslash {\bf S5}} & 5000 & 0.10 & 0.31 & 0.12 & 0.10 & 0.25 & 0.18 & 0.24 & 0.15 & -\\
\hline
\end{tabular}}
\caption{Settings {\bf S1} to {\bf S5} with $\tau=0.2$ and heteroscedastic errors. The columns indexed with \texttt{TSLS}, \texttt{BA} \texttt{TSHT} and \texttt{CIIV} represent the oracle TSLS CI with the knowledge of $\mathcal{V}$, the oracle bias-aware confidence interval, the CI by \texttt{TSHT}, and the CI by \texttt{CIIV}, respectively. Under the columns indexed with ``Searching'' (or ``Sampling''), the columns indexed with $\VTSHT$ and $\VCIIV$ represent our proposed searching (or sampling) CI in Algorithm \ref{algo: USS-plurality} with $\VTSHT$ and $\VCIIV$, respectively. The column indexed with ``Check" reports the proportion of simulations passing the Plurality rule check in Algorithm \ref{algo: USS-plurality}. The columns indexed with \texttt{Union} represent the union of TSLS estimators, which pass the Sargan test. The columns indexed with $\pz-1$ and $\lceil \pz/2\rceil$ correspond to the \texttt{Union} methods assuming two valid IVs and the majority rule, respectively.}
\label{tab: CI hetero 0.2}
\end{table}

\begin{table}[H]
\centering
\resizebox{0.8\linewidth}{!}{
\begin{tabular}[t]{|c|c|c c|c c|c c|c c|c|}
\multicolumn{11}{c}{Empirical Coverage of Confidence Intervals for $\tau=0.4$}\\
\hline
\multicolumn{2}{|c|}{ } & \multicolumn{2}{c|}{Oracle} & \multicolumn{2}{c|}{ } & \multicolumn{2}{c|}{Searching} & \multicolumn{2}{c|}{Sampling} & \multicolumn{1}{c|}{ } \\
\hline
Set & n & \texttt{TSLS} & \texttt{BA} & \texttt{TSHT} & \texttt{CIIV} & $\VTSHT$ & $\VCIIV$ & $\VTSHT$ & $\VCIIV$ & Check\\
\hline
 & 500 & 0.92 & 0.94 & 0.63 & 0.72 & 1.00 & 1.00 & 1.00 & 1.00 & 1.00\\
 & 1000 & 0.94 & 0.94 & 0.80 & 0.86 & 1.00 & 1.00 & 1.00 & 1.00 & 1.00\\
 & 2000 & 0.95 & 0.95 & 0.94 & 0.95 & 1.00 & 1.00 & 1.00 & 1.00 & 1.00\\
\multirow{-4}{*}{\centering\arraybackslash {\bf S1}} & 5000 & 0.94 & 0.94 & 0.94 & 0.93 & 1.00 & 1.00 & 1.00 & 1.00 & 1.00\\
\cline{1-11}
 & 500 & 0.95 & 0.96 & 0.68 & 0.60 & 0.97 & 0.98 & 0.97 & 0.96 & 1.00\\
 & 1000 & 0.92 & 0.95 & 0.77 & 0.78 & 0.98 & 0.98 & 0.98 & 0.97 & 1.00\\
 & 2000 & 0.94 & 0.92 & 0.88 & 0.94 & 0.99 & 1.00 & 0.99 & 1.00 & 1.00\\
\multirow{-4}{*}{\centering\arraybackslash {\bf S2}} & 5000 & 0.94 & 0.97 & 0.93 & 0.94 & 1.00 & 1.00 & 1.00 & 1.00 & 1.00\\
\cline{1-11}
 & 500 & 0.95 & 0.94 & 0.72 & 0.60 & 0.92 & 0.97 & 0.92 & 0.95 & 0.97\\
 & 1000 & 0.92 & 0.94 & 0.82 & 0.72 & 0.89 & 0.98 & 0.89 & 0.97 & 0.99\\
 & 2000 & 0.94 & 0.93 & 0.86 & 0.93 & 0.98 & 1.00 & 0.98 & 1.00 & 0.99\\
\multirow{-4}{*}{\centering\arraybackslash {\bf S3}} & 5000 & 0.94 & 0.97 & 0.92 & 0.94 & 1.00 & 1.00 & 1.00 & 1.00 & 1.00\\
\cline{1-11}
 & 500 & 0.94 & 0.93 & 0.52 & 0.38 & 0.91 & 0.86 & 0.91 & 0.84 & 0.92\\
 & 1000 & 0.96 & 0.92 & 0.69 & 0.70 & 0.98 & 0.93 & 0.97 & 0.91 & 0.92\\
 & 2000 & 0.94 & 0.98 & 0.89 & 0.91 & 0.98 & 0.96 & 0.97 & 0.96 & 0.98\\
\multirow{-4}{*}{\centering\arraybackslash {\bf S4}} & 5000 & 0.94 & 0.97 & 0.93 & 0.93 & 0.97 & 0.97 & 0.98 & 0.97 & 1.00\\
\cline{1-11}
 & 500 & 0.94 & 0.99 & 0.28 & 0.21 & 0.38 & 0.60 & 0.49 & 0.58 & 0.91\\
 & 1000 & 0.96 & 1.00 & 0.17 & 0.52 & 0.55 & 0.73 & 0.57 & 0.72 & 0.64\\
 & 2000 & 0.94 & 0.98 & 0.33 & 0.84 & 0.97 & 0.89 & 0.96 & 0.89 & 0.40\\
\multirow{-4}{*}{\centering\arraybackslash {\bf S5}} & 5000 & 0.94 & 0.83 & 0.77 & 0.93 & 0.98 & 0.97 & 0.98 & 0.97 & 0.82\\
\hline
\multicolumn{11}{c}{Average Lengths of Confidence Intervals for $\tau=0.4$}\\
\hline
\multicolumn{2}{|c|}{ } & \multicolumn{2}{c|}{Oracle} & \multicolumn{2}{c|}{ } & \multicolumn{2}{c|}{Searching} & \multicolumn{2}{c|}{Sampling} & \multicolumn{1}{c|}{ } \\
\hline
Set & n & \texttt{TSLS} & \texttt{BA} & \texttt{TSHT} & \texttt{CIIV} & $\VTSHT$ & $\VCIIV$ & $\VTSHT$ & $\VCIIV$ & Check\\
\hline
\cline{1-11}
 & 500 & 0.10 & 0.23 & 0.10 & 0.10 & 0.62 & 0.68 & 0.39 & 0.39 & -\\
 & 1000 & 0.08 & 0.13 & 0.08 & 0.07 & 0.42 & 0.45 & 0.25 & 0.26 & -\\
 & 2000 & 0.05 & 0.06 & 0.06 & 0.05 & 0.30 & 0.31 & 0.17 & 0.18 & -\\
\multirow{-4}{*}{\centering\arraybackslash {\bf S1}} & 5000 & 0.03 & 0.04 & 0.04 & 0.03 & 0.20 & 0.20 & 0.11 & 0.11 & -\\
\cline{1-11}
 & 500 & 0.15 & 0.51 & 0.34 & 0.14 & 0.62 & 0.68 & 0.49 & 0.45 & -\\
 & 1000 & 0.11 & 0.44 & 0.29 & 0.11 & 0.41 & 0.45 & 0.31 & 0.29 & -\\
 & 2000 & 0.08 & 0.26 & 0.14 & 0.08 & 0.29 & 0.31 & 0.19 & 0.20 & -\\
\multirow{-4}{*}{\centering\arraybackslash {\bf S2}} & 5000 & 0.05 & 0.06 & 0.05 & 0.05 & 0.19 & 0.19 & 0.12 & 0.12 & -\\
\cline{1-11}
 & 500 & 0.15 & 0.47 & 0.26 & 0.13 & 0.61 & 0.68 & 0.54 & 0.46 & -\\
 & 1000 & 0.11 & 0.43 & 0.30 & 0.11 & 0.41 & 0.46 & 0.34 & 0.31 & -\\
 & 2000 & 0.08 & 0.33 & 0.20 & 0.08 & 0.30 & 0.31 & 0.20 & 0.20 & -\\
\multirow{-4}{*}{\centering\arraybackslash {\bf S3}} & 5000 & 0.05 & 0.09 & 0.06 & 0.05 & 0.19 & 0.19 & 0.12 & 0.12 & -\\
\cline{1-11}
 & 500 & 0.32 & 0.99 & 0.37 & 0.26 & 0.79 & 0.68 & 0.72 & 0.57 & -\\
 & 1000 & 0.23 & 0.92 & 0.29 & 0.21 & 0.59 & 0.41 & 0.53 & 0.35 & -\\
 & 2000 & 0.16 & 0.40 & 0.18 & 0.16 & 0.30 & 0.27 & 0.26 & 0.22 & -\\
\multirow{-4}{*}{\centering\arraybackslash {\bf S4}} & 5000 & 0.10 & 0.13 & 0.10 & 0.10 & 0.17 & 0.17 & 0.14 & 0.14 & -\\
\cline{1-11}
 & 500 & 0.32 & 1.04 & 0.37 & 0.25 & 0.52 & 0.64 & 0.58 & 0.56 & -\\
 & 1000 & 0.23 & 0.97 & 0.23 & 0.21 & 0.55 & 0.42 & 0.57 & 0.35 & -\\
 & 2000 & 0.16 & 0.78 & 0.14 & 0.16 & 0.57 & 0.27 & 0.55 & 0.22 & -\\
\multirow{-4}{*}{\centering\arraybackslash {\bf S5}} & 5000 & 0.10 & 0.41 & 0.10 & 0.10 & 0.25 & 0.17 & 0.23 & 0.14 & -\\
\hline
\end{tabular}}
\caption{Settings {\bf S1} to {\bf S5} with $\tau=0.4$ and heteroscedastic errors. The columns indexed with \texttt{TSLS}, \texttt{BA} \texttt{TSHT} and \texttt{CIIV} represent the oracle TSLS CI with the knowledge of $\mathcal{V}$, the oracle bias-aware confidence interval, the CI by \texttt{TSHT}, and the CI by \texttt{CIIV}, respectively. Under the columns indexed with ``Searching'' (or ``Sampling''), the columns indexed with $\VTSHT$ and $\VCIIV$ represent our proposed searching (or sampling) CI in Algorithm \ref{algo: USS-plurality} with $\VTSHT$ and $\VCIIV$, respectively. The column indexed with ``Check" reports the proportion of simulations passing the Plurality rule check in Algorithm \ref{algo: USS-plurality}. The columns indexed with \texttt{Union} represent the union of TSLS estimators, which pass the Sargan test. The columns indexed with $\pz-1$ and $\lceil \pz/2\rceil$ correspond to the \texttt{Union} methods assuming two valid IVs and the majority rule, respectively.}
\label{tab: CI hetero 0.4}
\end{table}

\begin{figure}[H]
    \centering
    \includegraphics[scale=0.9]{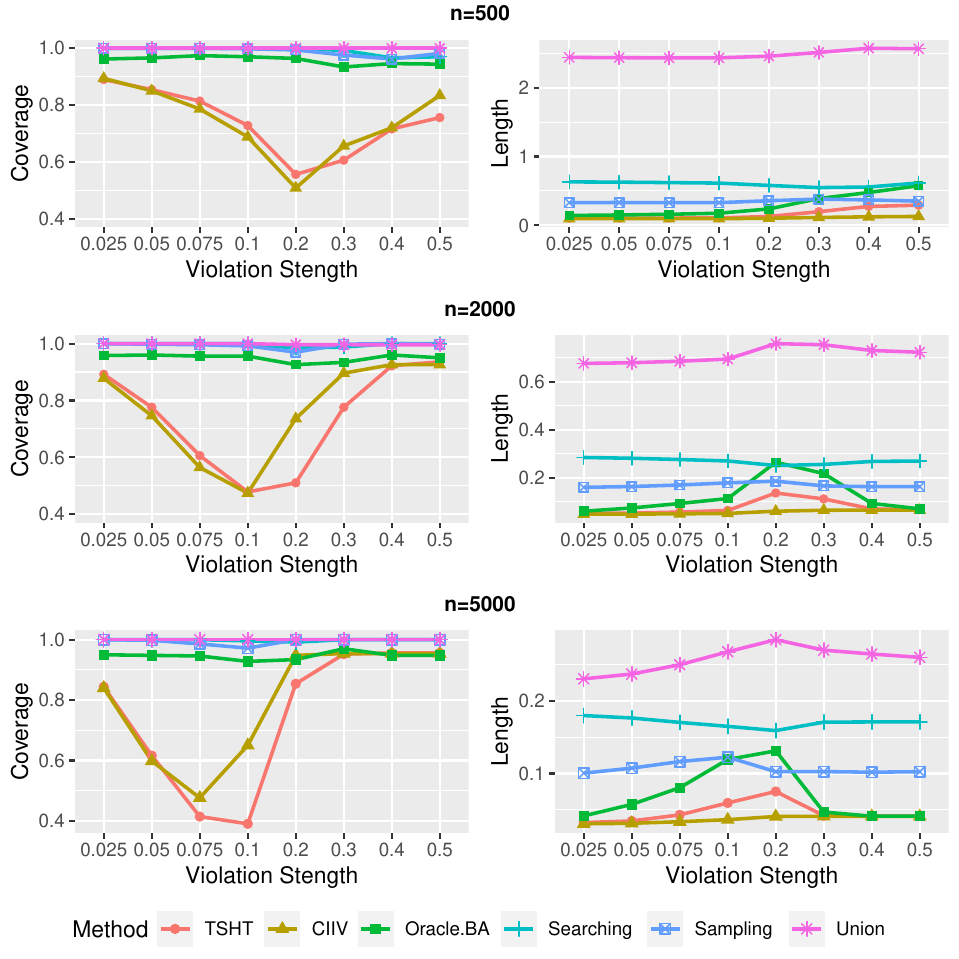}
     \caption{Empirical coverage and average lengths for the setting {\bf S2} with $\tau\in\{0.025,0.05,0.075,0.1,0.2,0.3,0.4,0.5\}$ and homoscadastic errors. $\TSHT$,  $\CIIV$, and \texttt{Union} stand for the CIs by \citet{guo2018confidence}, \citet{windmeijer2019confidence}, and \citet{kang2020two}, respectively. \texttt{Oracle BA} stands for the oracle bias-aware CI in \eqref{eq: bias aware}, which is the benchmark. \texttt{Searching} and \texttt{Sampling} correspond to our proposed searching and sampling CIs in Algorithm \ref{algo: USS-plurality} in the main paper.} 
    \label{fig: varying tau homo}
\end{figure}

In Figure \ref{fig: varying tau homo}, we investigate the performance of our proposed methods by varying the violation strength. We present the results for homoscadastic regression errors with different $\tau,$ which is the counterpart of the results reported in Figure \ref{fig: varying tau hetero} in the main paper.
We generate the errors $(e_i,\delta_i)^{\intercal}$ following bivariate normal with zero mean, unit variance and ${\rm Cov}(\epsilon_i,\delta_i)=0.8.$

{
\subsection{Simulations with violation levels scaled to $\sqrt{\log n/n}$}
\label{sec: local vio sim}
We consider the setting {\bf S2} with homoscadastic errors and set $\tau=({\rm VF}/\gamma_0)\cdot \sqrt{\log n/n}$ with $\text{Violation Factor}{\rm (VF)}$ varying across $\{0.05,0.1,0.2,0.3,0.4,0.5,0.8,1,2,3,4,5\}.$ For $n=500, 2000, 5000,$ we observe that \texttt{TSHT} and \texttt{CIIV} does not have coverage for $0.1\leq {\rm VF}\leq 2$. For ${\rm VF}=0.05$, even though the IV selection error exists, the bias due to including the locally invalid IVs is relatively small, which does not affect the coverage. When ${\rm VF}\geq 3,$ both \texttt{TSHT} and \texttt{CIIV} can  correctly separate valid and invalid IVs. 
\begin{figure}[H]
    \centering
    \includegraphics[scale=0.85]{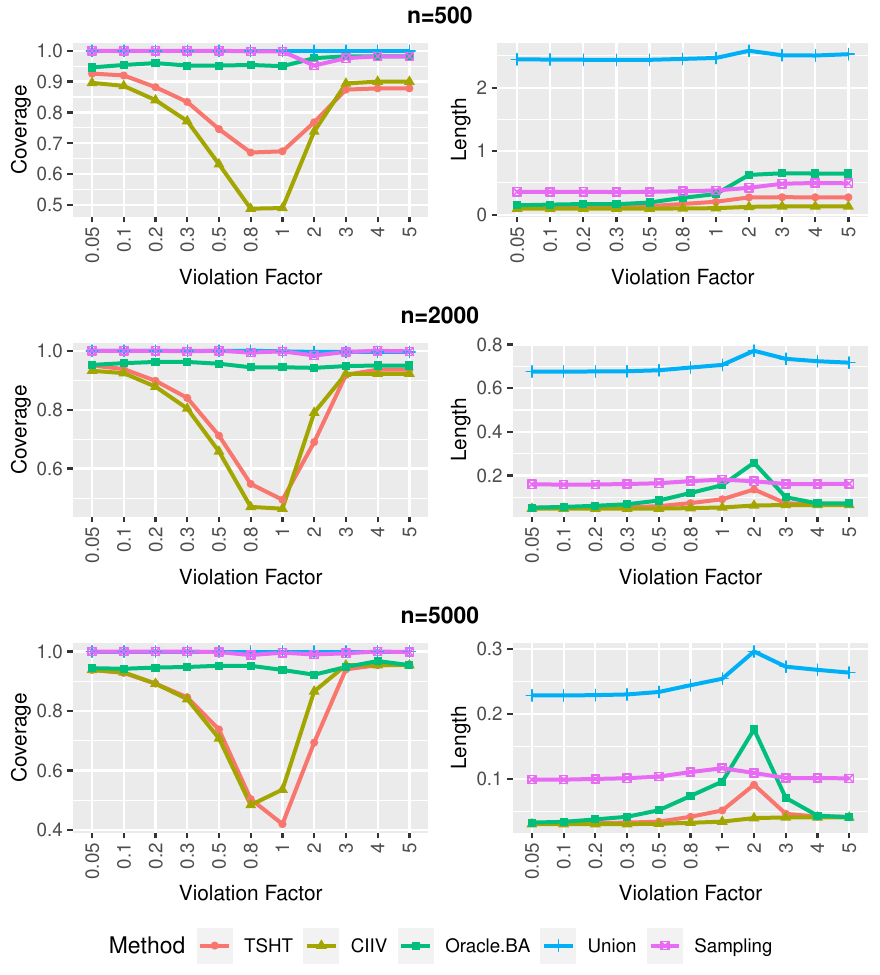}
     \caption{{Empirical coverage and average lengths for the setting {\bf S2} with homoscadastic errors and $\tau=({\rm VF}/\gamma_0)\cdot \sqrt{\log n/n}$ where $\text{Violation Factor}{\rm (VF)}$ is varied across $\{0.05,0.1,0.2,0.3,0.4,0.5,0.8,1,2,3,4,5\}.$ $\TSHT$,  $\CIIV$, and \texttt{Union} stand for the CIs by \citet{guo2018confidence}, \citet{windmeijer2019confidence}, and \citet{kang2020two}, respectively. \texttt{Oracle BA} stands for the oracle bias-aware CI in \eqref{eq: bias aware}, which is the benchmark. \texttt{Searching} and \texttt{Sampling} correspond to our proposed searching and sampling CIs in Algorithm \ref{algo: USS-plurality} in the main paper.} } 
    \label{fig: varying tau depending on n}
\end{figure}
}

\subsection{CIIV setting}
\label{sec: CIIV setting}

Similar to the setting in CIIV paper \citep{windmeijer2019confidence}, we further consider the following settings.
\begin{enumerate}
\item[] {\bf CIIV-1} (Plurality rule): set 
 $\gammap= 0.4\cdot{\bf 1}_{21}$ and  $\pip=({\bf 0}_{9},\tau \cdot {\bf 1}_{6},\frac{\tau}{2}\cdot{\bf 1}_{6})^{\intercal}$. 
\item[] {\bf CIIV-2} (Plurality rule): set $\gammap= 0.4\cdot{\bf 1}_{21}$ and  $\pip=({\bf 0}_{9},\tau \cdot {\bf 1}_{3},-\tau \cdot {\bf 1}_{3},\frac{\tau}{2}\cdot{\bf 1}_{3},-\frac{\tau}{2}\cdot{\bf 1}_{3})^{\intercal}$. 
\end{enumerate}
We vary $\tau$ across $\{0.2, 0.4\}$ where $\tau$ represents the invalidity level. The setting {\bf CIIV-1} with $\tau=0.4$ corresponds to the exact setting considered in \citet{windmeijer2019confidence}. For a small $\tau$ and sample size $n$, the setting {\bf CIIV-1} does not necessarily satisfy the finite-sample plurality rule (Condition \ref{cond: plurality-finite}) since $\tau$ and $\tau/2$ are close to each other for a small $\tau>0$. For the setting {\bf CIIV-2}, the invalidity levels are more spread out and the finite-sample plurality rule may hold more plausibly.

\begin{table}[htp!]
\centering
\scalebox{0.8}{
\begin{tabular}[t]{|c c|c c|c c|c c|c c|c | c|}
\multicolumn{12}{c}{Empirical Coverage of Confidence Intervals For {\bf CIIV-1}}\\
\hline
\multicolumn{2}{|c|}{ } & \multicolumn{2}{c|}{Oracle} & \multicolumn{2}{c|}{ } & \multicolumn{2}{c|}{Searching} & \multicolumn{2}{c|}{Sampling} & \multicolumn{1}{c|}{ } & \multicolumn{1}{c|}{\texttt{Union}} \\
\hline
$\tau$ & n & \texttt{TSLS} & \texttt{BA} & \texttt{TSHT} & \texttt{CIIV} & $\VTSHT$ & $\VCIIV$ & $\VTSHT$ & $\VCIIV$ & Check & $p_z-1$\\
\hline
 & 500 & 0.94 & 0.96 & 0.00 & 0.16 & 1.00 & 0.99 & 0.92 & 0.62 & 1.00 & 1.00\\
 & 1000 & 0.93 & 0.98 & 0.00 & 0.42 & 1.00 & 0.91 & 0.94 & 0.69 & 1.00 & 1.00\\
 & 2000 & 0.95 & 0.93 & 0.00 & 0.72 & 0.58 & 0.94 & 0.58 & 0.93 & 1.00 & 1.00\\
\multirow{-4}{*}{\centering\arraybackslash 0.2} & 5000 & 0.95 & 0.94 & 0.01 & 0.91 & 0.66 & 1.00 & 0.66 & 1.00 & 0.80 & 1.00\\
\cline{1-12}
 & 500 & 0.94 & 0.95 & 0.00 & 0.69 & 0.60 & 0.90 & 0.60 & 0.87 & 1.00 & 1.00\\
 & 1000 & 0.93 & 0.95 & 0.00 & 0.89 & 0.36 & 0.98 & 0.36 & 0.98 & 0.98 & 1.00\\
 & 2000 & 0.95 & 0.90 & 0.40 & 0.94 & 0.98 & 1.00 & 0.98 & 1.00 & 0.90 & 1.00\\
\multirow{-4}{*}{\centering\arraybackslash 0.4} & 5000 & 0.95 & 0.95 & 0.95 & 0.95 & 1.00 & 1.00 & 1.00 & 1.00 & 1.00 & 1.00\\
\hline
\multicolumn{12}{c}{} \\
\multicolumn{12}{c}{Average Lengths of Confidence Intervals for {\bf CIIV-1}}\\
\cline{1-12}
\multicolumn{2}{|c|}{ } & \multicolumn{2}{c|}{Oracle} & \multicolumn{2}{c|}{ } & \multicolumn{2}{c|}{Searching} & \multicolumn{2}{c|}{Sampling} & \multicolumn{1}{c|}{ } & \multicolumn{1}{c|}{\texttt{Union}} \\
\hline
$\tau$ & n & \texttt{TSLS} & \texttt{BA} & \texttt{TSHT} & \texttt{CIIV} & $\VTSHT$ & $\VCIIV$ & $\VTSHT$ & $\VCIIV$ & Check & $p_z-1$\\
\hline
 & 500 & 0.09 & 0.50 & 0.06 & 0.09 & 0.90 & 0.88 & 0.52 & 0.40 & - & 1.44\\
 & 1000 & 0.07 & 0.51 & 0.04 & 0.07 & 0.60 & 0.60 & 0.47 & 0.29 & - & 1.10\\
 & 2000 & 0.05 & 0.49 & 0.03 & 0.05 & 0.39 & 0.42 & 0.39 & 0.21 & - & 0.90\\
\multirow{-4}{*}{\centering\arraybackslash 0.2} & 5000 & 0.03 & 0.51 & 0.05 & 0.03 & 0.18 & 0.27 & 0.12 & 0.13 & - & 0.73\\
\cline{1-12}
 & 500 & 0.09 & 0.96 & 0.06 & 0.10 & 0.88 & 0.90 & 0.88 & 0.42 & - & 2.08\\
 & 1000 & 0.07 & 1.02 & 0.06 & 0.07 & 0.39 & 0.61 & 0.34 & 0.28 & - & 1.68\\
 & 2000 & 0.05 & 1.08 & 0.20 & 0.05 & 0.40 & 0.43 & 0.20 & 0.21 & - & 1.25\\
\multirow{-4}{*}{\centering\arraybackslash 0.4} & 5000 & 0.03 & 0.03 & 0.03 & 0.03 & 0.27 & 0.27 & 0.13 & 0.13 & - & 0.80\\
\hline
\end{tabular}
}
\caption{The setting {\bf CIIV-1}. The columns indexed with \texttt{TSLS}, \texttt{BA} \texttt{TSHT} and \texttt{CIIV} represent the oracle TSLS CI with the knowledge of $\mathcal{V}$, the oracle bias-aware confidence interval, the CI by \texttt{TSHT}, and the CI by \texttt{CIIV}, respectively. Under the columns indexed with ``Searching'' (or ``Sampling''), the columns indexed with $\VTSHT$ and $\VCIIV$ represent our proposed searching (or sampling) CI in Algorithm \ref{algo: USS-plurality} with $\VTSHT$ and $\VCIIV$, respectively. The column indexed with ``Check" reports the proportion of simulations passing the Plurality rule check in Algorithm \ref{algo: USS-plurality}. The column indexed with $\pz-1$ corresponds to the \texttt{Union} methods assuming two valid IVs.} 
\label{tab: CIIV repre}
\end{table}

In Table \ref{tab: CIIV repre}, we consider the setting {\bf CIIV-1} and compare different CIs in terms of empirical coverage and average lengths. In terms of coverage, our proposed searching and sampling CIs attain the desired coverage level (95\%); CIs by the \texttt{Union} method achieve the desired coverage level; \texttt{CIIV} achieves the desired 95\% coverage level for $\tau=0.2$ with $n=5000$ and $\tau=0.4$ with $n=2000,5000$; and \texttt{TSHT} achieves the desired 95\% coverage level only for $\tau=0.4$ with $n=5000.$ We shall point out that the searching and sampling CIs with inaccurate initial estimators $\VTSHT$ tend to perform badly in terms of coverage, see the settings with $\tau=0.2$ and $n=5000$ or $\tau=0.4$ and $n=2000.$ The corresponding searching and sampling CIs with the initial estimators $\VCIIV$ are more reliable in these settings.

In terms of interval lengths, the sampling and searching CIs are much shorter than the CIs by the \texttt{Union} method. The sampling CI can have a comparable length with the oracle bias-aware confidence interval. 

The results for the setting {\bf CIIV-2} is reported in Table \ref{tab: CIIV2}.
\begin{table}[htp!]
\centering
\scalebox{0.8}{
\begin{tabular}[t]{|c c|c c|c c|c c|c c|c | c|}
\multicolumn{12}{c}{Empirical Coverage of Confidence Intervals For {\bf CIIV-2}}\\
\hline
\multicolumn{2}{|c|}{ } & \multicolumn{2}{c|}{Oracle} & \multicolumn{2}{c|}{ } & \multicolumn{2}{c|}{Searching} & \multicolumn{2}{c|}{Sampling} & \multicolumn{1}{c|}{ } & \multicolumn{1}{c|}{\texttt{Union}} \\
\hline
$\tau$ & n & \texttt{TSLS} & \texttt{BA} & \texttt{TSHT} & \texttt{CIIV} & $\VTSHT$ & $\VCIIV$ & $\VTSHT$ & $\VCIIV$ & Check & $p_z-1$\\
\hline
 & 500 & 0.94 & 0.94 & 0.60 & 0.55 & 1.00 & 1.00 & 1.00 & 0.97 & 1.00 & 1.00\\
 & 1000 & 0.93 & 0.94 & 0.55 & 0.80 & 1.00 & 1.00 & 1.00 & 0.98 & 1.00 & 1.00\\
 & 2000 & 0.95 & 0.95 & 0.60 & 0.84 & 0.95 & 1.00 & 0.95 & 1.00 & 1.00 & 1.00\\
\multirow{-4}{*}{\centering\arraybackslash 0.2} & 5000 & 0.95 & 0.93 & 0.88 & 0.91 & 1.00 & 1.00 & 1.00 & 1.00 & 0.98 & 1.00\\
\cline{1-12}
 & 500 & 0.94 & 0.95 & 0.59 & 0.83 & 0.96 & 1.00 & 0.96 & 1.00 & 1.00 & 1.00\\
 & 1000 & 0.93 & 0.94 & 0.70 & 0.90 & 0.96 & 1.00 & 0.96 & 1.00 & 0.98 & 1.00\\
 & 2000 & 0.95 & 0.98 & 0.71 & 0.94 & 1.00 & 1.00 & 1.00 & 1.00 & 1.00 & 1.00\\
\multirow{-4}{*}{\centering\arraybackslash 0.4} & 5000 & 0.95 & 0.96 & 0.95 & 0.95 & 1.00 & 1.00 & 1.00 & 1.00 & 1.00 & 1.00\\
\hline
\multicolumn{12}{c}{} \\
\multicolumn{12}{c}{Average Lengths of Confidence Intervals for {\bf CIIV-2}}\\
\cline{1-12}
\multicolumn{2}{|c|}{ } & \multicolumn{2}{c|}{Oracle} & \multicolumn{2}{c|}{ } & \multicolumn{2}{c|}{Searching} & \multicolumn{2}{c|}{Sampling} & \multicolumn{1}{c|}{ } & \multicolumn{1}{c|}{\texttt{Union}} \\
\hline
$\tau$ & n & \texttt{TSLS} & \texttt{BA} & \texttt{TSHT} & \texttt{CIIV} & $\VTSHT$ & $\VCIIV$ & $\VTSHT$ & $\VCIIV$ & Check & $p_z-1$\\
\hline
 & 500 & 0.09 & 0.14 & 0.06 & 0.09 & 0.94 & 0.92 & 0.55 & 0.43 & - & 1.78\\
 & 1000 & 0.07 & 0.14 & 0.05 & 0.07 & 0.60 & 0.61 & 0.50 & 0.30 & - & 1.46\\
 & 2000 & 0.05 & 0.10 & 0.04 & 0.05 & 0.30 & 0.42 & 0.30 & 0.21 & - & 1.30\\
\multirow{-4}{*}{\centering\arraybackslash 0.2} & 5000 & 0.03 & 0.09 & 0.06 & 0.03 & 0.23 & 0.27 & 0.13 & 0.13 & - & 1.04\\
\cline{1-12}
 & 500 & 0.09 & 0.18 & 0.08 & 0.09 & 0.72 & 0.91 & 0.70 & 0.42 & - & 2.91\\
 & 1000 & 0.07 & 0.15 & 0.07 & 0.07 & 0.52 & 0.61 & 0.34 & 0.29 & - & 2.48\\
 & 2000 & 0.05 & 0.41 & 0.21 & 0.05 & 0.40 & 0.43 & 0.19 & 0.21 & - & 1.81\\
\multirow{-4}{*}{\centering\arraybackslash 0.4} & 5000 & 0.03 & 0.03 & 0.03 & 0.03 & 0.27 & 0.27 & 0.13 & 0.13 & - & 1.11\\
\hline
\end{tabular}
}
\caption{The setting {\bf CIIV-2}. The columns indexed with \texttt{TSLS}, \texttt{BA} \texttt{TSHT} and \texttt{CIIV} represent the oracle TSLS CI with the knowledge of $\mathcal{V}$, the oracle bias-aware confidence interval, the CI by \texttt{TSHT}, and the CI by \texttt{CIIV}, respectively. Under the columns indexed with ``Searching'' (or ``Sampling''), the columns indexed with $\VTSHT$ and $\VCIIV$ represent our proposed searching (or sampling) CI in Algorithm \ref{algo: USS-plurality} with $\VTSHT$ and $\VCIIV$, respectively. The column indexed with ``Check" reports the proportion of simulations passing the Plurality rule check in Algorithm \ref{algo: USS-plurality}. The column indexed with $\pz-1$ corresponds to the \texttt{Union} methods assuming two valid IVs.} 
\label{tab: CIIV2}
\end{table}

\subsection{High dimensional IVs and covariates}
\label{sec: high dim sim}
\begin{figure}[H]
    \centering
    \includegraphics[width=0.8\textwidth]{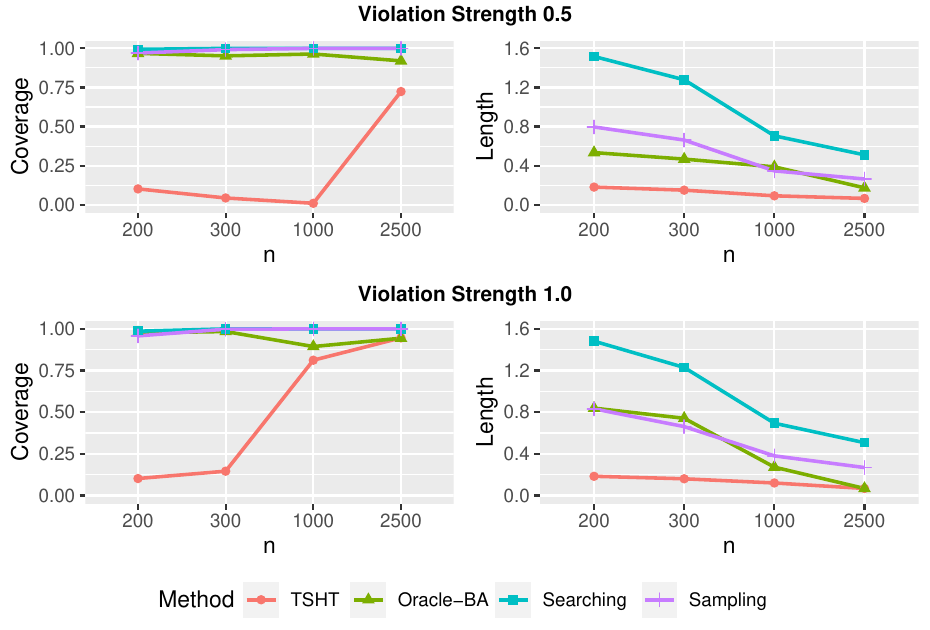}
    \caption{Empirical coverage and average lengths for high-dimensional setting with $\pz=200$ and $\px=150.$ We vary the violation strength $\tau$ across $\{0.5,1\}.$  \texttt{Oracle-BA} and \texttt{TSHT} represent the oracle bias-aware CI in \eqref{eq: bias aware} and the CI by  \citet{guo2018confidence}, respectively. The searching and sampling CIs are implemented as in Algorithm \ref{algo: USS-plurality}.}
    \label{fig: high-dim}
\end{figure}

We implement the high-dimensional version of Algorithm \ref{algo: USS-plurality}, as detailed in \eqref{sec: high dim}.
We follow the same high-dimensional setting as in Section 5.3 of  \citet{guo2018confidence}. Particularly, we set $p_x = 150, p_z = 200$ and generate the sparse vector $\gammap = 0.5\cdot ({\bf 1}^{\intercal}_7, {\bf 0}^{\intercal}_{\pz-7})^{\intercal}$, $\pip = ({\bf 0}^{\intercal}_5, \tau/2, \tau/2, {\bf 0}^{\intercal}_{\pz-7})^{\intercal}.$ That is, $\mathcal{S}=\{1,2,\cdots,7\}$, $\mathcal{V}=\{1,2,\cdots,5\}$, and the $6$-th and $7$-th IVs are invalid. As reported in Figure \ref{fig: high-dim}, the CIs by \texttt{TSHT} can only achieve the desired coverage for a sufficiently large sample size $n$ and invalidity level $\tau$. In contrast, our proposed searching and sampling CIs achieve the desired coverage across all settings. The sampling CIs are helpful in reducing the length of the searching CIs. The sampling CIs are in general longer than the oracle bias-aware CIs (the benchmark), but the sampling CIs have a comparable length to the oracle bias-aware CIs.

As reported in Table \ref{tab: high dim}, the CIs by \texttt{TSHT} can only achieve the desired coverage for a sufficiently large sample size $n$ and invalidity level $\tau$. In contrast, our proposed searching and sampling CIs achieve the desired coverage across all settings. The sampling CI is helpful in reducing the length of the searching CI. The sampling CI is in general longer than the oracle bias-aware CI (the benchmark), but  the sampling CI can have a comparable length to the oracle bias-aware CI in certain scenarios.
In Table \ref{tab: add HighDim}, we report the high-dimensional simulation results with $\pz=100$ and the results are similar to those in Table \ref{tab: high dim}.
\begin{table}[htp!]
\centering
\resizebox{\linewidth}{!}{
\begin{tabular}[t]{|c c|c c|c c|c c c|c c|c c|}
\hline
\multicolumn{2}{|c|}{ } & \multicolumn{2}{c|}{\texttt{Oracle TSLS}} & \multicolumn{2}{c|}{\texttt{Oracle BA}} & \multicolumn{3}{c|}{\texttt{TSHT}} & \multicolumn{2}{c|}{Searching} & \multicolumn{2}{c|}{Sampling} \\
\hline
$\tau$ & n & Cov & Len & Cov & Len & Bias & Cov & Len & Cov & Len & Cov & Len\\
\hline
 & 200 & 0.96 & 0.26 & 0.97 & 0.53 & 0.17 & 0.10 & 0.18 & 0.99 & 1.52 & 0.97 & 0.80\\
 & 300 & 0.94 & 0.21 & 0.95 & 0.47 & 0.16 & 0.04 & 0.15 & 1.00 & 1.28 & 0.99 & 0.66\\
 & 1000 & 0.95 & 0.11 & 0.96 & 0.39 & 0.15 & 0.01 & 0.09 & 1.00 & 0.71 & 1.00 & 0.35\\
\multirow{-4}{*}{\centering\arraybackslash 0.5} & 2500 & 0.95 & 0.07 & 0.92 & 0.17 & 0.02 & 0.72 & 0.07 & 1.00 & 0.51 & 1.00 & 0.26\\
\cline{1-13}
 & 200 & 0.96 & 0.26 & 0.98 & 0.84 & 0.25 & 0.10 & 0.18 & 0.99 & 1.48 & 0.96 & 0.83\\
 & 300 & 0.94 & 0.21 & 0.98 & 0.74 & 0.20 & 0.15 & 0.16 & 1.00 & 1.23 & 1.00 & 0.66\\
 & 1000 & 0.95 & 0.11 & 0.89 & 0.27 & 0.03 & 0.81 & 0.12 & 1.00 & 0.69 & 1.00 & 0.38\\
\multirow{-4}{*}{\centering\arraybackslash 1.0} & 2500 & 0.95 & 0.07 & 0.94 & 0.07 & 0.00 & 0.95 & 0.07 & 1.00 & 0.51 & 1.00 & 0.27\\
\cline{1-13}
 & 200 & 0.96 & 0.26 & 0.96 & 0.37 & 0.05 & 0.80 & 0.23 & 1.00 & 1.44 & 0.99 & 0.81\\
 & 300 & 0.94 & 0.21 & 0.95 & 0.24 & 0.03 & 0.88 & 0.19 & 1.00 & 1.21 & 1.00 & 0.67\\
 & 1000 & 0.95 & 0.11 & 0.96 & 0.13 & 0.01 & 0.93 & 0.12 & 1.00 & 0.70 & 1.00 & 0.38\\
\multirow{-4}{*}{\centering\arraybackslash 2.0} & 2500 & 0.95 & 0.07 & 0.94 & 0.07 & 0.00 & 0.95 & 0.07 & 1.00 & 0.51 & 1.00 & 0.27\\
\hline
\end{tabular}
}
\caption{High-dimensional setting with $\pz=200$ and $\px=150.$ The columns indexed with ``Cov", ``Len", and ``Bias" denote the empirical coverage, the average lengths, and the absolute bias, respectively. The columns indexed with \texttt{Oracle TSLS}, \texttt{Oracle BA}, and \texttt{TSHT} represent the oracle TSLS CI with the knowledge of $\mathcal{V}$ and relevant IV and covariates, the oracle bias-aware CI in \eqref{eq: bias aware}, the CI by  \citet{guo2018confidence}, respectively. }
\label{tab: high dim}
\end{table}

\begin{table}[htp!]
\centering
\resizebox{\linewidth}{!}{
\begin{tabular}[t]{|c c|c c|c c|c c c|c c|c c|}
\hline
\multicolumn{2}{|c|}{ } & \multicolumn{2}{c|}{\texttt{Oracle TSLS}} & \multicolumn{2}{c|}{\texttt{Oracle BA}} & \multicolumn{3}{c|}{\TSHT} & \multicolumn{2}{c|}{Searching} & \multicolumn{2}{c|}{Sampling} \\
\hline
$\tau$ & n & Cov & Len & Cov & Len & Bias & Cov & Len & Cov & Len & Cov & Len\\
\hline
 & 200 & 0.94 & 0.26 & 0.96 & 0.52 & 0.16 & 0.11 & 0.19 & 1.00 & 1.46 & 0.97 & 0.78\\
 & 300 & 0.95 & 0.21 & 0.97 & 0.48 & 0.15 & 0.05 & 0.15 & 1.00 & 1.30 & 0.99 & 0.69\\
 & 1000 & 0.94 & 0.11 & 0.99 & 0.40 & 0.13 & 0.08 & 0.09 & 1.00 & 0.68 & 1.00 & 0.36\\
\multirow{-4}{*}{\centering\arraybackslash 0.5} & 2500 & 0.94 & 0.07 & 0.92 & 0.12 & 0.01 & 0.85 & 0.07 & 1.00 & 0.48 & 1.00 & 0.26\\
\cline{1-13}
 & 200 & 0.94 & 0.26 & 0.98 & 0.82 & 0.23 & 0.14 & 0.19 & 1.00 & 1.42 & 0.97 & 0.83\\
 & 300 & 0.95 & 0.21 & 0.99 & 0.75 & 0.19 & 0.18 & 0.17 & 1.00 & 1.27 & 1.00 & 0.69\\
 & 1000 & 0.94 & 0.11 & 0.97 & 0.15 & 0.01 & 0.91 & 0.11 & 1.00 & 0.67 & 1.00 & 0.37\\
\multirow{-4}{*}{\centering\arraybackslash 1.0} & 2500 & 0.94 & 0.07 & 0.99 & 0.09 & 0.00 & 0.94 & 0.07 & 1.00 & 0.47 & 1.00 & 0.26\\
\cline{1-13}
 & 200 & 0.94 & 0.26 & 0.96 & 0.38 & 0.04 & 0.81 & 0.24 & 1.00 & 1.40 & 0.99 & 0.78\\
 & 300 & 0.95 & 0.21 & 0.97 & 0.27 & 0.02 & 0.87 & 0.20 & 1.00 & 1.27 & 1.00 & 0.70\\
 & 1000 & 0.94 & 0.11 & 0.96 & 0.12 & 0.01 & 0.93 & 0.11 & 1.00 & 0.68 & 1.00 & 0.37\\
\multirow{-4}{*}{\centering\arraybackslash 2.0} & 2500 & 0.94 & 0.07 & 0.96 & 0.07 & 0.00 & 0.94 & 0.07 & 1.00 & 0.48 & 1.00 & 0.26\\
\hline
\end{tabular}}
\caption{High-dimensional setting with $\pz=100$ and $\px=150.$ The columns indexed with \texttt{Oracle TSLS}, \texttt{Oracle BA}, and \texttt{TSHT} represent the oracle TSLS CI with the knowledge of $\mathcal{V}$ and relevant IV and covariates, the oracle bias-aware CI in \eqref{eq: bias aware}, the CI by  \citet{guo2018confidence}, respectively. }
\label{tab: add HighDim}
\end{table}

\subsection{Extra real data results}
We present the summary statistics of the candidate IVs and baseline covariates in Table \ref{tab: variable summary}. We shall provide some explanations on ``Statement about fairness" and ``Statement about Talent". 
The CFPS survey has a few questions about subjects' view on the fair competition and talent pay-off, which are measures of  subjects' personality traits. The statement on fair competition measures an individual's willingness to compete through ability or education. A strong sense of competing for opportunities through fair competition signals a willingness to improve personal skills and knowledge through education. The statement on the talent indicates if an individual thinks their educational endeavors will pay off. A high score on the pay-off manifests that one would like to strive for higher education because they believe that their education will reward with their income. 

\begin{figure}[H]
\centering
\includegraphics[scale=0.5]{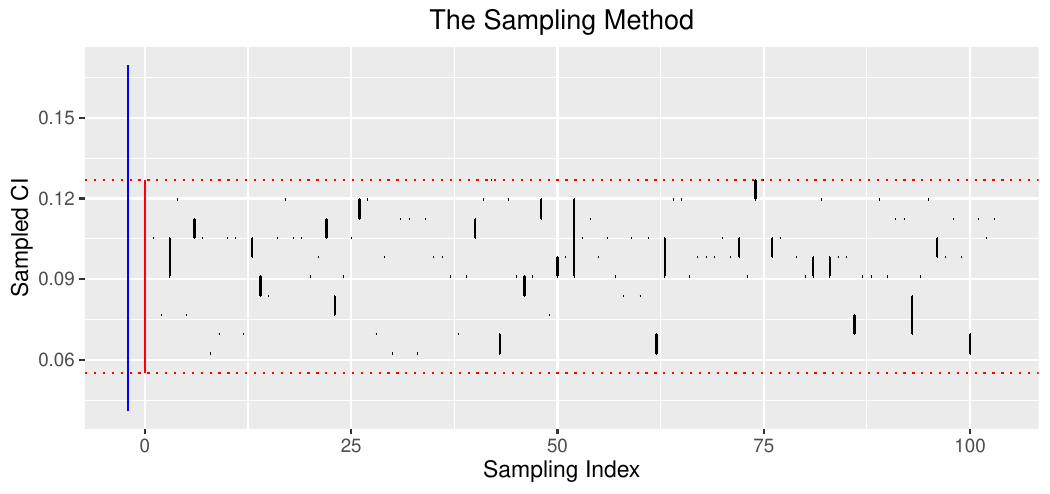}
\vspace{-2mm}
\caption{\small The axis corresponds to sampling indexes $\{1,2,\cdots,103\}$ (after re-ordering) and the y-axis reports the sampled CIs. Along the y-axis, the red interval  is $\CIsample=(0.0552, 0.1268)$ and the blue interval is $\CIsearch=(0.0409, 0.1698)$.}
\label{fig: sampling real}
\end{figure}

We plot the searching CI (in blue) and the sampling CI (in red) in Figure \ref{fig: sampling real}. Out of the 1000 sampled intervals, 103 of them are non-empty and the union of these 103 intervals is shorter than the searching CI.

\vspace{-2mm}

\begin{table}[htp!]
    \centering
    \begin{tabular}{ll}
        Variables & Statistics  \\
        \hline
        {Father's Education Level} &  \\
        \tabindent \tabindent Illiteracy & 1051 (27.97\%) \\
        \tabindent \tabindent Primary School & 1098 (29.22\%) \\
        \tabindent \tabindent Middle School & 996 (26.50\%) \\
        \tabindent \tabindent High School & 487 (12.96\%) \\
        \tabindent \tabindent College and Above & 126 (3.53\%) \\
        {Mother's Education Level} & \\
        \tabindent \tabindent Illiteracy & 1863 (49.57\%) \\
        \tabindent \tabindent Primary School & 948 (25.23\%) \\
        \tabindent \tabindent Middle School & 653 (17.38\%) \\
        \tabindent \tabindent High School & 258 (6.87\%) \\
        \tabindent \tabindent College and Above & 36 (0.96\%) \\
        { Spouse's Education Level} & \\
        \tabindent \tabindent Illiteracy & 347 (9.23\%) \\
        \tabindent \tabindent Primary School & 573 (15.25\%) \\
        \tabindent \tabindent Middle School & 1306 (34.75\%) \\
        \tabindent \tabindent Higher School & 687 (18.28\%) \\
        \tabindent \tabindent College and Above & 845 (22.49\%) \\
        { Group-level Education Years} & Mean (9.45) SD (2.47) \\
        { The Family Size} & Mean (4.30) SD (1.96) \\
        { Statement about Fairness} & \\
        \tabindent \tabindent Strongly Disagree & 49 (1.30\%) \\
        \tabindent \tabindent Disagree & 475 (12.64\%) \\
        \tabindent \tabindent Neutral & 2658 (70.73\%) \\
        \tabindent \tabindent Agree & 489 (13.01\%) \\
        \tabindent \tabindent Strongly Agree & 87 (2.32\%) \\
       { Statement about Talent} & \\
        \tabindent \tabindent Strongly Disagree & 23 (0.61\%) \\
        \tabindent \tabindent Disagree & 494 (13.15\%) \\
        \tabindent \tabindent Neutral & 2683 (71.39\%) \\
        \tabindent \tabindent Agree & 432 (11.50\%) \\
        \tabindent \tabindent Strongly Agree & 126 (3.35\%) \\
        { Reading books in the past 12 months} &\\
        \tabindent \tabindent No & 2396  (63.76\%)\\ 
        \tabindent \tabindent Yes & 1362 (36.24\%) \\
        { Log Education Expenditure in the past 12 months} & Mean (6.20) SD (4.03) \\
        { Urban} &\\
         \tabindent \tabindent Rural & 1400  (37.25\%) \\
        \tabindent \tabindent Urban  & 2358 (62.75\%) \\ 
        { Hukou} &\\
        \tabindent \tabindent Agricultural  &  2445 (65.06\%) \\
        \tabindent \tabindent Non-agricultural   & 1313 (34.94\%) \\ 
        { Gender} & \\
        \tabindent \tabindent Female  &  1425  (37.92\%) \\
        \tabindent \tabindent Male  & 2333  (62.08\%) \\ 
        \hline
    \end{tabular}
    \caption{The summary statistics of instrumental variables and baseline covariates. }
    \label{tab: variable summary}
\end{table}

\end{document}